\documentclass[aps,prd,preprintnumbers,groupedaddress,nofootinbib,amssymb,eqsecnum,notitlepage]{revtex4}
\usepackage{graphicx}
\usepackage{bm}
\usepackage{amsmath}
\usepackage{color}
\usepackage{amsfonts}
\usepackage{here}
\usepackage{graphicx}
\usepackage{amsmath,amsthm,amssymb}
\usepackage{bm}
\usepackage{comment}
\allowdisplaybreaks[1]

\usepackage{amsfonts}
\usepackage{dcolumn}
\usepackage{hyperref}

\begin{document}
\newcommand{\newc}{\newcommand}

\newcommand{\ben}{\begin{eqnarray}}
\newcommand{\een}{\end{eqnarray}}
\newc{\be}{\begin{equation}}
\newc{\ee}{\end{equation}}
\newc{\ba}{\begin{eqnarray}}
\newc{\ea}{\end{eqnarray}}
\newc{\bea}{\begin{eqnarray*}}
\newc{\eea}{\end{eqnarray*}}
\newc{\da}{\delta{A}}
\newc{\D}{\partial}
\newc{\ie}{{\it i.e.} }
\newc{\eg}{{\it e.g.} }
\newc{\etc}{{\it etc.} }
\newc{\etal}{{\it et al.}}
\newcommand{\nn}{\nonumber}
\newc{\ra}{\rightarrow}
\newc{\lra}{\leftrightarrow}
\newc{\lsim}{\buildrel{<}\over{\sim}}
\newc{\gsim}{\buildrel{>}\over{\sim}}
\newc{\aP}{\alpha_{\rm P}}
\newc{\Mpl}{M_{\rm pl}}
\newc{\tb}{\tilde{\beta}}
\newc{\bb}{\bar{\beta}}
\newcommand{\rk}[1]{\textcolor{blue}{#1}}

\title{Even-parity stability of hairy black holes in $U(1)$ gauge-invariant \\
scalar-vector-tensor theories}

\author{Chao Zhang$^{1, 2}$ and Ryotaro Kase$^{3}$}

\affiliation{
$^{1}$Basic Research Center for Energy Interdisciplinary, College of Science, China University of Petroleum-Beijing, Beijing 102249, China.\\
${}^{2}$Beijing Key Laboratory of Optical Detection Technology for Oil and Gas, China University of Petroleum-Beijing, Beijing, 102249, China.\\
$^{3}$
Department of Physics, Faculty of Science, 
Tokyo University of Science, 1-3, Kagurazaka,
Shinjuku-ku, Tokyo 162-8601, Japan.}

\date{\today}

\begin{abstract}
The $U(1)$ gauge-invariant scalar-vector-tensor theories, which catches five degrees of freedom, are valuable for its implications to inflation problems, generation of primordial magnetic fields, new black hole (BH) and neutron star solutions, etc. 
In this paper, we derive conditions for the absence of ghosts and Laplacian instabilities 
of nontrivial BH solutions dressed with scalar hair against both odd- and even-parity perturbations 
on top of the static and spherically symmetric background in the most general $U(1)$ gauge-invariant 
scalar-vector-tensor theories with second-order equations of motion. 
In addition to some general discussions, several typical concrete models are investigated. Specially, 
we show that the stability against even-parity perturbations is ensured outside the event horizon under 
certain constraints to these models. This is a crucial step to check the self-consistency of the theories 
and to shed light on the physically accessible models of such theories for future studies. 
\end{abstract}


\maketitle

\section{Introduction}
\label{introsec}

Over the past decades, general relativity (GR) has passed all the experimental tests with flying colors (see, e.g., Ref.~\cite{Ozel:2016oaf}). From the theoretical point of view, it is quite an elegant, robust and systematic theory to describe the gravity. 
Nonetheless, there are still many crucial open questions left by the framework of GR. For instance, the accelerated expansion of the universe, the origin and inherence of dark matter/energy, the possibility to construct a quantum theory of gravity, etc., 
are some  outstanding ones~\cite{Debono:2016vkp}.
To solve these problems and approach to the nature of gravity, many modified gravitational theories were constructed 
by introducing additional interactions and fields. 
For the purposes of justifying and confining those modified gravitational theories, they have to be 
subjected to experimental tests, including those from the solar system~\cite{Will:2014kxa} as well as strong-field 
regimes, e.g., the gravitational waves (GWs) led by compact celestial bodies~\cite{Tattersall:2018map,Barack:2018yly,Baibhav:2019rsa}.

The detection of the first GW from the coalescence of two massive black holes (BHs) by advanced LIGO/Virgo marked 
the beginning of a new era — the GW astronomy~\cite{LIGOScientific:2016aoc}. Following this observation, more than 
90 GW events have been identified by the LIGO/Virgo/KAGRA (LVK) scientific collaborations (see, e.g., 
Refs.~\cite{LIGOScientific:2018mvr, LIGOScientific:2019lzm, LIGOScientific:2020aai, LIGOScientific:2021djp}). 
In the future, more ground- and space-based GW detectors will be constructed~\cite{Moore:2014lga}, which will enable us 
to probe signals with a much wider frequency band and larger distances. As a result, more types of GW sources will be realized, 
including those from a final remnant BH~\cite{Shi:2019hqa}.  
BH is one of the most mysterious phenomena in the universe. The existence of BHs provides us a perfect 
way to test gravitational effects under extremely strong gravity as we pursued and mentioned above. On the other hand, from 
the theoretical point of view, BHs are also unique laboratories to test the deviation of modified theories of gravity from GR. 
The outbreak of interest on BHs has further gained momenta after the detection of the shadow of the supermassive compact
objects in the center of galaxy  M87 and Sagittarius~A* (Sgr~A*) in the center of the Milky Way galaxy, 
which are the most likely black hole candidates, with the Event Horizon Telescope (EHT)~\cite{EventHorizonTelescope:2019dse,EventHorizonTelescope:2019ggy,EventHorizonTelescope:2021srq, EventHorizonTelescope:2022xqj}.

The development of the GW astronomy as well as the interest on BHs have triggered the 
interest in the quasi-normal mode (QNM) of BHs, as GWs emitted in the ringdown phase 
can be considered as the linear combination of these individual modes~\cite{Berti:2018vdi}. 
From the classical point of view, QNMs are eigenmodes of dissipative systems. 
The information contained in QNMs provides the keys to revealing whether BHs are ubiquitous in our universe, 
and more importantly whether GR is the correct theory to describe gravity even in the strong field regime~\cite{Berti:2018cxi}. 
In addition to the observational purposes, QNM is also an important indicator of the stability of a specific 
spacetime~\cite{Lin:2016wci, Zhang:2023kzs}. Under certain circumstance, the results from QNMs will exhibit a manifest 
consistency with those from the Lagrangian-based stability analysis~\cite{Tsujikawa:2021typ, Zhang:2022fbz}. 

In GR, according to the no-hair theorem, an isolated and stationary BH is completely
characterized by only three quantities, mass, angular momentum, and electric charge. 
Astrophysically, we expect BHs to be electrically neutral, so they are uniquely described by the Kerr solution.
Nonetheless, in theories that beyond GR, the existence of additional degrees of freedom (DOFs) can give rise to 
new hairs to the field configuration and spacetime metric~\cite{Heisenberg:2018mgr}.  The theories containing 
a scalar field $\phi$ coupled to gravity besides two tensor polarizations arising from the gravity sector are 
dubbed scalar-tensor theories. In particular, Horndeski constructed most general scalar-tensor theories with 
second-order equations of motion~\cite{Horndeski:1974wa, Deffayet:2011gz, Kobayashi:2011nu}. 
On the other hand, for a vector field coupled to gravity, it is known that 
generalized Proca theories are the most general vector-tensor theories with second-order equations of motion~
\cite{Heisenberg:2014rta, Tasinato:2014eka, Tasinato:2014mia} (see also Refs.~\cite{ErrastiDiez:2019trb,ErrastiDiez:2019ttn}). 
These two important classes of field theories, Horndeski and generalized Proca theories, can be unified 
in the framework of scalar-vector-tensor (SVT) theories with second-order equations of motion~\cite{Heisenberg:2018acv}. 
The SVT theories can be classified into two cases depending on whether they respect the $U(1)$ gauge symmetry or not. 
In the presence of $U(1)$ gauge symmetry the longitudinal component of a vector field vanishes, so that the propagating DOFs are 
five in total (one scalar, two transverse vectors, two tensor polarizations). 
The breaking of $U(1)$ gauge symmetry leads to the propagation of the longitudinal scalar besides the five DOFs. 
Although a canonical scalar field in the Einstein-Maxwell theories cannot have non-trivial profile 
under the assumption of static and spherically symmetric configuration, i.e.,   
the no-hair theorem~\cite{Chase:1970omy, Bekenstein:1971hc, Bekenstein:1996pn} holds in the theories, 
this is not the case in SVT theories due to the existence of the coupling between scalar and vector DOFs. 
Indeed, one can construct a hairy, static and spherically symmetric BH solutions in $U(1)$ gauge-invariant 
SVT theories in which the scalar field can possess nontrivial profile\cite{Heisenberg:2018vti,Ikeda:2019okp}. 

In this paper, we study the stability of static and spherically symmetric BHs in the $U(1)$ gauge-invariant (GI) SVT theory. 
The one against the odd-parity perturbations has been studied in Ref.~\cite{Heisenberg:2018mgr}. This paper is a successor that turns to the even-parity sector \cite{Thompson:2016fxe, Kase:2020yjf, Kase:2021mix, Liu:2022csl}, which completes this kind of 
Lagrangian-based stability analysis. By demanding the self-consistency of the theory, Ref.~\cite{Heisenberg:2018mgr} has 
already brought some constraints to the phase space of the theory-dependent coupling parameters for certain models of 
the $U(1)$ GI SVT theory. As will be seen later, in combining with the even-parity analysis, the corresponding phase space 
will be further confined. 

The rest of the paper is organized as following: 
Sec.~\ref{modelsec} provides the necessary fundamental information of the $U(1)$ GI SVT theory. 
The corresponding background field equations are discussed there. 
After that, we quickly review the stability analysis against the odd-parity perturbations in Sec.~\ref{oddsec}. 
Taking advantages of the results led by the odd-parity sector, we further run the stability analysis for 
the even-parity case in Sec.~\ref{evensec}. Notice that, the analysis is divided into 3 parts according to 
$l\geq2$, $l=0$ and $l=1$. We apply our general stability conditions to the three typical models 
in Sec.~\ref{application} for the background solutions studied in Ref.~\cite{Heisenberg:2018vti}.   
Finally, some of the concluding remarks will be given in Sec.~\ref{concludesec}.

As a usual treatment, in the following we shall set the speed of light as well as the reduced Planck constant to one, 
viz., $c=\hbar=1$\footnote{Notice that, after this unit selection, there is still one DOF left for the unit system of 
$[L, M , T]$. As an example, one can further set the radii of metric horizon to the Planck mass ($r_h=M_{\text{pl}}$) 
to fix the unit system.}. All the Greek letters in indices run from 0 to 3.  Other usages of indices will be explained 
when it is necessary. The whole paper is working under the signature $(- + + +)$.

\section{Background equations in $U(1)$ gauge-invariant 
SVT theories}
\label{modelsec}

We consider the $U(1)$ gauge-invariant (GI) scalar-vector-tensor (SVT) theories 
described by the action
\be
\mathcal{S}=\int {\rm d}^4x \sqrt{-g}\left(\frac{M_{\rm pl}^2}{2}R
+\sum_{i=2}^4\mathcal{L}^i_{\rm SVT}\right)\,,
\label{action}
\ee
where $g$ is a determinant of the metric tensor $g_{\mu \nu}$, 
$M_{\rm pl}^2 R/2$ is the Einstein-Hilbert term composed of
the reduced Planck mass $M_{\rm pl}$ associated with Newton's 
gravitational constant $G$ as $M_{\rm pl}=1/\sqrt{8\pi G}$ and 
 $R$ denotes the Ricci scalar. The Lagrangians $\mathcal{L}^i_{\rm SVT}$ with $i=2,3,4$ 
representing the $U(1)$ GI SVT interactions~\cite{Heisenberg:2018acv}
between a scalar field $\phi$ and a vector field $A_{\mu}$ are given by
\ba
\mathcal{L}^2_{\rm SVT}&=&f_2(\phi,X,F,\tilde{F},Y)\,, 
\label{L2}\\
\mathcal{L}^3_{\rm SVT}&=&
\left[ f_3(\phi,X)g_{\rho\sigma}+\tilde{f}_3(\phi,X)
\nabla_\rho \phi\nabla_\sigma \phi \right] \tilde{F}^{\mu\rho}\tilde{F}^{\nu\sigma} \nabla_\mu \nabla_\nu \phi\,, 
\label{L3}\\
\mathcal{L}^{4}_{\rm SVT}&=&
f_4(\phi,X)L^{\mu\nu\alpha\beta}F_{\mu\nu}F_{\alpha\beta}
+\left[ 
\frac12f_{4,X}(\phi,X)+\tilde{f}_4(\phi) \right] 
\tilde{F}^{\mu\nu}\tilde{F}^{\alpha\beta}
\nabla_\mu\nabla_\alpha \phi\nabla_\nu\nabla_\beta\phi\,,
\label{L4}
\ea
where $\nabla_{\mu}$ is the covariant derivative operator. 
The function $f_2$ depends on $\phi$ and the following quantities, 
\be
X \equiv -\frac{1}{2} \nabla_{\mu} \phi  \nabla^{\mu} \phi\,,
\qquad
F \equiv -\frac{1}{4} F_{\mu \nu} F^{\mu \nu}\,,
\qquad
\tilde{F} \equiv -\frac{1}{4} F_{\mu \nu} \tilde{F}^{\mu \nu}\,,
\qquad
Y \equiv \nabla_{\mu}\phi \nabla_{\nu}\phi 
F^{\mu \alpha}{F^{\nu}}_{\alpha}\,, 
\ee
where 
\be
F_{\mu \nu} \equiv \nabla_{\mu}A_{\nu}-\nabla_{\nu}A_{\mu}\,,
\qquad
\tilde{F}^{\mu\nu} \equiv \frac{1}{2}
\mathcal{E}^{\mu\nu\alpha\beta} F_{\alpha\beta}\,.
\ee
Here, the anti-symmetric Levi-Civita tensor $\mathcal{E}^{\mu\nu\alpha\beta}$ 
satisfies the normalization $\mathcal{E}^{\mu\nu\alpha\beta}
\mathcal{E}_{\mu\nu\alpha\beta}=-4!$. 
We denote the derivative of $f_2$ as $f_{2,Z}=\partial f_2/\partial Z$ where $Z$ is any of $\phi$, $X$, $F$, $\tilde{F}$, and $Y$. 
Meanwhile, $f_3, \tilde{f}_3, f_4$ are functions of $\phi,X$ 
with the same notation such as $f_{3,\phi} = \partial f_3/\partial \phi$, $f_{3,X} = \partial f_3/\partial X$, etc.,
and $\tilde{f}_4$ depends on $\phi$ alone as seen from Eq.~\eqref{L4}.
The double dual Riemann tensor $L^{\mu \nu \alpha \beta}$ is defined by 
\be
L^{\mu\nu\alpha\beta} \equiv \frac{1}{4}
\mathcal{E}^{\mu\nu\rho\sigma}
\mathcal{E}^{\alpha\beta\gamma\delta} R_{\rho\sigma\gamma\delta}\,,
\ee
where $R_{\rho\sigma\gamma\delta}$ is the Riemann tensor.

In this paper, we study the stability of the hairy BH solutions 
in the $U(1)$ GI theories studied in Ref.~\cite{Heisenberg:2018vti} 
on top of the static and spherically symmetric background given by the line element [under the Boyer-Lindquist coordinate $(t, r, \theta, \varphi)$]
\be
{\rm d}s^2=-f(r) {\rm d}t^{2} +h^{-1}(r) {\rm d}r^{2}
+ r^{2} \left({\rm d}\theta^{2}+\sin^{2}\theta\,{\rm d}\varphi^{2} 
\right)\,,
\label{metric_bg}
\ee
where $f$ and $h$ depend on the radial coordinate $r$. 
According to the underlying symmetry of the spacetime, we consider 
the scalar field $\phi$ depending on $r$ alone at the level of background, 
such that 
\be
\phi=\phi(r)\,.
\label{phi_bg}
\ee
Similarly, the background components of $A_{\mu}$ are given as
\be
A_{\mu}=(A_0(r), 0, 0, 0)\,,
\label{Amu_bg}
\ee%
where the radial component is absent due to the $U(1)$ gauge-invariance  \cite{Heisenberg:2018mgr}. 
We note that one can introduce the magnetic charge $P$ by setting $A_{\varphi}=-P\cos\theta$ 
as in Refs.~\cite{Lee:1991jw,Fernandes:2019kmh,Taniguchi:2024ear}. However, we do not include 
such term in this paper since we focus on the stability analysis of the solutions given in Ref.~\cite{Heisenberg:2018vti} 
in which the magnetic charge is absent. 
Denoting the quantities $X$, $F$, $\tilde{F}$, and $Y$ evaluated on the background 
with the overbar, they reduce to 
\be
\label{barXFY}
\overline{X}=-\frac{h}{2} \phi'^2\,,\qquad
\overline{F}=\frac{h A_0'^2}{2f}\,,\qquad
\overline{\tilde{F}}=0\,,\qquad
\overline{Y}=4\overline{X}\, \overline{F}\,.
\ee
Here, a prime in the superscript denotes the derivative with respect to $r$. 
Since the dependence on $\tilde{F}$ and $Y$ in $f_2$ under a static and spherically 
symmetric background either vanishes or can be expressed in terms of $X$ and $F$, 
it can be omitted at the background level~\cite{Heisenberg:2017hwb,Kase:2017egk}.
However, since the above relations hold only at the background level, the dependence on 
$\tilde{F}$ and $Y$ in $f_2$ may give rise to specific effect on the dynamics of 
the odd- and even-parity perturbations. Thus, we keep the full dependence 
in $f_2$, i.e., $f_2=f_2(\phi,X,F,\tilde{F},Y)$, in this paper\footnote{In fact, 
as seen from Sec.~\ref{oddsec}, the quantity ${\tilde F}$ will hold the linear-order contributions in the odd-party sector. In contrast, ${\tilde F}$ is a quantity with a magnitude at most the third order of the gravitational perturbation in the even-parity sector and hardly affects the linear perturbation calculations as we will see in Sec.~\ref{evensec}.}. 
We omit the overbar in the following discussion before stimulating any confusions. 

By the variation of the action (\ref{action}) with respect to $f,h,\phi,A_0$, 
we obtain background equations of motion, respectively, as\footnote{We notice that the explicit dependence of $f_2$ on $Y$ was not considered in the counterparts of Eqs.~\eqref{be1}-\eqref{JA} in Ref.~\cite{Heisenberg:2018vti} by assuming that $Y$ has been spelled out as $\overline{Y}=4\overline{X}\, \overline{F}$ at the background as we can see in Eq.~\eqref{barXFY}. In this paper, we explicitly include such a dependence in $f_2$ to be consistent with the fact that the relation $\overline{Y}=4\overline{X}\, \overline{F}$ does not hold at the perturbation level. Thus, Eqs.~\eqref{be1}-\eqref{JA} look slightly different in comparing to the equations in Ref.~\cite{Heisenberg:2018vti}.}
\ba
{\cal E}_{00}&\equiv& 
M_{\rm pl}^2 rfh' -\Big[ M_{\rm pl}^2 f(1-h)
+r^2 \left\{ f f_2-hA_0'^2 (f_{2,F}-2h\phi'^2f_{2,Y}) \right\}
-2r h^2 \phi' A_0'^2f_3\notag\\
&&+hA_0'^2 \left\{ 
4(h-1)f_4-h^2\phi'^2 (f_{4,X}+2\tilde{f}_4) \right \}\Big]=0\,,
\label{be1}\\
{\cal E}_{11}&\equiv& 
M_{\rm pl}^2 rh f' -\Big[ M_{\rm pl}^2 f(1-h)
+r^2 \left\{ f f_2+fh \phi'^2 f_{2,X}-h A_0'^2 (f_{2,F}-4h\phi'^2f_{2,Y}) 
\right\}
-2rh^2 \phi'A_0'^2 \left(3f_3-h \phi'^2 f_{3,X} \right) 
\nonumber \\
& &+hA_0'^2 \left\{ 4(3h-1)f_4-h (9h-4)\phi'^2f_{4,X} 
+h^3 \phi'^4 f_{4,XX}-10h^2\phi'^2 \tilde{f}_4
\right\}\Big]=0\,,\label{be2}\\
{\cal E}_{\phi}&\equiv&
J_{\phi}' - {\cal P}_{\phi}=0\,,\label{be3}\\
{\cal E}_{A}&\equiv&J_{A}' =0\,,\label{be4}
\ea
where 
\ba
\hspace{-0.7cm}
J_{\phi} &\equiv& -\sqrt{\frac{h}{f}} \left[  
r^2 \left(f f_{2,X}+2hA_0'^2f_{2,Y} \right)\phi' -2h A_0'^2 \left(2h \tilde{f}_4+3hf_{4,X}-2f_{4,X}\right)\phi' 
+2rh^2 A_0'^2f_{3,X}\phi'^2 
\right.\notag\\
&&\left.
+h^3 A_0'^2 f_{4,XX}\phi'^3
-2rhA_0'^2 f_3 \right]\,,\label{Jphi}\\
\hspace{-0.7cm}
{\cal P}_{\phi} &\equiv&
\frac{1}{\sqrt{fh}} \left[ r^2ff_{2,\phi}+hA_0'^2
\left\{ 4f_{4,\phi}+2h (r\phi'f_{3,\phi}-2f_{4,\phi})
+h^2 \left(f_{4,X\phi}+2\tilde{f}_{4,\phi}\right)\phi'^2 
\right\} \right]\,,\\
\hspace{-0.7cm}
J_A &\equiv& \sqrt{\frac{h}{f}} A_0' \left[ 
r^2 \left(f_{2,F}-2h\phi'^2f_{2,Y} \right)+4rh\phi' f_3+8(1-h)f_4
+2h^2\phi'^2 \left(f_{4,X}+2\tilde{f}_4 \right) \right]\,.
\label{JA}
\ea
From Eq.~(\ref{be4}), the current $J_A$ of vector field is conserved by virtue of the $U(1)$ gauge symmetry. In addition, if we demand the shift symmetry of the scalar field, the current $J_\phi$ will also be conserved. In such a case, this current vanishes at the horizon and hence it should be zero everywhere as we will see in Sec.~\ref{modelsec}. Interested readers can check these field equations out in \cite{supplemental}.
We note that the coupling $\tilde{f}_3$ never appears in Eqs.~(\ref{be1})-(\ref{be4}) due to the underlying symmetry of background spacetime.

\section{stability conditions against odd-parity perturbations }
\label{oddsec}

In this section, we revisit the stability conditions against odd-parity perturbations 
in the $U(1)$ GI SVT theories which is studied in Ref.~\cite{Heisenberg:2018mgr}. 
We consider the perturbed metric $g_{\mu\nu}=\bar{g}_{\mu\nu}+h_{\mu\nu}$ where $\bar{g}_{\mu\nu}$ is the background metric 
defined by Eq.~(\ref{metric_bg}) and $h_{\mu\nu}$ is the perturbation satisfying $|h_{\mu\nu}|\ll|\bar{g}_{\mu\nu}|$. 
The perturbations $h_{\mu\nu}$ can be expanded in terms of the spherical harmonics $Y_{lm}(\theta,\varphi)$ due to 
the underlying symmetry of the background spacetime. In doing so, the perturbations are classified into 
odd- and even-parity modes where the former changes the sign as $(-1)^{l+1}$ under the parity transformation 
$(\theta,\varphi)\to(\pi-\theta,\varphi+\pi)$  and the latter as $(-1)^{l}$ \cite{PhysRevD.2.2141}. 
The components of $h_{\mu\nu}$ in the odd-parity perturbations are expressed as follows 
\ba
& &
h_{tt}=h_{tr}=h_{rr}=0\,,\nonumber \\
& &
h_{ta}=\sum_{l,m}Q_{lm}(t,r)E_{ab}
\nabla^bY_{lm}(\theta,\varphi)\,,
\qquad
h_{ra}=\sum_{l,m}W_{lm}(t,r)E_{ab} \nabla^bY_{lm}(\theta,\varphi)\,,\nonumber \\
& &
h_{ab}=
\frac{1}{2}\sum_{l,m}
U_{lm}(t,r) \left[
E_{a}^c \nabla_c\nabla_b Y_{lm}(\theta,\varphi)
+ E_{b}^c \nabla_c\nabla_a Y_{lm}(\theta,\varphi)
\right]\,,
\ea
where $a,b$ represent either $\theta$ or $\varphi$, $Q_{lm}$, $W_{lm}$, and $U_{lm}$ are functions of $t$ and $r$. 
The tensor $E_{ab}$ is associated with the anti-symmetric symbol $\varepsilon_{ab}$ satisfying $\varepsilon_{\theta\varphi}=1$ 
as $E_{ab} \equiv \sqrt{\gamma}\, \varepsilon_{ab}$, where $\gamma \equiv \sin^2 \theta$ is the determinant of the metric $\gamma_{ab}$ defined on 
the surface of two-dimension unit sphere. 
The scalar filed $\phi$ does not contribute to the odd-parity perturbations, i.e., it has contributions only at the background level as long as we consider odd-parity perturbations. The odd-parity vector field perturbations $\delta A_{\mu}$ on top of the background value 
$\bar{A}_{\mu}$ satisfying $|\delta A_{\mu}|\ll|\bar{A}_{\mu}|$ are given by \cite{Kase:2018voo,Kase:2018owh,Kase:2021mix}
\be
\da_{t}=\da_{r}=0\,,\qquad 
\da_{a}=\sum_{l,m}\da_{lm}(t,r)E_{ab}\,\partial^bY_{lm}(\theta,\varphi)\,,
\ee
where $\da_{lm}$ depends on $t$ and $r$.

Under a infinitesimal gauge transformation $x_{\mu} \to x_{\mu}+\xi_{\mu}$, where $\xi_t=0=\xi_r$, and $\xi_a=\sum_{l,m} \Lambda_{lm}(t,r) 
E_{ab} \nabla^b Y_{lm} (\theta, \varphi)$, the metric perturbations transform as $Q_{lm} \to Q_{lm}+\dot{\Lambda}_{lm}$, 
$W_{lm} \to W_{lm}+\Lambda_{lm}'-2\Lambda_{lm}/r$, and $U_{lm} \to U_{lm}+2\Lambda_{lm}$ where a dot on top represents the derivative with respect to $t$. 
We choose the Regge-Wheeler gauge \cite{Regge:1957td} satisfying $\Lambda_{lm}=-U_{lm}/2$ in which $h_{ab}$ in the odd-parity sector merely vanishes. 
In the following, we omit the labels $l$ and $m$ of the quantities $Q_{lm}$, $W_{lm}$, and $\da_{lm}$ for simplicity. 

We expand the action (\ref{action}) up to quadratic order in odd-parity perturbations and perform the integration 
with respect to $\theta$ and $\varphi$. In doing so, we can focus on the perturbations of $m=0$ mode 
without loss of generality by virtue of the underlying background symmetry~\cite{deRham:2020ejn}. 
After repetitively using the integration by parts and the integral formulas for spherical harmonics 
(see, e.g., Ref.~\cite{Zhang:2023kzs} and references therein), the second-order action reduces to 
\ba
{\cal S}_{\rm odd}^{(2)}&=&\sum_{l,m} L \int {\rm d}t {\rm d}r\, 
{\cal L}_{\rm odd}^{(2)}\,, 
\label{oddact}
\ea
where 
\be
L \equiv l(l+1)\,,
\ee
and  
\ba
{\cal L}_{\rm odd}^{(2)}&=&
\alpha_1\left(\dot{W}-Q'+\frac{2}{r}Q\right)^2+2\left( \alpha_2 \da'+\alpha_3 \da \right)
\left(\dot{W}-Q'+\frac{2}{r}Q\right)
\notag\\
&&
+\alpha_4 \dot{\da}^2
+\alpha_5 {\da}'^2+(L-2) \left( \alpha_6 W^2
+\alpha_7 Q^2 
+\alpha_8Q\da \right)
+L\alpha_9 \da^2\,.
\label{oddLag}
\ea
The coefficients $\alpha_i$ ($i=1,...,9$) are given by 
\ba
& &
\alpha_1 \equiv \frac{M_{\rm pl}^2\sqrt{h}}{4\sqrt{f}}\,,\qquad 
\alpha_2 \equiv \frac{h^{3/2}A_0'}{2r\sqrt{f}} \left[ 
r\phi' f_3-4f_4+h\phi'^2 (f_{4,X}+2\tilde{f}_4) \right]\,,
\nonumber \\
& &
\alpha_3 \equiv -\frac{\sqrt{h}A_0'}{2r^2\sqrt{f}} 
\left[ r^2(f_{2,F}-2h\phi'^2f_{2,Y})+4h (r \phi' f_3-2f_4)+2h^2 
\phi'^2 (f_{4,X}+2\tilde{f}_4) \right]\,,
\nonumber \\
& &
\alpha_4 \equiv \frac{1}{2r\sqrt{fh}} \left[ rf_{2,F}+ 
2h\phi'(f_3+h\phi'^2\tilde{f}_3)-4h'f_4
+(2h\phi''+\phi'h')\{rf_3+h\phi'(f_{4,X}+2\tilde{f}_4)\}
\right]\,,
\nonumber \\
& &
\alpha_5 \equiv -\frac{\sqrt{fh}}{2r} \left[ r (f_{2,F}-2h\phi'^2f_{2,Y})
+2h \phi' f_3+\frac{hf'}{f}\{r\phi'f_3-4f_4+h\phi'^2(f_{4,X}+2\tilde{f}_4)\} \right]\,,
\nonumber \\
& &
\alpha_6 \equiv -\frac{\sqrt{fh}}{4r^2} 
\left[ M_{\rm pl}^2 - \frac{hA_0'^2}{f}\{4f_4-h\phi'^2(f_{4,X}+2\tilde{f}_4)\} \right]\,,
\nonumber \\
& &
\alpha_7 \equiv \frac{1}{4r^2\sqrt{fh}}\left(M_{\rm pl}^2-\frac{4hA_0'^2f_4}{f}\right)
\,,\qquad
\alpha_8 \equiv \frac{4}{r^2}\left(\sqrt{\frac{h}{f}}A_0'f_4\right)'
\,,
\nonumber \\
& &
\alpha_9 \equiv -\frac{1}{2r^2\sqrt{fh}} \biggl[ 
ff_{2,F}-hA_0'^2f_{2,\tilde{F}\tilde{F}}
+2\sqrt{fh}(\sqrt{fh}\phi')'f_3
+h^{2}f'\phi'^3\tilde{f}_3
-4\sqrt{fh}\left(\sqrt{\frac{h}{f}}f'\right)'f_4
\nonumber \\
& &
\hspace{2.5cm}
+h^{3/2}f'\phi'(\sqrt{h}\phi')'(f_{4,X}+2\tilde{f}_4)
 \biggr]\,.
\label{alphai}
\ea
Notice that, in the absence of the dependence of $\tilde{F}$ and $Y$ in $f_2$, the above coefficients $\alpha_i$ ($i=1,...,9$) 
reduce to those defined in Ref.~\cite{Kase:2018voo} up to an overall factor, which are denoted as $\alpha^{\rm (pre)}_i$ ($i=1,...,9$), such that 
\be
\alpha_i=r^2\sqrt{\frac{f}{h}}\,\alpha^{\rm (pre)}_i\,.
\ee

The second-order action (\ref{oddact}) identically vanishes for the monopole mode $l=0$, i.e., $L=0$. 
For the dipole mode characterized by $l=1$ ($L=2$), the vector field perturbation $\da$ is 
the only propagating DOF, and it is shown that the stability conditions for 
$\da$ are the same as those for $l\geq2$ in Ref.~\cite{Heisenberg:2018mgr}. Thus, we focus only 
on the modes characterized by $l\geq2$ and revisit their stability conditions in this paper.  

The dynamical variables in the Lagrangian (\ref{oddLag}) are $W$ and $\da$. In order to integrate out 
the remaining non-dynamical variable $Q$, we introduce an auxiliary field $\chi$ to the Lagrangian as 
\ba
\hspace{-0.8cm}
{\cal L}_{\rm odd}^{(2)}&=&
\alpha_1\left[2\chi\left(\dot{W}-Q'
+\frac{2}{r}Q+\frac{\alpha_2\da'
+\alpha_3 \da}{\alpha_1}\right)-\chi^2\right]
-\frac{(\alpha_2\da'+\alpha_3\da)^2}
{\alpha_1}
\notag\\
\hspace{-0.8cm}
&&
+\alpha_4 \dot{\da}^2
+\alpha_5 {\da}'^2+(L-2) \left( \alpha_6 W^2
+\alpha_7 Q^2 
+\alpha_8Q\da \right)
+L\alpha_9 \da^2\,.
\label{oddLag2}
\ea
Varying the above Lagrangian with respect to $\chi$, we find that the auxiliary field $\chi$ 
satisfies
\be
\chi=\dot{W}-Q'
+\frac{2}{r}Q+\frac{\alpha_2\da'
+\alpha_3 \da}{\alpha_1}\,,
\label{chi1}
\ee
which guarantees the equivalence between Eqs.~(\ref{oddLag}) and (\ref{oddLag2}). 
Meanwhile, the variation of the Lagrangian (\ref{oddLag2}) with respect to $W$ and $Q$ leads 
\ba
&&
\alpha_1\dot{\chi}-(L-2)\alpha_6 W=0\,,
\label{eqW}\\
&&
\alpha_1 \chi'
+\frac{1}{r^2}(r^2\alpha_1)'\chi
+(L-2)\left( \alpha_7 Q+\frac{\alpha_8}{2}\da \right)=0\,,
\label{eqQ}
\ea
respectively. The above algebraic equations can be solved for $W$ and $Q$, respectively. 
We substitute these solutions into the Lagrangian (\ref{oddLag2}) and eliminate 
the perturbations $Q$ and $W$ from the second-order action. In doing so, 
the role of gravitational DOF carried by $Q$ and $W$ in the original Lagrangian 
(\ref{oddLag}) is transferred to the auxiliary field $\chi$ in the resultant Lagrangian  of the form 
\be
(L-2){\cal L}_{\rm odd}^{(2)}= 
\dot{\vec{\mathcal{X}}}^{t}{\bm K}\dot{\vec{\mathcal{X}}}
+\vec{\mathcal{X}}'^{t}{\bm G}\vec{\mathcal{X}}'
+\vec{\mathcal{X}}'^{t}{\bm S}\vec{\mathcal{X}}
+\vec{\mathcal{X}}^{t}{\bm M}\vec{\mathcal{X}}
\,,
\label{oddLag3}
\ee
where ${\bm {K,G,S,M}}$ are $2\times2$ matrices with the non-vanishing components given by 
\ba
& &
K_{11}=-\frac{\alpha_1^2}{\alpha_6}\,,\qquad 
K_{22}=(L-2)\alpha_4\,,
\notag\\
&&
G_{11}=-\frac{\alpha_1^2}{\alpha_7}\,,\qquad 
G_{22}=\frac{(L-2)(\alpha_1 \alpha_5-\alpha_2^2)}{\alpha_1}\,, \nonumber \\
& &
S_{12}=-S_{21}=-(L-2)\left(\alpha_2+\frac{\alpha_1\alpha_8}{2\alpha_7}\right)\,,
\notag\\
&&
M_{11}=-(L-2)\alpha_1-\frac{[(r^2\alpha_1)']^2}{r^4\alpha_7}
+\left[\frac{\alpha_1(r^2\alpha_1)'}{r^2\alpha_7}\right]'\,,
\notag\\
&&
M_{22}=-(L-2)\left[\frac{(L-2)\alpha_8^2}{4\alpha_7}-L\alpha_9
+\frac{\alpha_3^2}{\alpha_1}-\left(\frac{\alpha_2\alpha_3}{\alpha_1}\right)'\,\right]\,,
\notag\\
&&
M_{12}=M_{21}=(L-2)\left[\alpha_3-
\frac{\alpha_8(r^2\alpha_1)'}{2r^2\alpha_7}
-\frac{1}{2}
\left(\alpha_2-\frac{\alpha_1\alpha_8}{2\alpha_7}\right)'\,
\right]\,,
\label{KGSM}
\ea
and $\vec{\mathcal{X}}$ is the vector defined by 
\be
\vec{\mathcal{X}}^{t} \equiv \left(\chi,\da \right)\,.
\label{mathcalX}
\ee
The no-ghost conditions, $K_{11}>0$ and $K_{22}>0$, are guaranteed for 
\be
\alpha_6<0\,,\qquad \alpha_4>0\,. 
\label{oddnoghost}
\ee
By assuming the solution of the form $\vec{\mathcal{X}}^{t} \propto e^{i (\omega t-kr)}$ in the small-scale limit 
characterized by $k\to0$, the propagation speed of perturbations along the radial direction is expressed as 
$c_r=\omega/(\sqrt{fh}\,k)$ in proper time. Substitution of this expression into the dispersion relation, 
${\rm det} \left( \omega^2 {\bm K}+k^2{\bm G} \right)=0$, leads  
\ba
c_{r1{\rm ,odd}}^2
&=& -\frac{G_{11}}{fh K_{11}}
=-\frac{\alpha_6}{fh\,\alpha_7}\,,
\label{oddcr1}\\
c_{r2{\rm ,odd}}^2
&=& -\frac{G_{22}}{fh K_{22}}
=\frac{\alpha_2^2-\alpha_1 \alpha_5}
{fh\,\alpha_1 \alpha_4}\,, 
\label{oddcr2}
\ea
where $c_{r1{\rm ,odd}}$ and $c_{r2{\rm ,odd}}$ are the propagation speeds of the odd-parity perturbations arising from the gravity sector and the vector field perturbation, respectively. 
On the other hand, by assuming the solution of the form $\vec{\mathcal{X}}^{t} \propto e^{i (\omega t-l \theta)}$ 
in the limit that $L=l(l+1) \gg 1$, the squared propagation speed along the angular direction is expressed as 
$c_{\Omega}^2=r^2\omega^2/(l^2f)$ in proper time. We substitute this expression into the dispersion relation  of the form ${\rm det}(\omega^2{\bm K}+{\bm M})=0$. Picking up the dominant contributions for $l\gg1$, we obtain 
\ba
c_{\Omega 1{\rm ,odd}}^2&=&
-\frac{r^2 M_{11}}{l^2 f K_{11}}
=-\frac{r^2 \alpha_6}{f \alpha_1}\,,
\label{oddco1}\\
c_{\Omega 2{\rm ,odd}}^2&=&
-\frac{r^2 M_{22}}{l^2 f K_{22}}
=\frac{r^2 (\alpha_8^2-4\alpha_7 \alpha_9)}{4f\alpha_4 \alpha_7}\,.
\label{oddco2}
\ea
Here, $c_{\Omega1{\rm ,odd}}$ is the propagation speed along the angular direction for the perturbation 
arising from gravity sector, while $c_{\Omega2{\rm ,odd}}$ is that arising from the vector perturbation. 
We note that the expressions of Eqs.~\eqref{oddnoghost}-\eqref{oddco2} are all the same as those derived in Ref.~\cite{Heisenberg:2018mgr} 
while the dependence of $\tilde{F}$ and $Y$ in $f_2$ alters the coefficients $\alpha_3$, $\alpha_4$, $\alpha_5$, and $\alpha_9$. 

For the absence of Laplacian instabilities, we require the conditions $c_{r1{\rm ,odd}}^2\geq0$, 
$c_{r2{\rm ,odd}}^2\geq0$, $c_{\Omega 1{\rm ,odd}}^2\geq0$, and $c_{\Omega 2{\rm ,odd}}^2\geq0$. 
Among these conditions, Eq.~(\ref{oddco1}) shows that the condition $c_{\Omega 1{\rm ,odd}}^2\geq0$ 
is automatically satisfied under the no-ghost condition~(\ref{oddnoghost}) since $\alpha_1$ is non-negative by definition [cf., \eqref{alphai}].  
From Eqs.~(\ref{oddcr1}), (\ref{oddcr2}), (\ref{oddco2}), the other three conditions for the absence of Laplacian 
instabilities are translated to give
\ba
\alpha_7\geq0\,,\qquad
\alpha_2^2-\alpha_1 \alpha_5\geq0\,,\qquad
\alpha_8^2-4\alpha_7 \alpha_9\geq0\,.
\label{oddnoLap}
\ea
The above stability conditions will assist us below in understanding those for the even-parity sector. 

\section{stability conditions against even-parity perturbations }
\label{evensec}

We proceed to derive the stability conditions against even-parity perturbations. 
On top of the background spacetime characterized by Eq.~(\ref{metric_bg}), 
the components of metric perturbation $h_{\mu\nu}$ in the even-parity sector are given by 
\ba
&&
h_{tt}=f(r) \sum_{l,m}H_{0,lm}(t,r)Y_{lm}(\theta,\varphi)\,,\qquad
h_{tr}=h_{rt}=\sum_{l,m}H_{1,lm}(t,r)Y_{lm}(\theta,\varphi)\,,\notag\\
&&
h_{rr}=h(r)^{-1}\,\sum_{l,m}H_{2,lm}(t,r)Y_{lm}(\theta,\varphi)\,,\notag\\
&&
h_{ta}=h_{at}=\sum_{l,m}h_{0,lm}(t,r)\nabla_aY_{lm}(\theta,\varphi)\,,\qquad
h_{ra}=h_{ar}=\sum_{l,m}h_{1,lm}(t,r)\nabla_aY_{lm}(\theta,\varphi)\,,\notag\\
&&
h_{ab}=\sum_{l,m}\left[K_{lm}(t,r)g_{ab}Y_{lm}(\theta,\varphi)
+G_{lm}(t,r)\nabla_a\nabla_bY_{lm}(\theta,\varphi)\right]\,, 
\label{evenmetric}
\ea
where $H_{0,lm}$, $H_{1,lm}$, $H_{2,lm}$, $h_{0,lm}$, $h_{1,lm}$, $K_{lm}$, and $G_{lm}$ are 
scalar quantities depending on $t$ and $r$ [One should avoid confusing the $K$ and $G$ in here with those appearing in, e.g., \eqref{KGSM}]. 
Similarly, we consider the perturbations of scalar and vector fields on top of their background values as\footnote{
We note that we shall in general omit the overbar  in writing background quantities like 
$\bar \phi$ and ${\bar A}_\mu$ in the following, since the order of a quantity can be easily identified 
by the context and counting the perturbation terms, just like what we have done in, e.g., Eq.\eqref{be1}. }
\ba
\phi&=&\bar{\phi}+\sum_{l,m}\delta\phi_{lm}(t,r)Y_{lm}(\theta,\varphi)\,,\label{defdphi}\\
A_{\mu}&=&\bar{A}_{\mu}+\delta A_{\mu}\,,\label{defdA}
\ea
with
\be
\delta A_t=\sum_{l,m}\delta A_{0,lm}(t,r)Y_{lm}(\theta, \varphi)\,,\qquad
\delta A_r=\sum_{l,m}\delta A_{1,lm}(t,r)Y_{lm}(\theta, \varphi)\,,\qquad 
\delta A_a=\sum_{l,m}\delta A_{2,lm}(t,r)\nabla_{a}Y_{lm}(\theta, \varphi)\,, 
\label{vecpert}
\ee
where $\bar{\phi}$ and $\bar{A}_{\mu}$ are background values given by Eqs.~(\ref{phi_bg}) and 
(\ref{Amu_bg}), respectively. The scalar quantities $\delta\phi_{lm}$, $\delta A_{0,lm}$, 
$\delta A_{1,lm}$, and $\delta A_{2,lm}$ are functions of $t$ and $r$, which are, as usual, assumed to be much smaller than the background quantities. 

Let us consider the infinitesimal gauge transformation
$x_\mu\to x_\mu+\xi_\mu$ with 
\be
\xi_t=\sum_{l,m} {\cal T}_{lm}(t,r)Y_{lm}(\theta,\varphi)\,,\qquad 
\xi_r=\sum_{l,m} {\cal R}_{lm}(t,r)Y_{lm}(\theta,\varphi)\,,\qquad 
\xi_a=\sum_{l,m} \Theta_{lm}(t,r) \nabla_a Y_{lm}(\theta,\varphi)\,, 
\label{xivec2}
\ee
where ${\cal T}_{lm}$, ${\cal R}_{lm}$, and $\Theta_{lm}$ are the scalar quantities depending on $t$ and $r$. 
In the following we omit the subscripts $l$ and $m$ of the scalar quantities in Eqs.~(\ref{evenmetric}) and (\ref{xivec2}) for simplicity. 
Under the above transformation, the scalar quantities in Eq.~(\ref{evenmetric}) transform as 
\ba
& &
H_0 \to H_0+\frac{2}{f} \dot{{\cal T}}-\frac{f' h}{f}{\cal R}\,,\qquad 
H_1 \to H_1+\dot{{\cal R}}+{\cal T}'-\frac{f'}{f}{\cal T}\,,\qquad
H_2 \to H_2+2h{\cal R}'+h' {\cal R}\,,\label{H1tra} \notag \\
& &
h_0 \to h_0+{\cal T}+\dot{\Theta}\,,\qquad 
h_1 \to h_1+{\cal R}+\Theta'
-\frac{2}{r}\Theta\,,\qquad
K \to K+\frac{2}{r}h{\cal R}\,,\qquad 
G \to G+\frac{2}{r^2}\Theta\,,\notag\\
&&
\delta\phi\to\delta\phi-\phi'h{\cal R}\,,
\label{gaugetrans}
\ea
where we have dropped the $lm$ in subscripts for the corresponding quantities for simplicity. 
For $l\geq2$, we choose the gauge given by 
${\cal T}=-h_0+{r^2}\dot{G}/2$,
${\cal R}=-rK/({2h})$, and 
$\Theta=-{r^2}G/2$ so that the perturbations $h_0$, $K$, $G$ \footnote{One should avoid confusing the $G$ in here with the gravitational constant mentioned previously. } identically vanish. 
This gauge fixing is equivalent to setting 
\be
h_0=0\,,\qquad 
K=0\,,\qquad
G=0\,,
\label{gauge1}
\ee
in Eq.~(\ref{evenmetric}) from the beginning (This is sometimes referred as the EZ gauge \cite{Thompson:2016fxe}). 
We also consider the $U(1)$ gauge transformation, 
\be
\delta A_{\mu} \to \delta A_{\mu}+\partial_{\mu} \delta \chi
\qquad {\rm with} \qquad
\delta \chi=\sum_{l,m} \tilde{\chi}(t,r)Y_{lm}(\theta,\varphi)\,, 
\ee
under which the scalar quantities in Eq.~(\ref{vecpert}) transform as 
\be
\delta A_0\to\delta A_0+\dot{\tilde{\chi}}\,,\qquad
\delta A_1\to\delta A_1+\tilde{\chi}'\,,\qquad
\delta A_2\to\delta A_2+\tilde{\chi}\,.
\ee
For the gauge choice $\tilde{\chi}=-\delta A_2$, the quantity $\delta A_2$ identically vanishes. 
This corresponds to setting 
\be
\delta A_2=0\,,
\label{gauge2}
\ee
in Eq.~(\ref{vecpert}) from the beginning. Thus, with all the gauge choices mentioned above, we are now left with 7 variables, i.e., 
$\{\delta \phi, \delta A_0, \delta A_1, h_1, H_0, H_1, H_2\}$ (and at this point they themselves, as well as their combinations, represent gauge invariants).

\subsection{Second-order action and perturbation equations of motion}

We expand the action (\ref{action}) up to second order in terms of the even-parity perturbations 
given in Eqs.~(\ref{evenmetric}), (\ref{defdphi}), and (\ref{defdA}) under the gauge choices 
(\ref{gauge1}) and (\ref{gauge2}). In doing so, we can focus on the $m=0$ mode without loss of 
generality for the same reason as in the case of odd-parity perturbations. After lengthy but 
straightforward calculation, the second-order action in the even-parity sector reduces to 
\be
{\cal S}_{\rm even} = \sum_l\int {\rm d}t\, {\rm d}r 
\left({\cal L}_{u}+{\cal L}_{A}\right)\,,
\label{acteven}
\ee
where
\ba
{\cal L}_{u}
&\equiv&
H_0 \left[ a_1 \delta \phi'' +a_2 \delta \phi' +a_3 H_2'
+L a_4 h_1'+\left( a_5+L a_6 \right) \delta \phi 
+\left( a_7+L a_8 \right) H_2+L a_9 h_1 \right] 
\notag \\
&&
+L b_1 H_1^2
+H_1 \left( b_2 \dot{\delta \phi}'+b_3 \dot{\delta \phi}
+b_4 \dot{H}_2+L b_5 \dot{h}_1 \right)
+c_1 \dot{\delta \phi} \dot{H}_2
+H_2 \left[ c_2 \delta \phi'+ (c_3+L c_4) \delta \phi
+L c_5 h_1 \right]
\notag \\
&&
+c_6 H_2^2 
+L d_1 \dot{h}_1^2
+L h_1 \left( d_2 \delta \phi'
+d_3 \delta \phi \right)+L d_4 h_1^2
+e_1 \dot{\delta \phi}^2+e_2 \delta \phi'^2
+\left( e_3+L e_4 \right) \delta \phi^2
\,,
\label{Lu}
\\
{\cal L}_{A}
&\equiv&
v_1(\delta A_0'-\dot{\delta A}_1)^2
+(\delta A_0'-\dot{\delta A}_1)
\left(v_2H_0+v_3H_2+v_4\delta\phi'+v_5\delta\phi
+Lv_6 h_1 \right)
+\frac{1}{2}Lv_6h_1\dot{\delta A}_1
+v_7H_0^2
\notag \\
&&
+L(v_8h_1\delta A_0+v_9\delta A_0^2+v_{10}\delta A_1^2
+v_{11}H_1\delta A_1+v_{12}H_2\delta A_0+v_{13}\delta\phi\delta A_0)\,.
\label{LdA}
\ea
The coefficients $a_1$, $a_2$, ..., $v_9$ are given in Appendix~\ref{even_coeff}. 
In Refs.~\cite{Kase:2023kvq} as well as \cite{supplemental}, the even-parity perturbations in Horndeski theories 
with the interaction between scalar and Maxwell fields characterized by the Lagrangian 
$G_2(\phi,X,F)$ are studied. Compared to that, the presence of the $U(1)$ GI 
SVT interactions characterized by $f_3$, $\tilde{f}_3$, $f_4$, and $\tilde{f}_4$ in Eq.~(\ref{action}) 
gives rise to the new terms with the coefficients $v_6$, $v_{11}$, $v_{12}$, and $v_{13}$, in Eq.~(\ref{LdA}). 
For $f_2=G_2(\phi,X,F)$, $f_3=0$, $\tilde{f}_3=0$, $f_4=0$, and $\tilde{f}_4=0$, these coefficients 
identically vanish and the Lagrangian ${\cal L}_{A}$ coincides with that derived in Ref.~\cite{Gannouji:2021oqz,Kase:2023kvq}. 
Since the Lagrangian ${\cal L}_{A}$ possesses the similar structure to that in Ref.~\cite{Kase:2023kvq}, 
we can resort to the analogous method in order to identify the dynamical vector DOF in the even-parity sector. 
Let us introduce an auxiliary field $V(t,r)$ and rewrite Eq.~(\ref{LdA}) as follows, 
\ba
{\cal L}_{A}
&=&
v_1 \left\{2 V \left[ \delta A_0'-\dot{\delta A}_1 
+\frac{1}{2 v_1} \left( v_2H_0+v_3H_2+v_4\delta\phi'+v_5\delta\phi +Lv_6 h_1\right) \right] -V^2 \right\}
\notag \\
&&
-\frac{1}{4 v_1} \left( v_2H_0+v_3H_2+v_4\delta\phi'+v_5\delta\phi +Lv_6 h_1\right)^2
+\frac{1}{2}Lv_6h_1\dot{\delta A}_1
+v_7H_0^2
\notag \\
&&
+L(v_8h_1\delta A_0+v_9\delta A_0^2+v_{10}\delta A_1^2
+v_{11}H_1\delta A_1+v_{12}H_2\delta A_0+v_{13}\delta\phi\delta A_0)\,.
\label{LdA2}
\ea
By varying this action with respect to $V$, we find that 
\ba
V = \delta A_0'-\dot{\delta A}_1 +\frac{1}{2 v_1} \left( v_2H_0+v_3H_2+v_4\delta\phi'+v_5\delta\phi +Lv_6 h_1\right) \,,
\label{defV}
\ea
which shows the equivalence between Eqs.~(\ref{LdA}) and (\ref{LdA2}). 
Nevertheless, the auxiliary field $V$ plays a key role to find out the dynamical vector DOF 
on the point that the quadratic derivative terms $\delta A_0'^2$ and $\dot{\delta A}_1^2$ 
in Eq.~(\ref{LdA}) do not appear in Eq.~(\ref{LdA2}). The disappearance of these derivative terms 
in Eq.~(\ref{LdA2}) allows us to solve the perturbation equations of $\delta A_0$ and $\delta A_1$ 
for themselves explicitly in analogy with what we did in Eqs.~\eqref{eqW}-\eqref{eqQ}. 
On using these solutions, the dynamical property of vector field 
perturbations can be aggregated into the auxiliary field $V$ as we will see later. 
Moreover, the quadratic term $H_0^2$ in Eq.~(\ref{LdA2}) identically vanishes by virtue of 
the relation among the coefficients $v_1$, $v_2$, and $v_7$, of the form, 
\be
v_7=\frac{v_2^2}{4v_1}\,.
\label{conv7}
\ee
This shows that the perturbation $H_0$ appears only linearly in the total action~(\ref{acteven}) with 
Eqs.~(\ref{Lu}) and (\ref{LdA2}). In other words, the perturbation $H_0$ corresponds to a Lagrange 
multiplier. The variation of the action with respect to $H_0$ gives constraint on other perturbation 
variables, and $H_0$ simply disappears once the constraint is applied to the action. 

We vary the total action~(\ref{acteven}) represented by Eqs.~(\ref{Lu}) and (\ref{LdA2}) 
with respect to $H_0$, $H_1$, $H_2$, $h_1$, $\delta A_0$, $\delta A_1$, 
and $\delta\phi$, so as to obtain the following linear perturbation equations, 
\ba
\hspace{-0.7cm}
0 &=&  
a_3 H_2'+La_4 h_1'+\left(a_2-\frac{v_2 v_4}{2 v_1}\right) \delta \phi'
+ \left(La_6+a_5-\frac{v_2 v_5}{2 v_1}\right)\delta \phi
+ \left(La_8+a_7-\frac{v_2 v_3}{2 v_1}\right)H_2
\notag\\
&&
+L\left(a_9-\frac{v_2 v_6}{2 v_1} \right)h_1+v_2V \,,
\label{eqH0}\\
\hspace{-0.7cm}
0 &=&    
2 L b_1 H_1+b_3 \dot{\delta \phi }+b_4 \dot{H_2}+L b_5 \dot{h_1}+L v_{11} \delta A_1    \,,
\label{eqH1}\\
\hspace{-0.7cm}
0 &=&    
-b_4 \dot{H}_1
+\left(c_2-\frac{v_3 v_4}{2 v_1}\right) \delta \phi'
+\left(c_3+Lc_4-\frac{v_3 v_5}{2 v_1}\right)\delta \phi
+L\left(c_5 -\frac{ v_3 v_6}{2 v_1}\right)h_1
+ \left(2 c_6-\frac{v_3^2}{2 v_1}\right) H_2
\notag\\
&&
-a_3 H_0'
+\left(a_7-a_3'+La_8-\frac{v_2 v_3}{2 v_1}\right)H_0
+v_3 V+ L v_{12} \delta A_0\,,
\label{eqH2}\\
\hspace{-0.7cm}
0 &=&
-2 d_1 \ddot{h_1}
+\left(d_2-\frac{v_4 v_6}{2 v_1}\right) \delta \phi '
+\left(d_3-\frac{v_5 v_6}{2 v_1}\right)\delta \phi
+\left(2 d_4-\frac{L v_6^2}{2 v_1}\right) h_1
-a_4 H_0'
+\left(a_9-a_4'-\frac{v_2 v_6}{2 v_1}\right)H_0
\notag\\
&&
-b_5 \dot{H_1}
+\left(c_5 -\frac{ v_3 v_6}{2 v_1}\right)H_2
+v_6 V
+\frac{1}{2} v_6\dot{\delta A}_1 
+v_8\delta A_0 \,,
\label{eqh1}\\
\hspace{-0.7cm}
0 &=&
-2 (v_1 V)'
+L \left(v_8h_1
+2 v_9 \delta A_0 
+v_{12}H_2
+v_{13}\delta \phi\right)\,,
\label{eqA0}\\
\hspace{-0.7cm}
0 &=&  
2 v_1\dot{V}+L\left(
-\frac{1}{2} v_6 \dot{h_1}
+2 v_{10} \delta A_1
+ v_{11}H_1\right)
\,,
\label{eqA1}\\
0&=&
-2 e_1 \ddot{\delta\phi}-\left(2 e_2-\frac{v_4^2}{2v_1}\right) \delta\phi''
+ \left[2 e_3+2Le_4 +\left(\frac{v_4v_5}{2v_1}\right)'-\frac{v_5^2}{2v_1}\right] \delta\phi
- \left( a_2 -\frac{v_2v_4}{2v_1}\right) H_0'
\notag\\
&&
+ \left[ a_5+La_6-a_2'+\left(\frac{v_2v_4}{2v_1}\right)'-\frac{v_2v_5}{2v_1} \right] H_0
-b_3 \dot{H}_1-\left(c_2-\frac{v_3v_4}{2v_1}\right)H_2' 
\notag\\
&&
+ \left[ c_3+Lc_4-c_2' +\left(\frac{v_3v_4}{2v_1}\right)'-\frac{v_3v_5}{2v_1}\right] H_2
-L\left(d_2-\frac{v_4v_6}{2v_1}\right)h_1' 
+L \left[ d_3-d_2'+\left(\frac{v_4v_6}{2v_1}\right)'-\frac{v_5v_6}{2v_1} \right] h_1
\notag\\
&&
-\left[2 e_2'-\left(\frac{v_4^2}{2v_1}\right)' \right]\delta\phi'
-v_4V'+(v_5-v_4')V
+Lv_{13}\delta A_0
\,,
\label{eqdphi}
\ea
where we used the relation (\ref{conv7}) and substituted in the coefficients being 0 in Appendix~\ref{even_coeff}. 
In the following, we study the linear stability conditions for the three cases (1) $l\geq2$, (2) $l=0$, and (3) $l=1$, in turn.

\subsection{Linear stability conditions for $l\geq2$}

We introduce the following quantity
\ba
\psi &\equiv& H_2 - \frac{L}{r} h_1\,,
\label{psi}
\ea
which corresponds to the propagating DOF of gravitational sector as in the case of GR. 
Now, we have all the representations for the 3 DOFs, i.e., gravitational ($\psi$), vector ($V$) and scalar ($\delta \phi$), in hand.
We replace the $H_2$, $\dot{H}_2$, and $H_2'$ in \eqref{eqH0}-\eqref{eqA1} with $\psi$ and its derivatives. In doing so, 
the quantity $h_1'$ disappears from Eq.~(\ref{eqH0}) on account of the relation $a_3=-ra_4$, and the resultant equation 
can be explicitly solved for $h_1$ as a combination of $\psi$, $\delta\phi'$, $\delta\phi$, and $V$. Substituting this solution 
into Eqs.~(\ref{eqH1}), (\ref{eqA0}), (\ref{eqA1}) and combining them, we can also express $H_1$, $\delta A_0$, and $\delta A_1$ 
in terms of $\psi$, $\delta\phi$, $V$, and their derivatives. 
Finally, by substituting these solutions back into the total action~(\ref{acteven}) with Eqs.~(\ref{Lu}) and (\ref{LdA2}), 
the perturbations $H_1$, $H_2$, $h_1$, $\delta A_0$, $\delta A_1$ are removed. The quantity $H_0$ is also removed 
from the action as we discussed below Eq.~(\ref{conv7}). 
As a consequence, the resultant action is composed of the only 3 dynamical perturbations $\psi$, $\delta \phi$, and $V$. 
Denoting the reduced Lagrangian as ${\cal L}_{\rm even}^{(2)}$ by mimicking \eqref{oddLag3}, it can be written as 
\ba
{\cal L}_{\rm even}^{(2)}= 
\dot{\vec{\mathcal{Y}}}^{t}{\tilde {\bm K}} \dot{\vec{\mathcal{Y}}}
+\vec{\mathcal{Y}}'^{t} {\tilde{\bm G}} \vec{\mathcal{Y}}'
+\vec{\mathcal{Y}}'^{t} {\tilde{\bm S}} \vec{\mathcal{Y}}
+\vec{\mathcal{Y}}^{t} {\tilde{\bm M}} \vec{\mathcal{Y}}
\,,
\label{evenLag1}
\ea
where $\vec{\mathcal{Y}}$ is the vector defined by 
\be
\vec{\mathcal{Y}}^{t}=\left(\psi,\delta \phi, V \right)\,.
\label{mathcalY}
\ee
Here, the matrices ${\tilde{\bm K}}$, ${\tilde{\bm G}}$ and ${\tilde{\bm M}}$ are symmetric while ${\tilde{\bm S}}$ is anti-symmetric. 
Putting together the 2 DOFs from the odd-parity sector [cf., \eqref{mathcalX}] and the 3 DOFs from the even-parity sector [cf., \eqref{mathcalY}], we obtain 5 DOFs in total, as once promised. 

\subsubsection{No-ghost conditions}
\label{noghostsec}

The components of kinetic matrix ${\tilde {\bm K}}$ are given as 
\ba
&&
\tilde{K}_{11}=
\frac{1}{L}\left[{\cal K}_1+\left(\frac{a_4'}{a_4}
+\frac1r+\frac{L}{2rh}\right){\cal K}_2-\frac12{\cal K}_2'\right]\,,\notag\\
&&
\tilde{K}_{12}=
\frac{1}{2L}\left[\frac{fb_3}{ra_4}\left\{{\cal K}_1+\frac12\left(\frac{a_4'}{a_4}
+\frac1r+\frac{L}{2rh}\right){\cal K}_2\right\}
-\frac{1}{ra_4}\left(a_5+La_6-\frac{A_0'}{2}v_5\right){\cal K}_2
-\frac12\left(\frac{fb_3}{ra_4}{\cal K}_2\right)'\right]\,,\notag\\
&&
\tilde{K}_{22}=
e_1+\frac{1}{4L}\left[
\frac{fb_3}{r^2a_4^2}
\left\{fb_3{\cal K}_1
-2\left(a_5+La_6-\frac{A_0'}{2}v_5\right){\cal K}_2\right\}
-\frac12\left\{\left(\frac{fb_3}{ra_4}\right)^2{\cal K}_2\right\}'
\,\right]\,,\notag\\
&&
\tilde{K}_{13}=\frac{v_1}{4Lra_4}\left(\frac{fv_{6}}{v_{10}}{\cal K}_1-2A_0'{\cal K}_2\right)\,,\qquad
\tilde{K}_{23}=\frac{fb_3}{2ra_4}\tilde{K}_{13}\,,\qquad
\tilde{K}_{33}=\frac{fv_1^2}{2Lr^2a_4v_{10}}{\cal K}_1\,,
\label{Kcompts}
\ea
where we used the relations among the coefficients given in Appendix~\ref{even_coeff}, and introduced 
\be
{\cal K}_1 \equiv -\frac{2r^2a_4}{f}\left(1-\frac{fv_6^2}{8a_4v_{10}}\right)^{-1}\,,\qquad
{\cal K}_2 \equiv \frac{4r^2a_4}{f}\left(\frac{f'}{f}+\frac{L}{rh}-\frac2r+\frac{A_0'v_6}{2a_4}\right)^{-1}\,.
\ee
The ghost instabilities are absent if the kinetic matrix ${\tilde {\bm K}}$ is positive definite, 
which is translated to requiring the following three no-ghost conditions, 
\ba
{\tilde{K}}_{33}&>& 0\,,
\label{noghost1}\\
{\tilde{K}}_{11} {\tilde{K}}_{33} -{\tilde{K}}_{13}{\tilde{K}}_{31} &>& 0\,,
\label{noghost2}\\
\det{{\tilde{\bm K}}} &>& 0\,.
\label{noghost3}
\ea
On using the relations among the coefficients shown in Appendix~\ref{even_coeff}, 
the expression of the first no-ghost condition \eqref{noghost1} reduces to 
\be
{\tilde{K}}_{33} 
= \frac{\alpha_1 v_1^2}{L (\alpha_2^2-\alpha_1 \alpha_5)}>0\,.
\label{noghost1b}
\ee
Remembering that $\alpha_1$ is positive by definition, this shows that the first no-ghost condition automatically 
gets satisfied provided the odd-parity stability conditions given in Eq.~(\ref{oddnoLap}). 
Although the remaining two no-ghost conditions are slightly complicated, they can be simplified by the use of the following relation, 
\be
{\cal K}_2'=\frac{(r^2fh)'}{r^2fh}{\cal K}_2
-\frac{1}{8rf^2a_4}\left[L\left(\frac{f^3}{r^2h}\right)'+\frac{2f(rf'-2f)^2}{r^3}
-\frac{4f^2(2f^2\alpha_7+A_0'^2v_{10})}{ra_4}
\right]{\cal K}_2^2\,,
\ee
where we used Eq.~(\ref{v6prime}) in Appendix~\ref{even_coeff} for the derivation of this relation. 
Eliminating ${\cal K}_2'$ in Eq.~\eqref{Kcompts} by using the above relation and substituting the definition of 
${\cal K}_1$ and ${\cal K}_2$, the second and third no-ghost conditions \eqref{noghost2}-\eqref{noghost3} reduce to 
\ba
{\tilde{K}}_{11} {\tilde{K}}_{33} -{\tilde{K}}_{13}{\tilde{K}}_{31}
=\frac{8r^4hv_1^2a_4^3[L(2\sqrt{fh}{\cal P}_1+rA_0'v_6)-8r^2fh\alpha_7]}{L^2f^2(4Lra_4+2r^2hA_0'v_6+\sqrt{fh}{\cal P}_2)^2(\alpha_2^2-\alpha_1\alpha_5)}
&>&0\,,\label{noghost2a}
\\
\det{\tilde{\bm K}}=
\frac{32r^6hv_1^2a_4^3\alpha_7(L-2)(2\sqrt{fh}{\cal P}_1+rA_0'v_6-4r^2fh\alpha_7)}{L^2f^2\phi'^2(4Lra_4+2r^2hA_0'v_6+\sqrt{fh}{\cal P}_2)^2(\alpha_2^2-\alpha_1\alpha_5)}&>&0\,,\label{noghost3a}
\ea
respectively, where 
\ba
{\cal P}_1 \equiv \frac1f\left(\frac{r\sqrt{f}}{\sqrt{h}}\right)'a_4\,,\qquad 
{\cal P}_2 \equiv \frac{8r^3\sqrt{h}}{f}\left(\frac{\sqrt{f}}{r}\right)'a_4\,.
\label{defP1}
\ea
Under the stability conditions in the odd-parity sector given in Eq.~\eqref{oddnoLap}, 
the no-ghost conditions \eqref{noghost2a} and \eqref{noghost3a} are satisfied if 
$L(2\sqrt{fh}{\cal P}_1+rA_0'v_6)-8r^2fh\alpha_7>0$ and $2\sqrt{fh}{\cal P}_1+rA_0'v_6-4r^2fh\alpha_7>0$ hold, respectively. 
Since $L\geq2$ and $\alpha_7 \geq 0$, we notice that the former is always ensured as long as 
the latter holds. Hence, the second and third no-ghost conditions converge to 
\be
\mathcal{K} \equiv 2\left( {\cal P}_1+\frac{v_6 A_0' r}{2 \sqrt{f h}}\right) -4 \sqrt{fh} \alpha_7 r^2>0\,.
\label{noghost3b}
\ee
In conclusion, the no-ghost conditions in the even-parity sector give rise to 
only one additional constraint \eqref{noghost3b} to the stability conditions in the odd-parity sector. 

\subsubsection{Radial and angular Laplacian stability conditions}
By assuming the solution of the form $\vec{\mathcal{Y}}^{t} \propto e^{i (\omega t-kr)}$ in the small-scale limit 
characterized by $k\to0$, the propagation speed of perturbations along the radial direction is expressed as 
$c_r=\omega/(\sqrt{fh}\,k)$ in proper time. Substitution of this expression into the dispersion relation reads 
\be
{\rm det} \left( fhc_r^2 {\tilde{\bm K}}+ {\tilde{\bm G}} \right)=0\,. 
\label{dispersion1}
\ee
The components of matrix ${\tilde{\bm G}}$ are given as 
\ba
&&
\tilde{G}_{11}=-\frac{f^2{\cal K}_2^2}{16r^4a_4^2}
\left[\frac{1}{v_1}\left(A_0'v_1+\frac{\phi'}{2}v_4\right)^2-4(rc_5+c_6)-\frac{4r^2}{L}d_4+\frac{(2rhv_8+Lv_6)^2}{4Lh^2v_9}\right]\,,
\notag\\
&&
\tilde{G}_{12}=\frac{A_0'v_4-2a_2}{2ra_4}\tilde{G}_{11}-\frac{f{\cal K}_2}{4r^4a_4}\left[c_2+rd_2+\frac{v_4}{2v_1}\left(A_0'v_1+\frac{\phi'}{2}v_4\right)\right]\,,
\notag\\
&&
\tilde{G}_{22}=e_2-\frac{v_4^2}{4v_1}+\frac{A_0'v_4-2a_2}{ra_4}\left(\tilde{G}_{12}-\frac{A_0'v_4-2a_2}{4ra_4}\tilde{G}_{11}\right)\,,
\qquad
\frac{\tilde{G}_{13}}{\tilde{K}_{13}}=
\frac{\tilde{G}_{23}}{\tilde{K}_{23}}=
\frac{\tilde{G}_{33}}{\tilde{K}_{33}}=-\frac{fv_6^2-8a_4v_{10}}{8a_4v_9}\,.
\ea
By virtue of the relation between $\tilde{G}_{i3}$ and $\tilde{K}_{i3}$ ($i=1,2,3$), the propagation speed of the vector DOF 
decouples from the other two in the dispersion relation \eqref{dispersion1}. On using the relations among the coefficients 
given in Appendix~\ref{even_coeff}, it reduces to 
\be
c_{r3{\rm ,even}}^2=\frac{\alpha_2^2-\alpha_1\alpha_5}{fh\alpha_1\alpha_4}\,,\label{evencr3}
\ee
Although the remaining two propagation speeds look being coupled with each other in the dispersion relation \eqref{dispersion1}, 
they decouples on the use of the relations among coefficients given in Appendix~\ref{even_coeff} such that 
\ba
c_{r1{\rm ,even}}^2
&=&   -\frac{\alpha_6}{f h \alpha_7}\,,
\label{evencr1}\\
c_{r2{\rm ,even}}^2
&=& \frac{\phi ' \left[v_4 \left(2 A_0'v_1+ \phi' v_4\right)+4 v_1(c_2+4 r d_2)\right]}
{4 v_1 (2\sqrt{fh}{\cal P}_1+rA_0'v_6-4r^2fh\alpha_7)}\,.
\label{evencr2}
\ea
They correspond to the radial propagation speeds of the tenor and scalar mode, respectively. For the absence of 
Laplacian instabilities along the radial direction, we require 
\be
c_{r1{\rm ,even}}^2\geq0\,,\qquad
c_{r2{\rm ,even}}^2\geq0\,,\qquad
c_{r3{\rm ,even}}^2\geq0\,.
\label{evenrLap}
\ee
By observation, we notice that the squared radial propagation speeds of vector \eqref{evencr3} and tensor DOFs \eqref{evencr1} are equivalent to  those in the odd-parity sector given by \eqref{oddcr2} and \eqref{oddcr1}, respectively.  
Thus, the first and third radial Laplacian stability conditions are automatically satisfied as long as the odd-parity sector is stable. 

On the other hand, by assuming the solution of the form $\vec{\mathcal{Y}}^{t} \propto e^{i (\omega t-l \theta)}$ 
in the limit that $L=l(l+1) \gg 1$, the squared propagation speed along the angular direction  (which is dimensionless) is expressed as 
$c_{\Omega}^2=r^2\omega^2/(l^2f)$ in proper time. We substitute this expression into the dispersion relation of the form 
\be
{\rm det}\left(\frac{fl^2c_{\Omega}^2}{r^2} {\tilde{\bm K}}+ {\tilde{\bm M}}\right)=0\,.
\label{dispersion2}
\ee
We expand this equation for $L\gg1$ and pick up the dominant contribution so as to derive the propagation speeds along the angular direction. 
As we will see later, the leading order contribution possesses the linear dependence in $L$, but this contribution identically vanishes 
after the substitution of relations among coefficients. This shows that we need to extract the sub-leading order contribution from 
Eq.~\eqref{dispersion2}. In order to do so, we expand each component of matrices $\tilde{\bm K}$ and $\tilde{\bm M}$ for $L\gg1$ 
up to the sub-leading order and find that the components possess the following $L$-dependence, 
\ba
&&
\tilde{K}_{11}=\frac{\tilde{K}^{(0)}_{11}}{L}+\frac{\tilde{K}^{(1)}_{11}}{L^2}\,,\qquad
\tilde{K}_{12}=\frac{\tilde{K}^{(0)}_{12}}{L}+\frac{\tilde{K}^{(1)}_{12}}{L^2}\,,\qquad
\tilde{K}_{13}=\frac{\tilde{K}^{(0)}_{13}}{L}+\frac{\tilde{K}^{(1)}_{13}}{L^2}\,,\notag\\
&&
\tilde{K}_{22}=\tilde{K}^{(0)}_{22}+\frac{\tilde{K}^{(1)}_{22}}{L}\,,\qquad
\tilde{K}_{23}=\frac{\tilde{K}^{(0)}_{23}}{L}+\frac{\tilde{K}^{(1)}_{23}}{L^2}\,,\qquad
\tilde{K}_{33}=\frac{\tilde{K}^{(0)}_{33}}{L}\,,\nonumber \\
&&
\tilde{M}_{11}=\tilde{M}^{(0)}_{11}+\frac{\tilde{M}^{(1)}_{11}}{L}\,,\qquad
\tilde{M}_{12}=\tilde{M}^{(0)}_{12}+\frac{\tilde{M}^{(1)}_{12}}{L}\,,\qquad
\tilde{M}_{13}=\tilde{M}^{(0)}_{13}+\frac{\tilde{M}^{(1)}_{13}}{L}\,,\notag\\
&&
\tilde{M}_{22}=L\tilde{M}^{(0)}_{22}+\tilde{M}^{(1)}_{22}\,,\qquad
\tilde{M}_{23}=\tilde{M}^{(0)}_{23}+\frac{\tilde{M}^{(1)}_{23}}{L}\,,\qquad
\tilde{M}_{33}=\tilde{M}^{(0)}_{33}+\frac{\tilde{M}^{(1)}_{33}}{L}\,,
\label{KM}
\ea
where the quantities with superscripts $(0)$ and $(1)$ correspond to the coefficients of leading and sub-leading contributions for $L\gg1$, respectively, which do not contain $L$ themselves. 
We note that $\tilde{K}_{33}$ cannot be expanded further as seen from Eq.~\eqref{Kcompts}. 
Substituting these expressions into the dispersion relation \eqref{dispersion2} and expanding it, 
the leading order contribution is in proportion to $L$. As mentioned above, this contribution 
identically vanishes since the quantities $\tilde{M}^{(0)}_{ij}$ $(i, j=1,2,3)$ satisfy 
\be
\frac{\tilde{M}^{(0)}_{11}}{\tilde{K}^{(0)}_{11}}
=\frac{\tilde{M}^{(0)}_{13}}{\tilde{K}^{(0)}_{13}}
=\frac{\tilde{M}^{(0)}_{33}}{\tilde{K}^{(0)}_{33}}
=\frac{8a_4v_{10}-fv_6^2}{8a_4v_1}
\,,\qquad 
\tilde{M}^{(0)}_{11}\tilde{M}^{(0)}_{33}-\left(\tilde{M}^{(0)}_{13}\right)^2=0\,,
\label{comb1}
\ee
under the use of relations among the coefficients given in Appendix~\ref{even_coeff}. 
We then pick up the next-to leading order contribution proportional to $L^0$ in Eq.~\eqref{dispersion2}. 
Although the resultant equation is complicated, we can simplify it by using the following relations 
among coefficients appearing in Eq.~(\ref{KM}), 
\ba
&&
\frac{\tilde{K}^{(0)}_{23}}{\tilde{K}^{(0)}_{13}}=
\frac{\tilde{K}^{(1)}_{23}}{\tilde{K}^{(1)}_{13}}=\frac{fb_3}{2ra_4}\,,\qquad 
\tilde{K}^{(0)}_{12}\tilde{M}^{(0)}_{33}-\tilde{K}^{(0)}_{23}\tilde{M}^{(0)}_{13}
=rh\phi'\tilde{K}_{22}^{(0)}\tilde{K}_{33}^{(0)}\,,\notag\\
&&
\tilde{K}^{(1)}_{11}\tilde{M}^{(0)}_{33}-2\tilde{K}^{(1)}_{13}\tilde{M}^{(0)}_{13}
=r^2h^2\tilde{M}^{(0)}_{33}(\phi'^2\tilde{K}_{22}^{(0)}+4r^2\alpha_7)\,,
\label{comb2}
\ea
together with Eq.~(\ref{comb1}). We also introduce the following two quantities, 
\ba
M_1 &\equiv& \frac{ \tilde{M}_{11}^{(0)} {\tilde{M}}_{33}^{(1)} -2 \tilde{M}_{13}^{(0)} {\tilde{M}}_{13}^{(1)} 
+\tilde{M}_{33}^{(0)} {\tilde{M}}_{11}^{(1)} }
{4 r^2 f h^2 \alpha_7 \tilde{M}_{33}^{(0)}} \,,\label{defM1} \\
M_2 &\equiv& \frac{ \phi'(r h \phi' {\tilde{M}}_{22}^{(0)} \tilde{M}_{33}^{(0)}+ 
2\tilde{M}_{13}^{(0)} {\tilde{M}}_{23}^{(0)} -2 {\tilde{M}}_{12}^{(0)}\tilde{M}_{33}^{(0)}) }
{4 r f h v_1 \alpha_7 \tilde{M}_{33}^{(0)}} \,. 
\label{defM2}
\ea
Then, the next-to leading order contribution proportional to $L^0$ in Eq.~\eqref{dispersion2} 
can be expressed as 
\be
\left(c_{\Omega}^2+\frac{r^2\tilde{M}_{33}^{(0)}}{f\tilde{K}_{33}^{(0)}}\right)
\left[\left(c_{\Omega}^2 +M_1+M_2\right) 
\left( c_{\Omega}^2+\frac{r^2 \tilde{M}_{22}^{(0)}}{f \tilde{K}_{22}^{(0)}}\right) 
- \frac{r^2 \left( 4 f \alpha_7 M_2+ \phi'^2 \tilde{M}_{22}^{(0)} \right)^2}
{4 f^2 \phi'^2 \alpha_7 \tilde{K}_{22}^{(0)}}\right]=0\,,
\label{evencOeq}
\ee
where the matrix components appearing in the above expression and $M_1$, $M_2$ are given by 
\ba
&&
K_{22}^{(0)}=e_1\,,\qquad
M_{11}^{(0)}=-\frac{r^2v_6^2}{4v_1}\,,\qquad
M_{13}^{(0)}=-\frac{rv_6}{2}\,,\qquad
M_{33}^{(0)}=-v_1\,,\notag\\
&&
M_{11}^{(1)}
=r^2d_4-\frac{m_1^2}{4a_4^2v_9}+\frac{rh(m_1-ra_4v_8)m_2}{a_4v_1v_6}
+\frac14\left(\frac{rv_6m_1}{a_4v_9}-\frac{2r^2hm_2}{v_1}\right)'\,,\notag\\
&&
M_{33}^{(1)}
=-\frac{m_3^2}{a_4^2v_9}-\frac{hA_0'v_1m_4}{a_4}+\left(\frac{v_1m_3}{a_4v_9}\right)'\,,\notag\\
&&
M_{13}^{(1)}
=-\frac{m_1m_3}{2a_4^2v_9}-\frac{rhA_0'm_2}{2a_4}+\frac{h(m_1-ra_4v_8)m_4}{2a_4v_6}
+\frac14\left(\frac{v_1m_1+rv_6m_3}{a_4v_9}-rhm_4\right)'\notag\\
&&
M_{22}^{(0)}
=e_4+\frac{2hc_4a_6}{a_4}-\frac{m_5^2}{4a_4^2v_9}\,,\notag\\
&&
M_{12}^{(0)}
=-\frac{rd_3}{2}-\frac{hc_4(m_1-ra_4v_8)}{a_4v_6}-\frac{rha_6m_2}{2a_4v_1}
+\frac{rv_5v_6}{4v_1}+\frac{m_1m_5}{4a_4^2v_9}
+\frac14\left(2rhc_4+\frac{r(2d_2v_1-v_4v_6)}{2v_1}-\frac{rv_6m_5}{2a_4v_9}\right)'\,,\notag\\
&&
M_{23}^{(0)}
=\frac{hA_0'c_4v_1}{a_4}+\frac{v_5}{2}+\frac{m_3m_5}{2a_4^2v_9}-\frac{ha_6m_4}{2a_4}
-\frac14\left(v_4+\frac{v_1m_5}{a_4v_9}\right)'\,,
\ea
with the shortcut notations 
\ba
&&
m_1 \equiv ra_4v_8+\left(a_4+ra_9-\frac{rA_0'v_6}{2}\right)v_6\,,\qquad 
m_2  \equiv 2c_5v_1+\left(A_0'v_1+\frac{\phi'v_4}{2}\right)v_6\,,\notag\\
&&
m_3 \equiv a_4v_1'-\frac{A_0'v_1v_6}{2}\,,\qquad
m_4 \equiv 2A_0'v_1+\phi'v_4\,,\qquad 
m_5 \equiv a_6v_6+a_4v_{13}\,.
\ea

In a manner analogous to the radial propagation speeds, the angular propagation speed of the vector DOF decouples from the other two propagation speeds in Eq.~(\ref{evencOeq}). Denoting the vector propagating speed as $c_{\Omega 3{\rm ,even}}$, it is given by 
\be
c_{\Omega 3{\rm ,even}}^2
=-\frac{r^2\tilde{M}_{33}^{(0)}}{f\tilde{K}_{33}^{(0)}}
=\frac{r^2 \left(f v_6^2-8 a_4 v_{10}\right)}{8 f a_4 v_1}\geq0\,,
\label{evencO1}
\ee
whose positivity is required for the Laplacian stability of the perturbation $V$.  This forms our third angular Laplacian stability condition, as indicated in the subscript\footnote{This is referred as the third angular Laplacian stability condition since it belongs to the vector DOF, which appears as the third one in \eqref{mathcalY}. For a more convenient way of narration, this one is discussed before the first and second  angular Laplacian stability conditions.}. 
The remaining two propagation speeds associated with the gravitational and scalar DOFs generally couple with each other and 
can be expressed as 
\ba
c_{\Omega \pm{\rm ,even}}^2
=\frac12\left[-\left(M_1+M_2+\frac{r^2 \tilde{M}_{22}^{(0)}}{f \tilde{K}_{22}^{(0)}}\right)
\pm\sqrt{\left(M_1+M_2-\frac{r^2 \tilde{M}_{22}^{(0)}}{f \tilde{K}_{22}^{(0)}}\right)^2
+\frac{r^2 \left( 4 f \alpha_7 M_2+ \phi'^2 \tilde{M}_{22}^{(0)} \right)^2}
{ f^2 \phi'^2 \alpha_7 \tilde{K}_{22}^{(0)}}}\right]
\label{evencO23}
\ea
For the absence of Laplacian instabilities associate with the perturbation $\psi$ and $\delta\phi$, we require that the squared propagation speeds given in Eq.~\eqref{evencO23} are real and non-negative. These conditions [with the assistance of the Vieta's formula and Eq.\eqref{evencOeq}], with the nearest two below referred as the first and second angular Laplacian stability condition, respectively, are  translated to give
\ba
&& \mathcal{M}_1 \equiv  -\left( M_1 + M_2 + \frac{r^2 {\tilde{M}}_{22}^{(0)}}{f {\tilde{K}}_{22}^{(0)}} \right) \geq 0 \,, \label{evencO2}\\
&& \mathcal{M}_2 \equiv \frac{r^2 (M_1+M_2) {\tilde{M}}_{22}^{(0)}}{f {\tilde{K}}_{22}^{(0)}} - \frac{r^2 \left( 4f \alpha_7M_2 + \phi^{\prime 2} {\tilde{M}}_{22}^{(0)} \right)^2}{4f^2 \phi^{\prime 2} \alpha_7 {\tilde{K}}_{22}^{(0)}} \geq 0 \,, 
\label{evencO3}
\ea
and 
\be
\mathcal{M}_1^2-4\mathcal{M}_2=\left(M_1+M_2-\frac{r^2 \tilde{M}_{22}^{(0)}}{f \tilde{K}_{22}^{(0)}}\right)^2
+\frac{r^2 \left( 4 f \alpha_7 M_2+ \phi'^2 \tilde{M}_{22}^{(0)} \right)^2}
{ f^2 \phi'^2 \alpha_7 \tilde{K}_{22}^{(0)}}\geq0\,,
\label{evencO4}
\ee
with \eqref{evencO4} guaranteeing that $c_{\Omega \pm{\rm ,even}}^2$ are real. 
Imposing the Laplacian stability condition \eqref{oddnoLap} in the odd-parity sector, the quantity $\alpha_7$ 
in the third condition can not be negative. Moreover, on using the relations of coefficients in Appendix~\ref{even_coeff} 
and the definition of ${\cal P}_1$ given in Eq.~\eqref{defP1}, the quantity $\tilde{K}_{22}^{(0)}$ in the third condition 
can be written as 
\be
\tilde{K}_{22}^{(0)}=e_1=\frac{\cal K}{\sqrt{fh}\phi'^2}\,, 
\label{conK220}
\ee
which must be positive since ${\cal K}$ corresponds to the no-ghost condition derived in Eq.~\eqref{noghost3b}. 
These facts show that the last condition \eqref{evencO4} is automatically guaranteed as long as the Laplacian stability in the odd-parity sector and the no-ghost condition in the even-parity sector are satisfied.  

In spite of the complexity of stability conditions \eqref{noghost3b}, \eqref{evenrLap}, \eqref{evencO1}, \eqref{evencO2}, and \eqref{evencO3}, some of the stability conditions against even-parity perturbations can be omitted since they 
are redundant with those against odd-parity perturbations as we have already seen. 
Let us summarize that before closing this subsection. 
For the absence of ghosts, we found that only one additional condition, $\mathcal{K}>0$, 
to the odd mode stability is required where $\mathcal{K}$ is given in Eq.~\eqref{noghost3b}. 
The tensor, scalar, and vector propagation  squared speeds along the radial direction, i.e., 
$c_{r1{\rm ,even}}^2$, $c_{r2{\rm ,even}}^2$, and $c_{r3{\rm ,even}}^2$, are given in 
Eqs.~\eqref{evencr1}, \eqref{evencr2} and \eqref{evencr3}, respectively. Among them, we find 
that the tensor and vector propagation speeds are equivalent to the radial propagation speeds 
of corresponding DOFs against odd-parity perturbations given in Eqs.~\eqref{oddcr1} and 
\eqref{oddcr2}, respectively. Hence, the radial Laplacian stability condition against even-parity 
perturbations gives rise to only one additional condition $c_{r2{\rm ,even}}^2\geq0$ to 
the odd mode stability. On the other hand, the angular propagation speeds against even-parity 
perturbations are independent of the odd mode stability. Hence, the angular Laplacian stabilities 
require three conditions. The positivity of $c_{\Omega 3{\rm ,even}}^2$ is satisfied under 
the condition \eqref{evencO1}. The tensor and scalar propagation squared speeds, 
$c_{\Omega \pm{\rm ,even}}^2$, are guaranteed to be positive under the two conditions 
\eqref{evencO2} and \eqref{evencO3}. 

In summary, we need to consider five additional conditions~\eqref{noghost3b}, \eqref{evencr2}, 
\eqref{evencO1}, \eqref{evencO2}, and \eqref{evencO3}\footnote{The relative algebraic expressions of all the nine stability conditions mentioned in this section can be found in \cite{supplemental}.}, to the stability of odd-parity sector. 
Nevertheless, it would be still very hard to carry out 
the stability analysis for the most general case. Instead, we shall process to next section and 
consider several concrete models.  

\subsection{Linear stability conditions for $l=0$}

We consider the monopole perturbation characterized by $l=0$, i.e., $L=0$, in which case the quantities $h_0$, $h_1$, and $G$ identically vanish away from the second-order action of the even-parity sector \cite{Kase:2023kvq}. This means that one can use the gauge DOFs on ${\cal T}$ and $\Theta$ for our purpose other than to eliminate $h_0$ and $G$ as in Eq.~(\ref{gauge1}). However, the gauge DOF will not be completely fixed in such a case. Thus, we adopt the same gauge characterized by Eq.~(\ref{gauge1}) as in the case for $l\geq2$. Substituting $L=0$ into the second-order action \eqref{acteven} with Eqs.~\eqref{Lu} and \eqref{LdA2}, it reduces to 
\ba
{\cal S}_{\rm even}^{l=0} 
&=& \int {\rm d}t\, {\rm d}r 
\Bigg[
v_1\left\{2V\left(\delta A_0'-\dot{\delta A}_1+
\frac{v_3H_2+v_4\delta\phi'+v_5\delta\phi}{2v_1}\right)-V^2\right\}
-\frac{\left(v_3H_2+v_4\delta\phi'+v_5\delta\phi\right)^2}{4v_1}
\notag\\
&&
+\left(\Phi'+A_0'v_1V\right)H_0
-\frac{2}{f}\dot{\Phi}H_1
+\left( c_2 \delta \phi'+ c_3 \delta \phi\right)H_2
+c_6H_2^2
+e_1 \dot{\delta \phi}^2+e_2 \delta \phi'^2
+e_3 \delta \phi^2
\Bigg]
\,,
\label{acteven0}
\ea
where we introduced 
\be
\Phi \equiv a_1\delta\phi'+\left(a_2-a_1'-\frac12A_0'v_4\right)\delta\phi+a_3H_2\,, 
\label{defPhi}
\ee
and used the relations among coefficients in Appendix~\ref{even_coeff}. 
Compared to Eqs.~\eqref{Lu} and \eqref{LdA2} for $l\neq0$, the perturbations $H_1$, $\delta A_0$, $\delta A_1$ 
do not possess the quadratic terms in Eq.~\eqref{acteven0}. This fact shows that, in addition to $H_0$, these variables 
also reduce to Lagrange multipliers giving rise to constraints on the other variables for $l=0$. Indeed, the variation of 
the action (\ref{acteven0}) with respect to $\delta A_0$, $\delta A_1$, $H_0$, and $H_1$ lead to the following constraints 
\ba
&&(v_1V)'=0\,,\label{eqdA00}\\ 
&&\dot{V}=0\,,\label{eqdA10}\\ 
&&\Phi'+A_0'v_1V=0\,,\label{eqH00}\\ 
&&\dot{\Phi}=0\,,\label{eqH10}
\ea
respectively. Integrating Eqs.~\eqref{eqdA00} and \eqref{eqdA10}, we obtain 
\be
V=\frac{{\cal C}_1}{v_1}\,,
\label{solV0}
\ee
where ${\cal C}_1$ denoting an integration constant. Substituting this solution into 
Eq.~(\ref{eqH00}) and integrating it together with Eq.~\eqref{eqH10}, we obtain 
\be
\Phi={\cal C}_2-{\cal C}_1A_0\,,
\label{solPhi0}
\ee
where ${\cal C}_2$ is an integration constant. From the definition of $\Phi$ in Eq.~\eqref{defPhi}, 
the above solution leads a constraint on the perturbation $H_2$ as 
\be
H_2=\frac{1}{a_3}\left[{\cal C}_2-{\cal C}_1A_0
-a_1\delta\phi'-\left(a_2-a_1'-\frac12A_0'v_4\right)\delta\phi\right]\,.
\label{solH20}
\ee
We substitute Eqs.~\eqref{solV0}, \eqref{solPhi0}, and \eqref{solH20} into the second-order action (\ref{acteven0}) for $l=0$. 
This operation removes all the metric perturbations, i.e.,  $H_0$, $H_1$, and $H_2$, and the vector perturbation $V$ from the action. 
In other words, we find that the monopole perturbation is governed by the scalar perturbation $\delta\phi$ as a single DOF. 
Omitting the integration constants ${\cal C}_1$ and ${\cal C}_2$ irrelevant to the dynamics of perturbations, 
the resultant action is expressed as 
\be
{\cal S}_{\rm even}^{l=0} 
= \int {\rm d}t\, {\rm d}r 
\left[
e_1\dot{\delta\phi}^2+\left(e_2-\frac{v_4^2}{4v_1}\right)\delta\phi'^2
+M_0\delta\phi^2
\right]\,,
\ee
where $M_0$ is given by 
\ba
M_0 \equiv e_3+\frac{b_3(b_3c_6-b_4c_3)}{b_4^2}-\frac{(b_3v_3-b_4v_5)^2}{4b_4^2v_1}
+\frac12\left(\frac{b_3c_2}{b_4}-\frac{(b_3v_3-b_4v_5)v_4}{2b_4v_1}\right)'\,.
\ea
The above action shows that the no-ghost condition for the monopole perturbation is given by $e_1>0$. 
As we have seen in Eq.~(\ref{conK220}), this is equivalent to the no-ghost condition for the perturbations with $l\geq2$. 
The propagation speed square along the radial direction, $c_{r,{\rm even}}^2=-(e_2-v_4^2/v_1)/e_1$, also coincides with that for $l\geq2$ derived in Eq.~\eqref{evencr2}. Thus, we have shown that the stability conditions derived for $l\geq2$ also ensure the absence of 
instabilities in the monopole perturbation. 

\subsection{Linear stability conditions for $l=1$}

We proceed to study the dipole perturbations characterized by $l=1$ ($L=2$) in which case the metric perturbations 
$K$ and $G$ appear in the second-order action only through the combination of the form $G-K$ \cite{Kase:2023kvq}. 
This means that, instead of using the gauge DOFs of ${\cal R}$ and $\Theta$ to eliminate both $G$ and $K$ as we adopted in Eq.~(\ref{gauge1}), it is possible to keep one of them by eliminating the combination $G-K$ and fixing either ${\cal R}$ or $\Theta$. 
Since Eq.~\eqref{gaugetrans} shows that the complete gauge choice of $\Theta$ can be obtained only via the transformation of the perturbation $G$, we use this gauge DOF to eliminate $G$. The residual gauge DOF ${\cal R}$ can also be completely fixed via the transformation of $\delta\phi$. Thus, we choose the following gauge for the dipole perturbation, 
\be
h_0=0\,,\qquad 
\delta\phi=0\,,\qquad 
G=K\,.
\label{gauge3}
\ee
The elimination of non-dynamical variables in the second-order action for $l=1$ can be operated in the same way for $l\geq2$ 
which we discussed below Eq.~\eqref{psi}. After this process, the resultant action is composed of two dynamical variables, 
$\psi$ and $V$, showing that the dipole perturbation possesses one less propagating DOF than the case with $l\geq2$. 
In an analogous way to Sec.~\ref{noghostsec}, we find that the no-ghost conditions for these two DOFs coincide with Eqs.~\eqref{noghost1b} and \eqref{noghost2a} substituted $L=2$. The squared propagation speeds along the radial direction for $l=1$
coincide with Eqs.~\eqref{evencr3} and \eqref{evencr2}, i.e., the propagation speeds of vector and scalar field perturbations for $l\geq2$, respectively. These facts show that the dipole perturbation is governed by vector and scalar field perturbations. 
Consequently, we find that the stability of the dipole perturbation does not give rise to additional conditions to those derived for $l\geq2$. 

\section{Application to the concrete models}
\label{application}

In Ref.~\cite{Heisenberg:2018vti},  three different concrete models possessing hairy BH solutions on the spherically symmetric spacetime were proposed based on the $U(1)$ GI SVT theories.
The stability of such models against odd-parity perturbations are studied in Ref.~\cite{Heisenberg:2018mgr}. These models are characterized by the following choices of functions 
\ba
\label{model1}
&{\rm Model\ 1}:\quad
&f_3= \beta_3,\qquad 
f_4= 0\,,\\
\label{model2}
&{\rm Model\ 2}:\quad
&f_3= \beta_3,\qquad 
f_4= \beta_4\,,\\
\label{model3}
&{\rm Model\ 3}:\quad
&f_3= \beta_3,\qquad 
f_4= \beta_4X\,,
\ea
respectively, where $\beta_3$ and $\beta_4$ are arbitrary constants under the constraints led by the odd-parity stability analysis (and later will be further confined by the even-parity ones). The other functions are common to all three models and are chosen as $f_2=X+F$ plus ${\tilde f}_3={\tilde f}_4=0$. 
In the following, we shall discuss the stability conditions against odd-parity perturbations, Eqs.~\eqref{oddnoghost} and \eqref{oddnoLap}, and those against even-parity perturbations, Eqs.~\eqref{noghost3b}, \eqref{evenrLap}, \eqref{evencO1}, \eqref{evencO2} and \eqref{evencO3}, for each model. 
Notice that, for the convenience of readers, some of the main results are summarized at the end of this section in Table \ref{table1}. One can move to there directly when the concluding remarks are demanded. 
On the other hand, due to the horrible length of some of the mathematical expressions in this section, some of the selected ones are solely$\backslash$also shown in \cite{supplemental} and$\backslash$or Appendix~\ref{secVexpression}. 

\subsection{Stability analysis for the Model 1}
\label{applicationA}

In this model, the quantities $\alpha_1$, $\alpha_6$, and $\alpha_7$ associated with stability conditions against odd-parity perturbations reduce to  
\be
\alpha_1=\frac{M_{\rm pl}^2\sqrt{h}}{4\sqrt{f}}\,,\qquad
\alpha_6=-\frac{\sqrt{fh}M_{\rm pl}^2}{4r^2}\,,\qquad 
\alpha_7=\frac{M_{\rm pl}^2}{4r^2\sqrt{fh}}\,.
\ee
This shows that one of the no-ghost conditions, $\alpha_6<0$, is trivially satisfied in Eq.~(\ref{oddnoghost}). 
Moreover, the propagation speed of gravitational DOF along the radial and angular direction reduce to that of light, i.e., $c_{r1,{\rm odd}}^2=1=c_{\Omega1,{\rm odd}}^2$. This fact shows that the Laplacian stability conditions for the gravitational DOF in the odd-parity sector are also automatically satisfied. The stability conditions associated with the vector field perturbation remain nontrivial. Thus, in Eqs.~\eqref{oddnoghost} and \eqref{oddnoLap}, the nontrivial stability conditions for this model are 
\be
\alpha_4>0\,,\qquad
\alpha_2^2-\alpha_1\alpha_5\geq0\,,\qquad 
\alpha_8^2-4\alpha_7\alpha_9\geq0\,, 
\label{oddconM1}
\ee
which are consistent with the Eqs.~(3.23), (3.25) and (3.28) in Ref.~\cite{Heisenberg:2018mgr}. 
The above constraints can be numerically translated to the phase space of $\{\beta_3, \beta_4\}$ 
in the models~(\ref{model2}) and (\ref{model3}). 
An overall constraint on $\{\beta_3, \beta_4\}$  from the odd-parity sector can be found in 
Ref.~\cite{Heisenberg:2018mgr} by comprehensively considering all the three models discussed in here. 
We shall go back to this point in the following subsections. 

In the eve-parity sector, the no-ghost condition represented by Eq.~\eqref{noghost3b} reduces to  
\ba
\mathcal{K} &=& \frac{2 r_h^4 h^2 A_0^{\prime 4} {\tilde \beta}_3^2}{M_{\rm pl}^2 f^2} \,,
\label{noghost3M1}
\ea
where we have defined the dimensionless factor ${\tilde \beta}_3 \equiv M_{\rm pl} \beta_3/ r_h^2$ 
and $r_h$ stands for the radii of the event horizon so that we have $f(r_h)=h(r_h)=0$. 
Since Eq.~(\ref{noghost3M1}) shows that the quantity $\mathcal{K}$ is positive for any 
non-zero value of ${\tilde \beta}_3$, the third no-ghost condition will always get satisfied\footnote{
According to the (3.14) and (3.15) of \cite{Heisenberg:2018vti}, we will have $f/h \to 1$ as $r \to r_h$. Thus, $\mathcal{K}$ 
won't be zero at the event horizon so that it has to stay positive. In addition, notice that, here we are assuming $A_0' \neq 0$.}.
We note that the background equations~\eqref{be1}-\eqref{be4} have been used during simplifications 
and we shall apply the same operation without additional alerts in the following. 
Regarding Eqs.~\eqref{be3}-\eqref{be4}, we especially notice that the conditions $J_\phi(r)=0$ and 
$J_A(r)=\text{constant} \equiv r_h^2 \kappa$ hold as discussed in the end of Sec.~\ref{modelsec}. 
Thus, we have  
\ba
\phi' = \frac{2 h \tilde{\beta }_3 A_0'^2 r_h^2}{rf M_{\text{pl}}}\,,
\qquad A_0' = \frac{\kappa  \sqrt{f}r_h^2 M_{\text{pl}}}{r \sqrt{h} \left(4 h \tilde{\beta }_3 r_h^2 \phi '+r M_{\text{pl}}\right)}\,.
\label{beM1}
\ea
Here, $\kappa$ is temporarily borrowed to denote a dimensionless constant. 
With the polynomial solutions given by Eqs.~(3.14)-(3.16) in Ref.~\cite{Heisenberg:2018vti} 
and evaluation of \eqref{JA} at $r_h$, we notice that for Model 1 we have 
$\kappa = \sqrt{2 \mu} M_{\text{pl}}/r_h$, where $\mu \in (0, 1)$ so that 
$\left(r_h/ M_{\text{pl}}\right)\kappa \in (0, \sqrt{2})$~\cite{Heisenberg:2018mgr}.

On the other hand, for Model 1, the squared second propagation speed given by Eq.~\eqref{evencr2} reduces to
\ba
c_{r2{\rm ,even}}^2 = 1\,,
\label{rLap23M1}
\ea
which shows that the second Laplacian stability condition along the radial direction is automatically satisfied. 

Regarding the third Laplacian stability condition along the angular direction, we plug the full expressions 
of $v_6$, $v_{10}$, etc., into Eq.~\eqref{evencO1}, we find that 
\ba
c_{\Omega 3{\rm ,even}}^2 &=& {\tilde{ K}}_{33} \frac{2 h L \left[2 f^2 h (h+1) \tilde{\beta }_3^2 A_0'^2 r_h^4 M_{\text{pl}}^4-f h^2 r^2 \tilde{\beta }_3^2 A_0'^4 r_h^4 M_{\text{pl}}^2-12 h^4 \tilde{\beta }_3^4 A_0'^6 r_h^8+f^3 r^2 M_{\text{pl}}^6\right]^2}{r^2 (f h)^{3/2} M_{\text{pl}}^6 \left(8 h^2 \tilde{\beta }_3^2 A_0'^2 r_h^4+f r^2 M_{\text{pl}}^2\right)^3} \,.
\label{aLap1M1}
\ea
From its mathematical structure we notice that, $c_{\Omega 3{\rm ,even}}^2$ is definitely 
a non-negative quantity provided that $\tilde{K}_{33}>0$ holds under the stability of 
odd-parity sector. Thus, this angular Laplacian stability condition gets satisfied. 

We proceed to the first angular Laplacian stability condition in which the story becomes 
a little sophisticated. To deal with this problem, we first convert $\mathcal{M}_1$ given in Eq.~(\ref{evencO2}) 
into a dimensionless form by 
(i) solving 
Eqs.~\eqref{beM1} for $A_0'$ \footnote{Notice that, this set of algebraic equations have 3 roots originally. 
However, only one of them is real so we keep it as our solution for $A_0'$.} and substituting it into $\mathcal{M}_1$; 
(ii) replacing the model parameter $\beta_3$ in $\mathcal{M}_1$ with the dimensionless ${\tilde \beta}_3$; 
(iii) changing the variable $r$ to the dimensionless one $\xi \equiv r_h/r$; 
(iv) taking advantage of the freedom for choosing unit system, setting $M_{\text{pl}}=r_h$ (cf., footnote \#1). 
As a result, a dimensionless form of $\mathcal{M}_1$ is obtained as a function of 
$\{f, h, \xi, \kappa, {\tilde \beta}_3\}$. We postpone its lengthy explicit expression 
to Appendix~\ref{secVexpression}. Knowing the fact that $f, h \in [0, 1]$, $\xi \in (0, 1]$ 
since we have $r \in [r_h, +\infty)$, $\kappa \in (0, \sqrt{2})$ and $| {\tilde \beta}_3 | \lesssim {\cal O}(1)$\cite{Heisenberg:2018mgr}, we can attempt to determine the range of $\mathcal{M}_1$ in the 5-dimension phase space spanned by $\{f, h, \xi, \kappa, {\tilde \beta}_3\}$. In practice, all of these could be fulfilled numerically on {\it Mathematica}. It turns out that the minimum of $\mathcal{M}_1$ in this phase space is indeed positive, 
which is about\footnote{Strictly speaking, the true minimum should be bigger than this, since $f$ and $h$ are correlated so that certain regions of this phase space won't be reached.} $2.001$.
This result guarantees this angular Laplacian stability condition. 

The second angular Laplacian stability condition (\ref{evencO3}) is even more complicated. 
To analyze $\mathcal{M}_2$, new strategies have to be selected.  After some attempts, we know 
{\it a posteriori} that new constraints on ${\tilde \beta}_3$ have to be performed for this stability 
condition to hold. Now the mission is to determine the new constraint. For this goal and by referring 
the analysis to the first angular Laplacian stability condition in which the minimum of 
$\mathcal{M}_1$ is achieved near the $r=r_h$, as well as the stability conditions in odd-parity sector 
studied in Ref.~\cite{Heisenberg:2018mgr} in which the previously mentioned constraint on 
${\tilde \beta}_3$ is obtained near the $r=r_h$, we shall focus on the behavior of $\mathcal{M}_2$ 
near the horizon $r \to r_h$. 
As usual, a dimensionless form of $\mathcal{M}_2$ could be obtained by following 
the same procedure mentioned earlier. However, since we are focusing on the $r \to r_h$ limit, 
the phase space now reduces to 2 dimension, spanned by $\{\mu, {\tilde \beta}_3\}$. 
Now the $\mathcal{M}_2$ could be written as 
\ba
\left. \mathcal{M}_2 \right|_{r \to r_h} &=&  \Big[ 6144 (\mu -1)^2 \mu ^5 (3 \mu -4) \tilde{\beta }_3^8-3200 (\mu -1)^2 \mu ^4 \tilde{\beta }_3^6+16 \mu ^2 \left(9 \mu ^2-10 \mu +1\right) \tilde{\beta }_3^4 \notag\\
&& -8192 \mu ^6 \left(2 \mu ^2-5 \mu +3\right)^2 \tilde{\beta }_3^{10}-8 (\mu -1) \mu  \tilde{\beta }_3^2+1 \Big] {\left[1-4 (\mu -1) \mu  \tilde{\beta }_3^2\right]^{-2}}  \,.
\label{aLap3M1}
\ea
We plot it out in Fig.~\ref{plot1} of Appendix~\ref{plotsanalysis}. By looking at the contour of 
the allowed region for $\mathcal{M}_2$, we can read off the new upper bound for legal 
$|{\tilde \beta}_3 |$, which is approximately given by $|{\tilde \beta}_3| \lesssim 0.441$. 
In addition, it's worth mentioning here that, by adopting this new upper bound and the numerical method for analyzing   $\mathcal{M}_1$ with {\it Mathematica}, the minimum of  $\mathcal{M}_2$ in the allowed region of phase space $\{f, h, \xi, \kappa, {\tilde \beta}_3\}$ is indeed found to be positive. Thus, we conclude that, under the new constraint $|{\tilde \beta}_3| \lesssim 0.441$, the second angular Laplacian stability condition is guaranteed. 

\subsection{Stability analysis for the  Model 2}
\label{applicationB}

As before, we first consider the third no-ghost condition \eqref{noghost3b}. 
For Model 2 characterized by Eq.~\eqref{model2}, $\mathcal{K}$ becomes 
\ba
\mathcal{K} &=& \frac{2 h^2 \tilde{\beta }_3^2 A_0'^4 r_h^4}{f^2 M_{\text{pl}}^2} \,,
\label{noghost3M2}
\ea
 which happens to be identical to Eq.~\eqref{noghost3M1}. Thus, for any non-zero ${\tilde \beta}_3$, the third no-ghost condition will always get satisfied.
As mentioned in last subsection, the background equations \eqref{be1}-\eqref{be4} have been 
used during simplifications. By the same procedure to obtain Eq.~\eqref{beM1}, 
Eqs.~\eqref{be3}-\eqref{be4} lead to 
\ba
\phi' = \frac{2 h \tilde{\beta }_3 A_0'^2 r_h^2}{f r M_{\text{pl}}}\,,\qquad 
A_0' = \frac{\kappa\sqrt{f}  r_h^2 M_{\text{pl}}}{\sqrt{h} \left(-8 h \tilde{\beta }_4 r_h^2 M_{\text{pl}}+8 \tilde{\beta }_4 r_h^2 M_{\text{pl}}+4 h r \tilde{\beta }_3 r_h^2 \phi '+r^2 M_{\text{pl}}\right)}\,,
\label{beM2}
\ea
where we have defined the dimensionless factor ${\tilde \beta}_4 \equiv r_h^{-2} \beta_4$.
With the polynomial solutions given in Eqs.~(4.5)-(4.7) of Ref.~\cite{Heisenberg:2018vti} 
and evaluate Eq.~\eqref{JA} at $r_h$, we notice that for Model 2 we have 
$\kappa = \sqrt{2 \mu (1+{\tilde \beta}_4 )} M_{\text{pl}}/r_h$, where $\mu \in (0, 1)$. 
Since the odd-mode stability conditions give the constraint ${\tilde \beta}_4 \in (-0.125, 0.251)$ 
(see also Eqs.~(4.12)-(4.14), (4.20) and (4.21) in Ref.~\cite{Heisenberg:2018mgr}), 
we obtain that $\left(r_h/ M_{\text{pl}}\right)\kappa \in (0, 1.59)$.
To manifest the constraints on ${\tilde \beta}_4$, we plot out the allowed region given by 
the stability conditions against odd-parity perturbations, i.e., Eqs.~(4.12)-(4.14), (4.20) and (4.21) 
in Ref.~\cite{Heisenberg:2018mgr} in the 3-dimension phase space 
$\{{\tilde \beta}_3, {\tilde \beta}_4, \mu\}$ in the panel (a) of Fig.~\ref{plotbeta4} 
in Appendix~\ref{plotsanalysis}, where the constraints $\mu \in (0, 1)$ and 
$| {\tilde \beta}_3 | \lesssim 0.441$ are assumed. From this figure we learn that, 
to meet these constrains for any legal arbitrary $\mu$, there must be upper and lower 
bounds on ${\tilde \beta}_4$. We further observe that, the most stringent constraints are 
from the $\mu \approx 0$ case. Thus, we plot out the allowed region in 
$\{{\tilde \beta}_3, {\tilde \beta}_4\}$ phase space in the panel (b) of Fig.~\ref{plotbeta4} 
in Appendix~\ref{plotsanalysis} by setting $\mu = 0$. Finally, from this panel we read off 
the constraints of ${\tilde \beta}_4$, which is about ${\tilde \beta}_4 \in (-0.125, 0.251)$\footnote
{Interestingly, an upper bound to ${\tilde \beta}_4$, which is quite similar to the one found in here, could also be obtained by solely doing the even-parity stability analysis. Ignoring the tolerable numerical errors, we conclude that the results from odd- and even-parity sectors are consistent. }. 

On the other hand, for Model 2, the squared second radial propagation speed 
given in Eq.~\eqref{evencr2} reduces to 
\ba
c_{r2{\rm ,even}}^2 = 1\,,
\label{rLap23M2}
\ea
which shows that the second radial Laplacian stability condition is automatically satisfied. 

Next, let us substitute the full expressions of $v_6$, $v_{10}$, etc., into Eq.~\eqref{evencO1} 
to obtain
\ba
c_{\Omega 3{\rm ,even}}^2 &=& \frac{{\tilde{ K}}_{33}}{r^2 (f h)^{3/2} M_{\text{pl}}^6 \left[8 h^2 \tilde{\beta }_3^2 A_0'^2 r_h^4+f M_{\text{pl}}^2 \left(8 \tilde{\beta }_4 r_h^2 (1-h)+r^2\right)\right]^3}  \, \notag\\
&& \times 2 h L \Big\{ 2 f^2 h A_0'^2 r_h^2 M_{\text{pl}}^4 \left[\left(\tilde{\beta }_3^2+8 \tilde{\beta }_4^2\right) r_h^2+h \left(\tilde{\beta }_3^2-8 \tilde{\beta }_4^2\right) r_h^2+r^2 \tilde{\beta }_4\right] \notag\\
&& -f h^2 \tilde{\beta }_3^2 A_0'^4 r_h^4 M_{\text{pl}}^2 \left(-32 h \tilde{\beta }_4 r_h^2+8 \tilde{\beta }_4 r_h^2+r^2\right)-12 h^4 \tilde{\beta }_3^4 A_0'^6 r_h^8+f^3 M_{\text{pl}}^6 \left(4 (h-1) \tilde{\beta }_4 r_h^2+r^2\right)\Big\}^2. \notag\\
\label{aLap1M2}
\ea
Similar to Eq.~(\ref{aLap1M1}), this result shows that $c_{\Omega 3{\rm ,even}}^2$ is definitely non-negative 
as long as $\tilde{K}_{33}>0$ holds, so that this angular Laplacian stability is guaranteed. 

Moving to the first angular Laplacian stability condition, the story is even more sophisticated 
than that of Model 1 due to the presence of ${\tilde \beta}_4$. Because of this, the analysis of 
$\mathcal{M}_1$ will be divided into two parts. The first attempt is about the behavior around $r_h$. 
As usual, a dimensionless form of $\mathcal{M}_1$ is needed. The basic steps for obtaining that is 
the same as what we did in last subsection. The only difference is that the resultant $\mathcal{M}_1$ 
is now a function of six variables, i.e., $\{f, h, \xi, \kappa, {\tilde \beta}_3, {\tilde \beta}_4\}$. 
On top of that, it became more accessible to expand $\mathcal{M}_1$ around the metric horizon. 
To the lowest order of $(r-r_h)$, that leads to
\ba
\left. \mathcal{M}_1 \right|_{r \to r_h}  &=&  2 \Big\{ 4 \kappa ^{10} \tilde{\beta }_3^4 \tilde{\beta }_4+\kappa ^8 \tilde{\beta }_3^4 \left(-128 \tilde{\beta }_4^2+8 \tilde{\beta }_4+3\right) \notag\\
&& -2 \kappa ^6 \left(8 \tilde{\beta }_4+1\right)^2 \left[\left(3-8 \tilde{\beta }_4\right) \tilde{\beta }_3^4+2 \tilde{\beta }_4 \left(64 \tilde{\beta }_4^2+48 \tilde{\beta }_4+5\right) \tilde{\beta }_3^2+8 \tilde{\beta }_4^3 \left(8 \tilde{\beta }_4+1\right)^2\right] \notag\\
&&-2 \kappa ^2 \left(8 \tilde{\beta }_4+1\right)^6 \left[\left(8 \tilde{\beta }_4+1\right) \tilde{\beta }_3^2+2 \tilde{\beta }_4 \left(80 \tilde{\beta }_4^2+14 \tilde{\beta }_4-1\right)\right] \notag\\
&&+4 \tilde{\beta }_4 \left[2 \left(16 \tilde{\beta }_4+7\right) \tilde{\beta }_3^2+\tilde{\beta }_4 \left(256 \tilde{\beta }_4^2+88 \tilde{\beta }_4+7\right)\right] \left(8 \kappa  \tilde{\beta }_4+\kappa \right)^4+\left(4 \tilde{\beta }_4-1\right) \left(8 \tilde{\beta }_4+1\right)^9 \Big\} \notag\\
&& \times \left(8 \tilde{\beta }_4+1\right)^{-4} \left(-4 \left(\kappa ^2-4\right) \tilde{\beta }_4+64 \tilde{\beta }_4^2+1\right)^{-1} \notag\\
&& \times \left[\kappa ^4 \tilde{\beta }_3^2-2 \kappa ^2 \left(8 \tilde{\beta }_4+1\right) \left(\tilde{\beta }_3^2+8 \tilde{\beta }_4^2+\tilde{\beta }_4\right)+\left(4 \tilde{\beta }_4-1\right) \left(8 \tilde{\beta }_4+1\right)^3\right]^{-1}\,.
\label{aLap2M2}
\ea
On using this expression, we can determine the minimum of $\mathcal{M}_1$ in the phase space of 
$\{\mu, {\tilde{\beta }_3}, {\tilde{\beta }_4}\}$ with the built-in functions of {\it Mathematica}. 
Keeping in mind that $\mu \in (0, 1)$, ${\tilde{\beta }_3} \in (-0.441, 0.441)$ and 
${\tilde{\beta }_4} \in (-0.125, 0.251)$, it turns out that the minimum of $\mathcal{M}_1$ is a positive 
number, which supports this angular Laplacian stability condition for Model 2.  
A similar treatment to $\mathcal{M}_2$ brings us a result in the form of Eq.~\eqref{aLap2M2}. 
Because of its tedious expression, we put that in Appendix \ref{secVexpression}. 
The result shows that the minimum of  $\mathcal{M}_2$ in the phase space of 
$\{\mu, {\tilde{\beta }_3}, {\tilde{\beta }_4}\}$ is also determined to be a positive number. 
Therefore, it seems like the currently known constraints are sufficient to support  this angular Laplacian stability condition.

We note that the above discussion is based on just an analysis at the $r \to r_h$ limit. 
That brings us to the second part of analysis. To determine the ranges of $\mathcal{M}_{1, 2}$ 
in the whole exterior space $r\in [r_h, +\infty)$, we shall mimic \cite{Heisenberg:2018mgr} and 
plot them out for chosen parameters. 
To do so, one has to first solve for the background quantities $f$ and $h$ on the exterior space 
with the algebraic expressions \eqref{beM2} of $\phi'$ and $A_0'$ in terms of $f$, $h$ 
as well as their derivatives. 
They are achieved by numerically integrating Eqs.~\eqref{be1} and \eqref{be2} from 
$\xi=1-\epsilon_1$ to $\xi=\epsilon_2$, where factors $\epsilon_1$ and $\epsilon_2$ are 
chosen to be small enough to cover the desired region (see, e.g., Ref.~\cite{Zhang:2020too} 
for more details about the relative techniques used in doing the numerical integration and 
searching for background solutions). Of course, in practice we were using the dimensionless 
forms of Eqs.~\eqref{be1} and \eqref{be2}, and a new quantity similar to the $\mathfrak{Z}(\xi)$ 
defined in Eq.~\eqref{goZ} in Appendix~\ref{secVexpression} is introduced to simplify 
the expressions. Since this kind of treatment is straightforward and we have already briefly 
described the steps for similar scenarios in last subsection, we omit the details here. 
We note that the Taylor expansion 
\ba
{\mathfrak{f}}(\xi=1-\epsilon_1)=\left. {\mathfrak{f}} \right|_{\xi=1} + \frac{1}{1!}\left. \frac{d{\mathfrak{f}}}{d\xi} \right|_{\xi=1} (-\epsilon_1) + \frac{1}{2!}\left. \frac{d^2{\mathfrak{f}}}{d\xi^2} \right|_{\xi=1} (-\epsilon_1)^2\,,
\label{Taylor}
\ea
where ${\mathfrak{f}}$ stands for $f$ and $h$, was applied in generating the boundary conditions 
at $\xi=1-\epsilon_1$ keeping in mind that ${\mathfrak{f}}(\xi=1)=0$. The derivatives appearing in 
Eq.~\eqref{Taylor} are obtained by using Eqs.~\eqref{be1}, \eqref{be2} as well as their derivatives. 
It's worth mentioning here that, the polynomial solutions given in Ref.~\cite{Heisenberg:2018vti} 
were not utilized directly in generating the boundary conditions since there is a remaining parameter 
corresponding to an overall factor of $f$, which is unknown before obtaining the numerical solutions. 
This factor will be fixed by normalizing $f$ according to the requirement $f \to 1$ at the spatial infinity. 

The solution for $f$ and $h$ are shown in the panel (a) of Fig.~\ref{plot2} in Appendix \ref{plotsanalysis}. 
Since the upper limit of $ {\tilde{\beta }_3}$ is updated to $\approx 0.441$ compared to 
Ref.~\cite{Heisenberg:2018mgr}, we adopted it with ${\tilde{\beta }_4}=0.1$ and $\mu=0.5$. 
We can find deviation between Model 2 and that of GR. 
With the solutions of $f$ and $h$ in hand, together with the previously mentioned algebraic 
expressions for $A_0'$ and $\phi'$, all the background quantities are known. 
Inserting them into ${\mathcal{M}}_{1}$ and ${\mathcal{M}}_{2}$, we further obtain their 
numerical values. This part of results are exhibited in the panel (b) of Fig.~\ref{plot2} 
in which ${\mathcal{M}}_{1}$ and ${\mathcal{M}}_{2}$ are plotted as functions of $r_h/r$ 
instead of $r/r_h$ to show a larger scope. Fig.~\ref{plot2} shows that, both these two quantities 
stay positive in the whole space outside the horizon. This result further supports that the first 
and second angular Laplacian stability conditions for Model 2\footnote{
An ideal treatment is to test the positiveness of ${\mathcal{M}}_{1}$ and ${\mathcal{M}}_{2}$ 
using the Monte Carlo method and try to cover the whole allowed region in 
$\{\mu, {\tilde{\beta }_3}, {\tilde{\beta }_4}, \xi\}$ phase space.  Nevertheless, as seen from, 
e.g., Eqs.~\eqref{aLap2M1}-\eqref{aLap3M2}, the complexity of our problem has already made 
the calculations quite time-consuming and it is preventing us from running a more complete analysis. 
In principle, much more computational resources are needed for executing such an analysis.
}. 

\subsection{Stability analysis for the  Model 3}
\label{applicationC}

We first consider the third no-ghost condition \eqref{noghost3b}. For Model 3, $\mathcal{K}$ 
reduces to 
\ba
\mathcal{K} &=& \frac{2 h r^2 \tilde{\beta }_3^2 A_0'^4 r_h^4 \left[4 (1-h) \tilde{\beta }_4 A_0'^2 \sqrt{f h^7} r_h^2+f^{3/2} h^{5/2} r^2\right]}{(f h)^{3/2} M_{\text{pl}}^2 \left(6 h^2 \tilde{\beta }_4 A_0'^2 r_h^2-4 h \tilde{\beta }_4 A_0'^2 r_h^2-f r^2\right)^2} \,,
\label{noghost3M3}
\ea
Thus, by looking at both the numerator and denominator of the above expression, for any non-zero ${\tilde \beta}_3$, the third no-ghost condition will always get satisfied. As mentioned in previous subsections, the background equations \eqref{be3}-\eqref{be4} lead to 
\ba
\phi' &=& \frac{2 h r \tilde{\beta }_3 A_0'^2 r_h^2}{M_{\text{pl}} \left(-6 h^2 \tilde{\beta }_4 A_0'^2 r_h^2+4 h \tilde{\beta }_4 A_0'^2 r_h^2+f r^2\right)}\,,\notag\\
 A_0' &=& \frac{\kappa \sqrt{f} r_h^2 M_{\text{pl}}}{\sqrt{h} \left(6 h^2 \tilde{\beta }_4 r_h^2 M_{\text{pl}} \phi'^2-4 h \tilde{\beta }_4 r_h^2 M_{\text{pl}} \phi'^2+4 h r \tilde{\beta }_3 r_h^2 \phi '+r^2 M_{\text{pl}}\right)}\,.
\label{beM3}
\ea
With the polynomial solutions given by Eqs.~(4.15)-(4.17) in Ref.~\cite{Heisenberg:2018vti} 
and evaluation of \eqref{JA} at $r_h$, we notice that for Model 3 we have 
$\kappa = \sqrt{2 \mu } M_{\text{pl}}/r_h$, where $\mu \in (0, 1)$. 
It is quite necessary for one to notice that, the system given by Eq.~\eqref{beM3} forms 
a fifth-order algebraic equation of $A_0'$. As a result, there is no analytic expression of $\phi'$ or $A_0'$ in terms of $f$, $h$, etc. according to the well-known
Abel-Ruffini theorem. This is one of the key factors which makes the Model 3 much 
intricate than that of Model 2. 

On the other hand, for Model 3, the squared second  radial propagation speed 
given by Eq.~\eqref{evencr2} becomes
\ba
c_{r2{\rm ,even}}^2 &=& \frac{3 h^2 \tilde{\beta }_4 A_0'^2 r_h^2 \phi'^2+f M_{\text{pl}}^2}{2 h^2 \tilde{\beta }_4 A_0'^2 r_h^2 \phi'^2+f M_{\text{pl}}^2}\,,
\label{rLap2M3}
\ea
which shows that the second radial Laplacian stability condition holds. 

Next, the angular propagation speed given in Eq.~\eqref{evencO1} for the Model 3 is 
\ba
c_{\Omega 3{\rm ,even}}^2 &=& \frac{ {{\tilde{ K}}_{33}} }{M_{\text{pl}}^6 r^4 \sqrt{f h} \left(2 h (2-3 h) \tilde{\beta }_4 A_0'^2 r_h^2+f r^2\right)^2} \notag\\
&& \times  2 L \Big\{ 16 h^5 (3 h-2) r^2 r_h^{10} \tilde{\beta }_4^2 \tilde{\beta }_3^2 \left(2 \tilde{\beta }_4 M_{\text{pl}}^2+h \left(2 r_h^2 \tilde{\beta }_3^2-3 M_{\text{pl}}^2 \tilde{\beta }_4\right)\right) A_0'^{10}  \notag\\
&&  +4 f h^4 r_h^8 \tilde{\beta }_4 \Big[ 108 h^4 \tilde{\beta }_4^2 \left(3 M_{\text{pl}}^2 \tilde{\beta }_4-2 r_h^2 \tilde{\beta }_3^2\right) M_{\text{pl}}^4-72 h^3 \tilde{\beta }_4^2 \left(12 M_{\text{pl}}^2 \tilde{\beta }_4-5 r_h^2 \tilde{\beta }_3^2\right) M_{\text{pl}}^4  \notag\\
&&  +4 \tilde{\beta }_4 \left(8 M_{\text{pl}}^2 \tilde{\beta }_4 \left(2 \tilde{\beta }_4 M_{\text{pl}}^2+r_h^2 \tilde{\beta }_3^2\right)-3 r^4 \tilde{\beta }_3^2\right) M_{\text{pl}}^2  \notag\\
&&  -8 h \left(\left(2 r_h^2 \tilde{\beta }_3^4-3 M_{\text{pl}}^2 \tilde{\beta }_4 \tilde{\beta }_3^2\right) r^4+8 M_{\text{pl}}^4 \tilde{\beta }_4^2 \left(6 \tilde{\beta }_4 M_{\text{pl}}^2+r_h^2 \tilde{\beta }_3^2\right)\right)  \notag\\
&& + 3 h^2 \left(\left(2 r_h^2 \tilde{\beta }_3^4-3 M_{\text{pl}}^2 \tilde{\beta }_4 \tilde{\beta }_3^2\right) r^4+8 M_{\text{pl}}^4 \tilde{\beta }_4^2 \left(36 M_{\text{pl}}^2 \tilde{\beta }_4-5 r_h^2 \tilde{\beta }_3^2\right)\right)\Big] A_0'^8  \notag\\
&&  -4 f^2 h^3 r^2 r_h^6 \Big[6 h^2 \tilde{\beta }_4^2 \left(13 r_h^2 \tilde{\beta }_3^2-72 M_{\text{pl}}^2 \tilde{\beta }_4\right) M_{\text{pl}}^4+18 h^3 \tilde{\beta }_4^2 \left(12 M_{\text{pl}}^2 \tilde{\beta }_4-5 r_h^2 \tilde{\beta }_3^2\right) M_{\text{pl}}^4  \notag\\
&&  +\tilde{\beta }_4 \left(3 r^4 \tilde{\beta }_3^2-8 M_{\text{pl}}^2 \tilde{\beta }_4 \left(8 \tilde{\beta }_4 M_{\text{pl}}^2+3 r_h^2 \tilde{\beta }_3^2\right)\right) M_{\text{pl}}^2  \notag\\
&&  +3 h \left(\left(r_h^2 \tilde{\beta }_3^4-M_{\text{pl}}^2 \tilde{\beta }_4 \tilde{\beta }_3^2\right) r^4+8 M_{\text{pl}}^4 \tilde{\beta }_4^2 \left(12 \tilde{\beta }_4 M_{\text{pl}}^2+r_h^2 \tilde{\beta}_3^2\right)\right)\Big] A_0'^6  \notag\\
&&  +f^3 h^2 r^4 M_{\text{pl}}^2 r_h^4 \Big[-\tilde{\beta }_3^2 r^4-288 h M_{\text{pl}}^4 \tilde{\beta }_4^2+24 M_{\text{pl}}^2 \tilde{\beta }_4 \left(4 \tilde{\beta }_4 M_{\text{pl}}^2+r_h^2 \tilde{\beta }_3^2\right)  \notag\\
&&  +24 h^2 M_{\text{pl}}^2 \tilde{\beta }_4 \left(9 M_{\text{pl}}^2 \tilde{\beta }_4-2 r_h^2 \tilde{\beta }_3^2\right)\Big] A_0'^4  \notag\\
&&  +2 f^4 h r^6 M_{\text{pl}}^4 r_h^2 \left(8 \tilde{\beta }_4 M_{\text{pl}}^2+r_h^2 \tilde{\beta }_3^2+h \left(r_h^2 \tilde{\beta }_3^2-12 M_{\text{pl}}^2 \tilde{\beta }_4\right)\right) A_0'^2+f^5 r^8 M_{\text{pl}}^6 \Big\}^2 \notag\\
&& \times \Big\{4 f h r^2 A_0'^2 r_h^2 \left[h \left(2 \tilde{\beta }_3^2 r_h^2-3 \tilde{\beta }_4 M_{\text{pl}}^2\right)+2 \tilde{\beta }_4 M_{\text{pl}}^2\right] \notag\\
&& -4 h^2 (3 h-2) \tilde{\beta }_4 A_0'^4 r_h^4 \left(h \left(2 \tilde{\beta }_3^2 r_h^2-3 \tilde{\beta }_4 M_{\text{pl}}^2\right)+2 \tilde{\beta }_4 M_{\text{pl}}^2\right)+f^2 r^4 M_{\text{pl}}^2\Big\}^{-3}\,.
\label{aLap1M3}
\ea
The terms in the first line of Eq.~\eqref{aLap1M3} is non-negative as long as ${{\tilde{ K}}_{33}}>0$. 
The positiveness of $c_{\Omega 3{\rm ,even}}^2$ is solely determined by the part to the $-3$ power, i.e., 
the last two lines in Eq.~\eqref{aLap1M3}.  
By treating $f$, $h$, ${\tilde \beta}_3$, ${\tilde \beta}_4$ and $A_0'$ as free parameters, 
and considering the known constraints on them, one can determine its minimum by 
using numerical methods. Taking advantage of its dimensionless form, this calculation could be 
fulfilled by using the built-in functions of {\it Mathematica}. 
This manipulation shows that the condition $c_{\Omega 3{\rm ,even}}^2>0$ is always guaranteed. 

We proceed to the first and second angular Laplacian stability conditions in which 
the story becomes astonishingly sophisticated, in comparing to that of Model 1. 
To conquer this problem, we shall first run the analysis around the $r=r_h$ 
where the algebraic equations in Eq.~\eqref{beM3} can be solved for $A_0'$ as 
$A_0'(r=r_h)=\sqrt{2 \mu} = \kappa$ by choosing the unit system so that $M_{\text{pl}}=r_h$. 
On using this solution at the horizon, we start to expand $\mathcal{M}_{1}$ and $\mathcal{M}_{2}$ 
step by step around the horizon. After a sequence of time-consuming but straightforward 
manipulations, their expressions at the horizon converted into the dimensionless form 
were obtained. The one for $\mathcal{M}_{1}$ is given below
\ba
\left. \mathcal{M}_1 \right|_{r \to r_h}  &=& \frac{1}{2} \Big\{ 10 \kappa ^{12} \tilde{\beta }_3^4 \tilde{\beta }_4+4 \kappa ^{10} \tilde{\beta }_3^2 \tilde{\beta }_4 \left(2 \tilde{\beta }_3^2+7 \tilde{\beta }_4\right)+\kappa ^8 \left[\left(12-56 \tilde{\beta }_4\right) \tilde{\beta }_3^4+\tilde{\beta }_4 \left(7-72 \tilde{\beta }_4\right) \tilde{\beta }_3^2-16 \tilde{\beta }_4^3\right] \notag\\
&& -2 \kappa ^6 \left[12 \tilde{\beta }_3^4+3 \tilde{\beta }_4 \left(16 \tilde{\beta }_4+3\right) \tilde{\beta }_3^2+4 \tilde{\beta }_4^2 \left(24 \tilde{\beta }_4+1\right)\right]  -\kappa ^4 \tilde{\beta }_4 \left(56 \tilde{\beta }_3^2+160 \tilde{\beta }_4+1\right) \notag\\
&& -4 \kappa ^2 \left(2 \tilde{\beta }_3^2+11 \tilde{\beta }_4\right)-4 \Big\}   \left(4 \kappa ^2 \tilde{\beta }_4+1\right)^{-2} \left[\kappa ^4 \tilde{\beta }_3^2-2 \kappa ^2 \left(\tilde{\beta }_3^2+2 \tilde{\beta }_4\right)-1\right]^{-1}\,, 
\label{aLap2M3}
\ea
while the one for $\mathcal{M}_{2}$ is shown in Appendix~\ref{secVexpression} due to its intricate structure. 
With the known constraints on ${\tilde \beta}_3$, ${\tilde \beta}_4$ and $\mu$ (so does $\kappa$), we can determine the minimum of $\left. \mathcal{M}_{1} \right|_{r \to r_h}$ and 
$\left. \mathcal{M}_{2} \right|_{r \to r_h}$ on the use of the built-in functions of {\it Mathematica},
which shows that both of them are positive so that the first as well as second angular Laplacian stability conditions 
are satisfied for the given parameter regions. 

To further confirm these stability conditions, $\mathcal{M}_{1}$ and $\mathcal{M}_{2}$ will also be 
plotted out on $r\in [r_h, +\infty)$ with chosen coupling constants. This could be fulfilled by choosing 
${\mu}=0.5$, ${\tilde \beta}_3=0.441$ and ${\tilde \beta}_4=0.1$
as well as being given the corresponding numerical background solutions shown in the panel (a) 
of Fig.~\ref{plot4} in Appendix~\ref{plotsanalysis}. 
The fundamental steps and techniques of obtaining these numerical solutions are the same as 
what we used to draw the panel (a) of Fig.~\ref{plot2}. The only difference is that, for Model 3 
there is no analytic expression for $A_0'$. Thus, $f$, $h$, and  $A_0'$ need to be numerically solved 
simultaneously. For this reason, all of these three variables are plotted in the panel (a) of Fig.~\ref{plot4}. 
Again, to show a larger scope, all the quantities are plotted as functions of $\xi$. We note that 
$A_0'$ is plotted in its dimensionless form $r_h^{-1} dA_0/d\xi$, which, as shown in the plot, 
equals $-\xi^{-2} A_0'$. It's worth mentioning here that, $A_0'$ behaves as ${\cal O}(r^{-2})$ 
at the spatial infinity ($\xi \to 0$). That's why the quantity $-\xi^{-2} A_0'$ won't blow up as $\xi \to 0$. 
The results of $\mathcal{M}_{1}$ and $\mathcal{M}_{2}$ are exhibited in the panel (b) of 
Fig.~\ref{plot4}. 
It is clear that, both these two quantities stay positive at any distances outside the horizon. 
As a result, we conclude that this figure further supports the first and second angular Laplacian 
stability conditions in Model 3 inside the given parameter space. 

\begin{table*}[b]
\begin{tabular}{|c c c | c c| c c | c c  |} 
\hline  
		   &  &   & ~ Model 1:~ & &  ~ Model 2:~   & &  ~ Model 3:~ &
		\\
		\hline  
		~ Category~  & ~$\#$~ &  ~Described by ~ & Type & ~See~~~ &  Type  & ~See~~~   &  Type  &   ~See~~~ 
		\\
		\hline
  		\hline  
		  & first &  \eqref{noghost1b} & Type I &  N/A &   Type I  & N/A&   Type I   & N/A
		\\
  		No-ghost  & second &  \eqref{noghost2a} & Type I & N/A  & Type I  &  N/A    &  Type I   & N/A 
		\\
  		  & third &  \eqref{noghost3b} & Type II & \eqref{noghost3M1}  & Type II &  \eqref{noghost3M2}  & Type II   & \eqref{noghost3M3}
		\\
		\hline
  	Radial  & first &  \eqref{evencr1} & Type I &  N/A &   Type I  &  N/A  &   Type I  & N/A 
		\\
  		Laplacian  & second &  \eqref{evencr2} & Type II & \eqref{rLap23M1}  & Type II & \eqref{rLap23M2}   & Type II  &  \eqref{rLap2M3}
		\\
  		  & third &  \eqref{evencr3} & Type I & N/A  & Type I  &   N/A   &   Type I  &  N/A 
		\\
		\hline
    	Angular  & third &  \eqref{evencO1} & Type II &  \eqref{aLap1M1} & Type II & \eqref{aLap1M2}    &  Type III  & \eqref{aLap1M3}
 		\\
  		Laplacian  & first &  \eqref{evencO2} & Type III & \eqref{aLap2M1}  & Type IV & ~\eqref{aLap2M2} \& Fig.\ref{plot2}  ~ &  Type IV & ~\eqref{aLap2M3} \& Fig.\ref{plot4}~
		\\
  		  & second &  \eqref{evencO3} & Type III & ~\eqref{aLap3M1} \& Fig.\ref{plot1}~ & Type IV & \eqref{aLap3M2} \& Fig.\ref{plot2}& Type IV  & \eqref{aLap3M3} \& Fig.\ref{plot4}
		\\
		\hline
	\end{tabular}
	\caption{\label{table1}
Summarize the stability analysis and corresponding conclusions for different categories of stability conditions as well as different models. Type I: This stability condition will automatically get satisfied, provided the other stability conditions (including those from the odd-parity sector); Type II: This stability condition can be determined immediately by observing the corresponding mathematical  structure; Type III: The analytic investigation will be quite complicated so that semi-analytic or numerical techniques have to be applied in confirming this stability condition. Notice that, the constraints on coupling parameters, i.e., ${\tilde{\beta }_3} \in (-0.441, 0.441)$ and ${\tilde{\beta }_4} \in (-0.125, 0.251)$ will be considered; Type IV: A complete numerical analysis is running out of the computational resources. Thus, the analysis will be carried out by doing expansions around the event horizon ($r=r_h$) and also by plotting out those key quantities for certain chosen coupling parameters (under the constraints mentioned above). }  
	\end{table*}

To summary, here we have dealt with 27 cases in total (combining Secs.~\ref{evensec} and \ref{application}), 
i.e., 9 stability conditions (3 no-ghost conditions + 3 radial Laplacian stability conditions + 3 angular Laplacian 
stability conditions) times 3 distinct models. Basically,
no instability was recognized up to now, given the known$\symbol{92}$updated  constraints  ${\tilde{\beta }_3} \in (-0.441, 0.441)$ and ${\tilde{\beta }_4} \in (-0.125, 0.251)$. For the convenience of readers, we summarize those main equations, results, analysis, etc. for these 27 cases in Table \ref{table1}. 
Notice that, the main results for each case are categorized in four different types, denoted by 
Type I, II, III and IV, respectively, mainly according to the explicitness of the stability in each case.  
Type I means the corresponding stability condition automatically get satisfied, at least by considering 
the known constraints to the theory; Type II means the corresponding stability condition deserves 
a little analysis. Nonetheless, the stability is quite explicit so that we can tell that just by looking at 
the mathematical structure of the corresponding expression; Type III is a little sophisticated, so that 
pure analytic investigation will not work. Even in such case, we could still confirm these stability 
conditions by semi-analytic or numerical methods and with the assistance of computers; Type  IV is a little 
vague from certain point of view. The problems are really complicated so we have to expediently check the corresponding stability conditions 
mainly by plotting out the results for  chosen coupling parameters. A more complete analysis may be executed in the future for this type of problems. 

\section{Conclusions}
\label{concludesec}

In this paper, the stability of static and spherically symmetric BHs with nontrivial scalar hair 
in the $U(1)$ GI SVT theory \eqref{action} is studied. For the background ansatz \eqref{metric_bg}, \eqref{phi_bg}, 
and \eqref{Amu_bg}, we considered the even-parity perturbations given by Eqs.~\eqref{evenmetric}, 
\eqref{defdphi}, and \eqref{defdA}. As a successor of the previous work~\cite{Heisenberg:2018mgr}, 
we investigated three types of stability conditions, i.e., the no-ghost, radial Laplacian and 
angular Laplacian ones. In addition to the general analysis, some of the stability conditions are 
discussed by considering three distinct concrete models and by applying suitable analytic, 
semi-analytic as well as numerical techniques. 

To make it more complete, the background field equations \eqref{be1}-\eqref{be4} and the stability conditions 
against the odd-parity perturbations are revisited in the presence of the quantities $Y$ and $\tilde{F}$ which 
are omitted in the previous work. 

On top of that, we first run the general analysis. The original Lagrangian is expanded to second order for 
the even-parity perturbations as Eq.~\eqref{acteven}, which contains 7 apparent DOFs. After that, with 
the help of integration by parts, the integral properties of spherical harmonics, and by introducing 
suitable new variables [given in Eqs.~\eqref{defV} and \eqref{psi}], all the non-dynamical terms 
disappear and we are left with a dramatically simplified Lagrangian \eqref{evenLag1}, 
adopting 3 DOFs as expected\cite{Heisenberg:2018mgr}. On using the resultant Lagrangian, 
we derived 9 stability conditions, i.e., 3 no-ghost conditions \eqref{noghost1}-\eqref{noghost3} 
+ 3 radial Laplacian stability conditions \eqref{evenrLap} + 3 angular Laplacian stability conditions  \eqref{evencO1}, \eqref{evencO2}-\eqref{evencO3}. 

It is then found that the first [cf., \eqref{noghost1b}] and second [cf., \eqref{noghost2a}] no-ghost 
conditions as well as the first [cf., \eqref{evencr1}] and third [cf., \eqref{evencr3}] radial Laplacian 
stability conditions get satisfied automatically, provided the other stability conditions, including 
those from the odd-parity sector. We then move to three specific concrete models characterized by 
Eqs.~\eqref{model1}-\eqref{model3}] to run the analysis in Sec.~\ref{application}. 
By inserting the corresponding functions of each model into the mathematical expression of the third no-ghost condition, it is found that this condition hold in all the three models 
[cf., \eqref{noghost3M1}, \eqref{noghost3M2} and \eqref{noghost3M3}].  
Similarly, the second radial Laplacian condition also hold for all these three models 
[cf., \eqref{rLap23M1}, \eqref{rLap23M2} and \eqref{rLap2M3}].  

The situation becomes a little complicated when moving to the angular Laplacian conditions. 
First of all, we can easily confirm the third angular Laplacian condition in an analytic way for 
Model 1 and Model 2 as in Eqs.~\eqref{aLap1M1} and \eqref{aLap1M2}, respectively. 
However, for the first and second angular Laplacian conditions of Model 1, semi-analytic 
as well as numerical techniques become necessary. It is found that these stability conditions 
are guaranteed by applying the constraints to the coupling parameters obtained from the odd-parity 
sector in Ref.~\cite{Heisenberg:2018mgr} together with the additional constraints indicated in Fig.~\ref{plot1} 
(see Appendix~\ref{plotsanalysis}), namely, ${\tilde{\beta }_3} \in (-0.441, 0.441)$ and 
${\tilde{\beta }_4} \in (-0.125, 0.251)$. With these constraints, the third angular Laplacian condition 
for Model 3 can also be confirmed [cf., \eqref{aLap1M3}]. Finally, we notice that the first and 
second angular Laplacian conditions for Model 2 and Model 3 are quite difficult to investigate 
(especially for the Model 3). To conquer this difficulty, by mimicking Ref.~\cite{Heisenberg:2018mgr}, 
specific values for the coupling parameters are chosen for these two models. As a result, we are 
able to plot out the discriminants [cf., \eqref{evencO2} and \eqref{evencO3}] for each case. 
These two stability conditions are then confirmed in Figs.~\ref{plot2} and \ref{plot4} 
(cf., Appendix~\ref{plotsanalysis}). The main formulas and analysis about these stability conditions 
are summarized for the three models in Table~\ref{table1}. 

Our current work can be enlarged to several other directions. For instance, we can dig further 
into the angular Laplacian conditions in order to put more stringent constraints on the phase 
space of the coupling parameters as well as consider more types of models for analysis. 
It is also of interest to put constraints on the model parameters through the observational data of 
BH shadow given by the EHT and the other observations. 
The constraints on the charges of the supermassive compact objects being black hole candidates 
are studied in Refs.~\cite{EventHorizonTelescope:2019dse,EventHorizonTelescope:2019ggy} 
assuming that the observed rings correspond to the photon sphere, as well as in 
Ref.~\cite{Tsukamoto:2024gkz} assuming that they correspond to the lensing ring.  
These procedures enable one to put constraints on the model parameter at the background level. 
On the other hand, the second-order Lagrangian analysis is closely related to the quasi-normal 
mode problems (see, e.g., Ref.~\cite{Berti:2018vdi}). We can also try to do some relative calculations 
based on Eq.~\eqref{evenLag1} and put the further constraint on the model parameters 
at the perturbation level.

\section*{Acknowledgements}

The authors are deeply grateful to Shinji Tsujikawa for useful discussion. 
CZ is supported by the National Natural Science Foundation of China under Grant No.~12205254 and the Science Foundation of China University of Petroleum, Beijing under Grant No.~2462024BJRC005. 
RK is supported by the Grant-in-Aid for Early-Career Scientists 
of the JSPS No.~20K14471 and the Grant-in-Aid for Scientific 
Research (C) of the JSPS No.~23K034210. 

\appendix

\section{Coefficients in the second-order action 
of even-parity perturbations}
\label{even_coeff}

The coefficients in Eqs.~(\ref{Lu}) and (\ref{LdA}) are given by 
\ba
&&
a_1 \equiv 0\,,\qquad
a_2=\frac{r(f'h-fh')}{fh\phi'}a_4+\frac{A_0'}{2}v_4+\frac{rA_0'}{2\phi'}v_6\,,\qquad
a_3=-ra_4\,,\qquad
a_4=\frac{\sqrt{fh}\Mpl^2}{2} =2 f \alpha_1\,,
\notag\\
&&
a_5=a_2'-\frac{A_0''}{2}v_4+\frac{A_0'}{2}(v_5-v_4')\,,\qquad
a_6=\frac{r(f'h+fh')-2fh}{2rfh^2\phi'}a_4-\frac{1}{h\phi'}a_4'-\frac{A_0'}{4h\phi'}v_6+\frac{2rf}{\phi'}\alpha_7\,,
\notag\\
&&
a_7=-(ra_4)'-\frac{A_0'}{4}(2A_0'v_1+\phi'v_4)-\frac{rA_0'}{2}v_6\,,\qquad
a_8=-\frac{a_4}{2h}\,, \qquad
a_9 = a_4'-\frac{r f'-2f}{2 r f}a_4 +\frac{A_0'}{4} v_6\,,
\notag\\
&&
b_1=\frac{a_4}{2f}\,,\qquad 
b_2=0\,,\qquad 
b_3=-\frac{2}{f}a_2+\frac{A_0'}{f}v_4\,,\qquad
b_4=\frac{2ra_4}{f}\,,\qquad 
b_5=-\frac{a_4}{f}\,,\qquad 
c_1 \equiv 0\,,
\notag\\
&&
c_2 \equiv \frac{r^2\sqrt{fh}\phi'(f_{2,X}-h\phi'^2f_{2,XX})}{2}
-\frac{r^2h^{5/2}\phi'A_0'^4(4h\phi'^2f_{2,YY}-f_{2,FY})}{f^{3/2}}
-\frac{h^{3/2}A_0'^2}{2\sqrt{f}}\times
\notag\\
&&
\hspace{.75cm}
\Big[h^3\phi'^5f_{4,XXX}+2rh^2\phi'^4f_{3,XX}+h\phi'^3\{6r^2f_{2,XY}+(4-13h)f_{4,XX}\}-12rh\phi'^2f_{3,X}
\notag\\
&&
\hspace{.75cm}
-\phi'\{r^2(6f_{2,Y}+f_{2,XF})+6(2-5h)f_{4,X}-20h\tilde{f}_4\}+6rf_3\Big]\,,
\notag\\
&&
c_3=-\frac{1}{2\sqrt{fh}}\frac{\partial{\cal E}_{11}}{\partial\phi}\,,\qquad 
c_4 \equiv -\frac{\sqrt{h}A_0'^2}{2r\sqrt{f}}\left[h^2\phi'^3f_{4,XX}+rh\phi'^2f_{3,X}-3h\phi'(f_{4,X}+2\tilde{f}_4)-rf_3\right]\,,
\notag\\
&&
c_5=-\frac{f'}{2f}a_4 -\frac{\phi'}{2}d_2-\frac{A_0'}{2}v_6+2r\alpha_6\,,\qquad
c_6=\frac{rf'}{2f}a_4-\frac{\phi'}{4}c_2+\frac{r\phi'}{4}d_2+\frac{A_0'^2}{4}v_1+\frac{\phi'A_0'}{8}v_4+\frac{rA_0'}{2}v_6-r^2\alpha_6\,,
\notag\\
&&
d_1=\frac{a_4}{2f}\,,\qquad
d_2 \equiv -\frac{h^{3/2}A_0'^2}{r\sqrt{f}}\left[h^2\phi'^3f_{4,XX}+rh\phi'^2f_{3,X}-2h\phi'(3f_{4,X}+2\tilde{f}_4)-rf_3\right]\,,
\notag\\
&&
d_3=\frac{2(f'h-fh')}{rfh\phi'}a_4+\frac{A_0'}{r\phi'}v_6+\frac{A_0'}{2}\frac{\partial v_6}{\partial\phi}+\frac{4}{h\phi'}\alpha_6+\frac{4f}{\phi'}\alpha_7\,,\qquad 
d_4=-2\alpha_6\,,
\notag\\
&&
e_1=\frac{1}{fh\phi'}\left(a_2-2rha_6-\frac{A_0'}{2}v_4\right)\,,\qquad 
e_2=-\frac{1}{\phi'}\left(c_2+rd_2+\frac{A_0'}{2}v_4\right)\,,\qquad 
e_3=\frac12\frac{\partial{\cal E}_{\phi}}{\partial\phi}\,,
\notag\\
&&
e_4=\frac{1}{r^2h\phi'^2}\left(1-\frac{3rf'}{2f}+\frac{rh'}{2h}\right)a_4+\frac{h\phi''+h'\phi'}{h\phi'^2}c_4+\frac{c_4'}{\phi'}
+\frac{fA_0''-2f'A_0'}{4fh\phi'^2}v_6+\frac{A_0'}{4h\phi'^2}v_6'\notag\\
&&
\hspace{.75cm}
-\frac{A_0'^2}{\phi'^2}v_9-\frac{A_0'^2}{fh\phi'^2}v_{10}-\frac{A_0'}{\phi'}v_{13}-\frac{2rh'+2(1-3h)}{h^2\phi'^2}\alpha_6+\frac{2r}{h\phi'^2}\alpha_6'+\frac{2(rf'h-f)}{h\phi'^2}\alpha_7\,,
\notag\\
&&
v_1 \equiv \frac{r^2h^{3/2}A_0'^2}{2f^{3/2}}\left[f_{2,FF}-4h\phi'^2(f_{2,FY}-h\phi'^2f_{2,YY})\right]
\notag\\
&&
\hspace{.75cm}
-\frac{\sqrt{h}}{2\sqrt{f}}\left[2h\phi'^2\{r^2f_{2,Y}-h(f_{4,X}+2\tilde{f}_4)\}-4rh\phi'f_3-8(1-h)f_4-r^2f_{2,F}\right]\,,
\notag\\
&&
v_2=A_0'v_1\,,\qquad
v_3=-A_0'v_1-\frac{\phi'}{2}v_4-rv_6\,,
\notag\\
&&
v_4 \equiv \frac{2r^2h^{5/2}\phi'A_0'^3(2h\phi'^2f_{2,YY}-f_{2,FY})}{f^{3/2}}
+\frac{h^{3/2}A_0'}{\sqrt{f}}\Big[2h\phi'^3(r^2f_{2,XY}-hf_{4,XX})-4rh\phi'^2f_{3,X}
\notag\\
&&
\hspace{0.75cm}
-r^2\phi'(4f_{2,Y}+f_{2,XF})+4\phi'\{(3h-2)f_{4,X}+2h\tilde{f}_4\}+4rf_3\Big]\,,
\notag\\
&&
v_5 \equiv \frac{\sqrt{h}A_0'}{\sqrt{f}}\left[2h\phi'^2\{h(f_{4,\phi X}+2\tilde{f}_{4,\phi})-r^2f_{2,\phi Y}\}
+4rh\phi'f_{3,\phi}+8(1-h)f_{4,\phi}+r^2f_{2,\phi F}\right]\,,
\notag\\
&&
v_6=4 \alpha_2\,,\qquad
v_7=\frac{A_0'^2}{4}v_1\,,\qquad
v_8=\frac{1}{2rf}\left[(rf'-2f)v_6+4rA_0'v_{10}\right]\,,
\notag\\
&&
v_9= \alpha_4\,,
\qquad
v_{10}=\alpha_5\,,
\qquad
v_{11}=-\frac{v_6}{2}\,,\qquad
v_{12}=-\frac{v_6}{2h}\,,
\notag\\
&&
v_{13} \equiv -\frac{2h^{3/2}\phi''(h\phi'^2f_{4,XX}+r\phi'f_{3,X}-f_{4,X}-2\tilde{f}_4)}{r\sqrt{f}}
+\frac{2\sqrt{h}A_0''[h\phi'(f_{4,X}+2 \tilde{f}_4)+rf_3]}{r\sqrt{f}} \notag\\
&&
\hspace{.75cm}
-\frac{\sqrt{h}f'A_0'[h\phi'(f_{4,X}+2\tilde{f}_4)+rf_3]}{rf^{3/2}} 
-\frac{h'A_0'[h^2\phi'^3f_{4,XX}+rh\phi'^2f_{3,X}-3h\phi'(f_{4,X}+2\tilde{f}_4)-rf_3]}{r\sqrt{fh}}
\notag\\
&&
\hspace{.75cm}
-\frac{2\sqrt{h}A_0'[h\phi'^2(\tilde{f}_3-2\tilde{f}_{4,\phi}-f_{4,\phi X})+r\phi'(f_{2,Y}-f_{3,\phi})-f_3]}{r\sqrt{f}}\,, 
\label{a2v}
\ea
where ${\cal E}_{11}$ and ${\cal E}_{\phi}$ are the background equations given in Eqs.~(\ref{be1}) and  (\ref{be3}), respectively, 
and $\alpha_i$ $(i=1,2,4,5,6,7)$ are the coefficients in the second-order action for the odd-parity perturbations defined in Eq.~(\ref{alphai}).
Interested readers may check them out in \cite{supplemental}.

On the other hand, it's worth mentioning here that, the coefficient $v_6$ adopts a useful feature that 
\ba
v_6' =  \frac{a_4 \left[2 f^2 \left(r h'+2 h\right)+h r^2 \left(f'\right)^2-f r \left(r f' h'+2 h \left(r f''+f'\right)\right)\right]-f h r^2 \left[A_0' \left(4 v_{10} A_0'+v_6 f'\right)+f v_6 A_0''+8 \alpha _7 f^2\right]}{f^2 h r^2 A_0'}\,. \notag\\
\label{v6prime}
\ea
Notice that, for the economy of notations, here we are writing $\bar \phi$ and $\bar A_0$ 
appearing in Eqs.~\eqref{defdphi} and \eqref{defdA} simply as $\phi$ and $A_0$, respectively. 
It should stimulate no confusions since all the above coefficients are for the perturbation terms 
and they themselves are, definitely, of the zeroth order. 

\section{Expressions for certain quantities appearing in Sec.\ref{application}}
\label{secVexpression}

The dimensionless form of $\mathcal{M}_1$  for Model 1 (cf. Sec.\ref{applicationA}) is given below
\ba
\mathcal{M}_1 &=& \bigg\{  2 f^{37/2} \mathfrak{Z}^2 \xi ^6 \left(f (78 h+4)-h \mathfrak{Z}^2 \xi ^2\right) \tilde{\beta }_3^2 h^{3/2}+56623104 f^{15/2} \mathfrak{Z}^{24} \xi ^{64} \left(h \mathfrak{Z}^2 \xi ^2+8 f (4 h-1)\right) \tilde{\beta }_3^{20} h^{39/2} \notag\\
&& +679477248 f^{13/2} \mathfrak{Z}^{28} \xi ^{72} \tilde{\beta }_3^{22} h^{45/2}   +1179648 f^{17/2} \mathfrak{Z}^{20} \xi ^{56} \Big[16 \left(5 \sqrt{f h^3}-40 \sqrt{f h^5}+53 \sqrt{f h^7}\right) f^{3/2}  \notag\\
&&  -43 h^{7/2} \mathfrak{Z}^4 \xi ^4+32 h \mathfrak{Z}^2 \xi ^2 \sqrt{f} \left(11 \sqrt{f h^5}-2 \sqrt{f h^3}\right)\Big] \tilde{\beta }_3^{18} h^{15}  \notag\\
&&  +393216 f^{19/2} \mathfrak{Z}^{16} \xi ^{48} \Big[4 \mathfrak{Z}^2 \xi ^2 \left(11 \sqrt{f h^3}-100 \sqrt{f h^5}+233 \sqrt{f h^7}\right) f^{3/2}+16 \left(19 h^3-30 h^2+12 h-1\right) \sqrt{h} f^3   \notag\\
&&  -25 h^{7/2} \mathfrak{Z}^6 \xi ^6+h \mathfrak{Z}^4 \xi ^4 \sqrt{f} \left(29 \sqrt{f h^3}-41 \sqrt{f h^5}\right) \Big] \tilde{\beta }_3^{16} h^{13}  \notag\\
&&  +2048 f^{21/2} \mathfrak{Z}^{14} \xi ^{42} \Big[ 16 \mathfrak{Z}^2 \xi ^2 \left(-31 \sqrt{f h^3}+20 \sqrt{f h^5}+1865 \sqrt{f h^7}\right) f^{3/2}  \notag\\
&&  +256 \left(155 h^3-123 h^2+24 h-2\right) \sqrt{h} f^3-57 h^{7/2} \mathfrak{Z}^6 \xi ^6+16 h \mathfrak{Z}^4 \xi ^4 \sqrt{f} \left(79 \sqrt{f h^3}-376 \sqrt{f h^5}\right)\Big] \tilde{\beta }_3^{14} h^{11}   \notag\\
&&  +32 f^{25/2} \mathfrak{Z}^{10} \xi ^{30} \Big[-16 f^2 \left(577 h^2-64 h+17\right) \mathfrak{Z}^2 \xi ^2 h^{3/2}-128 f (32 h+5) \mathfrak{Z}^4 \xi ^4 h^{5/2}+351 \mathfrak{Z}^6 \xi ^6 h^{7/2}  \notag\\
&&  +512 f^{5/2} \left(\sqrt{f h}+\sqrt{f h^3}-78 \sqrt{f h^5}+295 \sqrt{f h^7}\right)\Big] \tilde{\beta }_3^{10} h^7   \notag\\
&& + 16 f^6 \mathfrak{Z}^8 \xi ^{24} \Big[-8 f^{19/2} \left(797 h^2+12 h-5\right) \mathfrak{Z}^2 \xi ^2 h^{3/2}-2 f^{17/2} (279 h+41) \mathfrak{Z}^4 \xi ^4 h^{5/2}+27 f^{15/2} \mathfrak{Z}^6 \xi ^6 h^{7/2} \notag\\
&&  +32 \left(f^{21/2} \left(1197 h^2+4 h+30\right) h^{3/2}+\sqrt{f^{21} h}\right)\Big] \tilde{\beta }_3^8 h^5  \notag\\
&&  +2 f^5 \mathfrak{Z}^6 \xi ^{18} \Big[ 3 f^{19/2} \mathfrak{Z}^6 \xi ^6 h^{5/2}-4 \left(46 f^{21/2} h^{3/2}+3 \sqrt{f^{21} h}\right) \mathfrak{Z}^4 \xi ^4 h  \notag\\
&&  +64 f^2 \left(f^{21/2} (489 h+83) h^{3/2}+6 \sqrt{f^{21} h}\right)+4 f \mathfrak{Z}^2 \xi ^2 \left(f^{21/2} (20-981 h) h^{3/2}+3 \sqrt{f^{21} h}\right)\Big] \tilde{\beta }_3^6 h^4  \notag\\
&&  +512 \mathfrak{Z}^{12} \xi ^{36} \Big[ 64 f^{29/2} \left(802 h^3-369 h^2-3 h+2\right) h^{19/2}+219 f^{23/2} \mathfrak{Z}^6 \xi ^6 h^{25/2}  \notag\\
&&  -32 f^{13} \left(19 \sqrt{f h^3}-90 \sqrt{f h^5}-298 \sqrt{f h^7}\right) \mathfrak{Z}^2 \xi ^2 h^9+8 \left(4 f^{25/2} h^{23/2}-381 (f h)^{25/2}\right) \mathfrak{Z}^4 \xi ^4\Big] \tilde{\beta }_3^{12}  \notag\\
&&  +\left(4 f^{35/2} (2-57 h) \mathfrak{Z}^6 \xi ^{14} h^{7/2}-6 f^{33/2} \mathfrak{Z}^8 \xi ^{16} h^{9/2}+8 f^8 \left(f^{21/2} (552 h+71) h^{3/2}+\sqrt{f^{21} h}\right) \mathfrak{Z}^4 \xi ^{12} h^2\right) \tilde{\beta }_3^4  \notag\\
&&  +2 \sqrt{f^{41} h}  \bigg\} \notag\\
&& \times \bigg\{ 339738624 f^{17/2} h^{41/2} \xi ^{64} \mathfrak{Z}^{24} \tilde{\beta }_3^{20}+2 f^{37/2} h^{3/2} \xi ^6 \mathfrak{Z}^2 \tilde{\beta }_3^2 \left(f (38 h+2)-h \xi ^2 \mathfrak{Z}^2\right)  \notag\\
&&  +28311552 f^{19/2} h^{35/2} \xi ^{56} \mathfrak{Z}^{20} \tilde{\beta }_3^{18} \left(4 f (h-1)+3 h \xi ^2 \mathfrak{Z}^2\right)  \notag\\
&&  +1024 \xi ^{36} \mathfrak{Z}^{12} \tilde{\beta }_3^{12} \Big[16 f^{29/2} \left(319 h^2-158 h-17\right) h^{21/2}+16 f^{13} h^{11} \xi ^2 \mathfrak{Z}^2 \left(139 \sqrt{f h^3}+35 \sqrt{f h}\right)  \notag\\
&&  -149 \xi ^4 \mathfrak{Z}^4 (f h)^{25/2}\Big]+589824 f^{21/2} h^{29/2} \xi ^{48} \mathfrak{Z}^{16} \tilde{\beta }_3^{16} \left(16 f^2 (h-1)^2+32 f h (4 h-1) \xi ^2 \mathfrak{Z}^2-5 h^2 \xi ^4 \mathfrak{Z}^4\right)  \notag\\
&&  +196608 f^{23/2} h^{25/2} \xi ^{42} \mathfrak{Z}^{14} \tilde{\beta }_3^{14} \left(f^2 \left(52 h^2-56 h+4\right)+f h (107 h+13) \xi ^2 \mathfrak{Z}^2-12 h^2 \xi ^4 \mathfrak{Z}^4\right)  \notag\\
&&  +f^7 h^2 \xi ^{12} \mathfrak{Z}^4 \tilde{\beta }_3^4 \left(f^{19/2} h^{5/2} \xi ^4 \mathfrak{Z}^4-4 h \xi ^2 \mathfrak{Z}^2 \left(37 f^{21/2} h^{3/2}+\sqrt{f^{21} h}\right)+4 f \left(f^{21/2} (561 h+62) h^{3/2}+\sqrt{f^{21} h}\right)\right)  \notag\\
&&   +8 f^6 h^4 \xi ^{18} \mathfrak{Z}^6 \tilde{\beta }_3^6 \left(9 f^{19/2} h^{5/2} \xi ^4 \mathfrak{Z}^4+8 f^{23/2} (527 h+78) h^{3/2}+24 \sqrt{f^{23} h}-6 h \xi ^2 \mathfrak{Z}^2 \left(83 f^{21/2} h^{3/2}+5 \sqrt{f^{21} h}\right)\right)  \notag\\
&&   +16 f^5 h^6 \xi ^{24} \mathfrak{Z}^8 \tilde{\beta }_3^8 \Big[109 f^{19/2} h^{5/2} \xi ^4 \mathfrak{Z}^4+48 f^{23/2} (381 h+38) h^{3/2}+112 \sqrt{f^{23} h}  \notag\\
&&   -32 h \xi ^2 \mathfrak{Z}^2 \left(87 f^{21/2} h^{3/2}+8 \sqrt{f^{21} h}\right)\Big] \notag\\
&&    +256 f^4 h^8 \xi ^{30} \mathfrak{Z}^{10} \tilde{\beta }_3^{10} \Big[45 f^{19/2} h^{5/2} \xi ^4 \mathfrak{Z}^4-32 f^{11} \left(-191 \sqrt{f h^5}+20 \sqrt{f h^3}+3 \sqrt{f h}\right)  \notag\\
&&   +4 h \xi ^2 \mathfrak{Z}^2 \left(3 \sqrt{f^{21} h}-91 f^{21/2} h^{3/2}\right)\Big] +\sqrt{f^{41} h} \bigg\}^{-1} \,,
\label{aLap2M1} 
\ea
where
\ba
\mathfrak{Z}(\xi) &\equiv& r_h^{-1} \frac{d A_0}{d \xi} \notag\\
&=& \frac{\sqrt[3]{6} \left[\sqrt{6} \sqrt{f^3 h^9 \left(54 h \kappa ^2 \xi ^6 \tilde{\beta }_3^2+1\right)}-18 f^{3/2} h^5 \kappa  \xi ^3 \tilde{\beta }_3\right]^{2/3}-6^{2/3} f h^3}{12 h^{5/2} \xi ^3 \tilde{\beta }_3 \left[{\sqrt{6} \sqrt{f^3 h^9 \left(54 h \kappa ^2 \xi ^6 \tilde{\beta }_3^2+1\right)}-18 f^{3/2} h^5 \kappa  \xi ^3 \tilde{\beta }_3}\right]^{1/3}}.
\label{goZ} 
\ea

The dimensionless form of $\mathcal{M}_2$ at the $r \to r_h$ limit for Model 2 (cf. Sec.\ref{applicationB}) is given below
\ba
\left. \mathcal{M}_2 \right|_{r \to r_h}  &=&  \Big\{  \left(1-4 \tilde{\beta }_4\right)^2 \left(8 \tilde{\beta }_4+1\right)^{18}-4 \kappa ^2 \tilde{\beta }_4 \left(4 \tilde{\beta }_4-1\right) \left(128 \tilde{\beta }_4^2+32 \tilde{\beta }_4-1\right) \left(8 \tilde{\beta }_4+1\right)^{15}   \notag\\
&&  -64 \kappa ^6 \tilde{\beta }_4^3 \left(8 \tilde{\beta }_4+1\right)^{14}+12 \kappa ^4 \tilde{\beta }_4^2 \left(128 \tilde{\beta }_4^2+16 \tilde{\beta }_4-5\right) \left(8 \tilde{\beta }_4+1\right)^{14}  \notag\\
&&  +64 \tilde{\beta }_4^4 \left(8 \tilde{\beta }_4 \kappa +\kappa \right)^8 \left(8 \tilde{\beta }_4+1\right)^4  \notag\\
&&  -2 \kappa ^8 \tilde{\beta }_3^6 \left(\kappa ^2-16 \tilde{\beta }_4-2\right)^2 \left(8 \tilde{\beta }_4+1\right)^4 \notag\\
&&  \times  \left[155648 \tilde{\beta }_4^4-1024 \left(8 \kappa ^2-11\right) \tilde{\beta }_4^3+16 \left(7 \kappa ^4+12 \kappa ^2-180\right) \tilde{\beta }_4^2+8 \left(19 \kappa ^2-4\right) \tilde{\beta }_4+25\right]   \notag\\
&&  +8 \kappa ^{10} \tilde{\beta }_3^8 \left(\kappa ^2-16 \tilde{\beta }_4-2\right)^2  \notag\\
&&  \times \left[20 \tilde{\beta }_4 \kappa ^4+\left(-1088 \tilde{\beta }_4^2-64 \tilde{\beta }_4+9\right) \kappa ^2+24 \left(8 \tilde{\beta }_4+1\right)^2 \left(10 \tilde{\beta }_4-1\right)\right] \left(8 \tilde{\beta }_4+1\right)^2  \notag\\
&&  +2 \tilde{\beta }_3^2 \Big[-2 \kappa ^2 \left(4 \tilde{\beta }_4-1\right) \left(8 \tilde{\beta }_4+1\right)^{14}+\kappa ^4 \left(3072 \tilde{\beta }_4^4+3840 \tilde{\beta }_4^3-400 \tilde{\beta }_4^2-40 \tilde{\beta }_4-1\right) \left(8 \tilde{\beta }_4+1\right)^{10}  \notag\\
&&  -64 \kappa ^{10} \tilde{\beta }_4^3 \left(8 \tilde{\beta }_4+1\right)^8-2 \kappa ^6 \tilde{\beta }_4 \left(4096 \tilde{\beta }_4^4+15360 \tilde{\beta }_4^3+1216 \tilde{\beta }_4^2-120 \tilde{\beta }_4-5\right) \left(8 \tilde{\beta }_4+1\right)^8  \notag\\
&&  +64 \kappa ^{12} \tilde{\beta }_4^4 \left(8 \tilde{\beta }_4+1\right)^6+4 \tilde{\beta }_4^2 \left(8 \tilde{\beta }_4 \kappa +\kappa \right)^8 \left(256 \tilde{\beta }_4^2+272 \tilde{\beta }_4-1\right)\Big] \left(8 \tilde{\beta }_4+1\right)^2  \notag\\
&&  -32 \kappa ^{12} \tilde{\beta }_3^{10} \left(\kappa ^4-5 \left(8 \tilde{\beta }_4+1\right) \kappa ^2+6 \left(8 \tilde{\beta }_4+1\right)^2\right)^2  \notag\\
&&   +\tilde{\beta }_3^4 \Big[ 4 \kappa ^4 \left(8 \tilde{\beta }_4+1\right)^{14}+4 \kappa ^6 \left(4608 \tilde{\beta }_4^3-1792 \tilde{\beta }_4^2+96 \tilde{\beta }_4-5\right) \left(8 \tilde{\beta }_4+1\right)^{10}  \notag\\
&&  -16 \kappa ^{12} \tilde{\beta }_4^2 \left(72 \tilde{\beta }_4-23\right) \left(8 \tilde{\beta }_4+1\right)^7+512 \kappa ^{10} \tilde{\beta }_4^3 \left(1216 \tilde{\beta }_4^2-384 \tilde{\beta }_4+15\right) \left(8 \tilde{\beta }_4+1\right)^6 \notag\\
&&  +128 \kappa ^{10} \tilde{\beta }_4^2 \left(1216 \tilde{\beta }_4^2-384 \tilde{\beta }_4+15\right) \left(8 \tilde{\beta }_4+1\right)^6 +8 \kappa ^{10} \tilde{\beta }_4 \left(1216 \tilde{\beta }_4^2-384 \tilde{\beta }_4+15\right) \left(8 \tilde{\beta }_4+1\right)^6   \notag\\
&&  +\left(8 \tilde{\beta }_4 \kappa +\kappa \right)^8 \left(-196608 \tilde{\beta }_4^4+39936 \tilde{\beta }_4^3+4416 \tilde{\beta }_4^2-384 \tilde{\beta }_4+9\right)\Big]  \Big\} \notag\\
&& \times \Big\{ \left(8 \tilde{\beta }_4+1\right)^{14} \left(-4 \left(\kappa ^2-4\right) \tilde{\beta }_4+64 \tilde{\beta }_4^2+1\right) \left[2 \left(\kappa ^2+2\right) \tilde{\beta }_4-32 \tilde{\beta }_4^2+1\right]^2   \notag\\
&& +\kappa ^4 \tilde{\beta }_3^4 \left(8 \tilde{\beta }_4+1\right)^{10} \left(-16 \tilde{\beta }_4+\kappa ^2-2\right)^2 \left[-4 \left(\kappa ^2-4\right) \tilde{\beta }_4+64 \tilde{\beta }_4^2+1\right]   \notag\\
&&  -2 \tilde{\beta }_3^2 \left(8 \tilde{\beta }_4+1\right)^{10} \left(8 \kappa  \tilde{\beta }_4+\kappa \right)^2 \left[-4 \left(\kappa ^2-4\right) \tilde{\beta }_4+64 \tilde{\beta }_4^2+1\right]   \notag\\
&& \times  \left(2 \kappa ^4 \tilde{\beta }_4-64 \kappa ^2 \tilde{\beta }_4^2+512 \tilde{\beta }_4^3-24 \tilde{\beta }_4+\kappa ^2-2\right) \Big\}^{-1}\,.
\label{aLap3M2}
\ea
Notice that, \eqref{aLap3M2} will reduce to \eqref{aLap3M1} by taking ${\tilde \beta}_4=0$.

The dimensionless form of $\mathcal{M}_2$ at the $r \to r_h$ limit for Model 3 (cf. Sec.\ref{applicationC}) is given below
\ba
\left. \mathcal{M}_2 \right|_{r \to r_h}  &=& \frac{1}{2}  \Big\{ 16 \tilde{\beta }_3^6 \tilde{\beta }_4^3 \left(20 \tilde{\beta }_3^2-31 \tilde{\beta }_4\right) \kappa ^{24}-8 \tilde{\beta }_3^4 \tilde{\beta }_4^2 \left(4 \left(76 \tilde{\beta }_4-17\right) \tilde{\beta }_3^4+\tilde{\beta }_4 \left(31-140 \tilde{\beta }_4\right) \tilde{\beta }_3^2-368 \tilde{\beta }_4^3\right) \kappa ^{22}   \notag\\
&&  -\tilde{\beta }_3^2 \Big[64 \left(1-4 \tilde{\beta }_4\right)^2 \tilde{\beta }_3^8-4 \tilde{\beta }_4 \left(1216 \tilde{\beta }_4^2-752 \tilde{\beta }_4+29\right) \tilde{\beta }_3^6+\tilde{\beta }_4^2 \left(-2624 \tilde{\beta }_4^2+1904 \tilde{\beta }_4+31\right) \tilde{\beta }_3^4   \notag\\
&&  +32 \tilde{\beta }_4^4 \left(248 \tilde{\beta }_4-69\right) \tilde{\beta }_3^2+4864 \tilde{\beta }_4^6\Big] \kappa ^{20}   \notag\\
&&  +2 \tilde{\beta }_3^2 \Big[64 \left(32 \tilde{\beta }_4^2-28 \tilde{\beta }_4+5\right) \tilde{\beta }_3^8-4 \tilde{\beta }_4 \left(64 \tilde{\beta }_4^2-304 \tilde{\beta }_4+3\right) \tilde{\beta }_3^6
  \notag\\
&& 
-7 \tilde{\beta }_4^2 \left(576 \tilde{\beta }_4^2-720 \tilde{\beta }_4+83\right) \tilde{\beta }_3^4-4 \tilde{\beta }_4^3 \left(832 \tilde{\beta }_4^2+192 \tilde{\beta }_4-69\right) \tilde{\beta }_3^2-128 \tilde{\beta }_4^5 \left(14 \tilde{\beta }_4+19\right)\Big] \kappa ^{18}
  \notag\\
&& -2 \Big[32 \left(64 \tilde{\beta }_4^2-128 \tilde{\beta }_4+37\right) \tilde{\beta }_3^{10}+8 \left(256 \tilde{\beta }_4^3-560 \tilde{\beta }_4^2+221 \tilde{\beta }_4-9\right) \tilde{\beta }_3^8   \notag\\
&& 
+\tilde{\beta }_4 \left(-2304 \tilde{\beta }_4^3+4768 \tilde{\beta }_4^2-1474 \tilde{\beta }_4+77\right) \tilde{\beta }_3^6-\tilde{\beta }_4^2 \left(9728 \tilde{\beta }_4^3-9408 \tilde{\beta }_4^2+912 \tilde{\beta }_4+23\right) \tilde{\beta }_3^4 
  \notag\\
&& 
+16 \tilde{\beta }_4^4 \left(-576 \tilde{\beta }_4^2+304 \tilde{\beta }_4+57\right) \tilde{\beta }_3^2-2048 \tilde{\beta }_4^7\Big] \kappa ^{16}
  \notag\\
&& 
-4 \Big[192 \left(8 \tilde{\beta }_4-5\right) \tilde{\beta }_3^{10}+8 \left(384 \tilde{\beta }_4^2-335 \tilde{\beta }_4+30\right) \tilde{\beta }_3^8+\tilde{\beta }_4 \left(512 \tilde{\beta }_4^2-786 \tilde{\beta }_4+63\right) \tilde{\beta }_3^6   \notag\\
&& 
-8 \tilde{\beta }_4^2 \left(128 \tilde{\beta }_4^3+720 \tilde{\beta }_4^2-327 \tilde{\beta }_4+22\right) \tilde{\beta }_3^4   \notag\\
&& 
+4 \tilde{\beta }_4^3 \left(-1024 \tilde{\beta }_4^3-1536 \tilde{\beta }_4^2+468 \tilde{\beta }_4+19\right) \tilde{\beta }_3^2-512 \tilde{\beta }_4^6 \left(8 \tilde{\beta }_4+3\right)\Big] \kappa ^{14}
  \notag\\
&& 
+\Big[-2304 \tilde{\beta }_3^{10}-64 \left(140 \tilde{\beta }_4-33\right) \tilde{\beta }_3^8-4 \left(2072 \tilde{\beta }_4^2-714 \tilde{\beta }_4+25\right) \tilde{\beta }_3^6   \notag\\
&& 
+ \tilde{\beta }_4 \left(6144 \tilde{\beta }_4^3+7424 \tilde{\beta }_4^2-1832 \tilde{\beta }_4+69\right) \tilde{\beta }_3^4   \notag\\
&& 
+ \tilde{\beta }_4^2 \left(28672 \tilde{\beta }_4^3+12032 \tilde{\beta }_4^2-2528 \tilde{\beta }_4-19\right) \tilde{\beta }_3^2+256 \tilde{\beta }_4^5 \left(128 \tilde{\beta }_4+15\right)\Big] \kappa ^{12}
  \notag\\
&& 
+\Big[-1536 \tilde{\beta }_3^8+\left(400-3472 \tilde{\beta }_4\right) \tilde{\beta }_3^6+32 \tilde{\beta }_4^2 \left(112 \tilde{\beta }_4-1\right) \tilde{\beta }_3^4 
  \notag\\
&& 
+2 \tilde{\beta }_4^2 \left(10240 \tilde{\beta }_4^2+1216 \tilde{\beta }_4-199\right) \tilde{\beta }_3^2+256 \tilde{\beta }_4^4 \left(108 \tilde{\beta }_4+5\right)\Big] \kappa ^{10}
  \notag\\
&& 
+\left[-400 \tilde{\beta }_3^6+2 \left(512 \tilde{\beta }_4^2-174 \tilde{\beta }_4+9\right) \tilde{\beta }_3^4+24 \tilde{\beta }_4 \left(320 \tilde{\beta }_4^2+3 \tilde{\beta }_4-1\right) \tilde{\beta }_3^2+80 \tilde{\beta }_4^3 \left(160 \tilde{\beta }_4+3\right)\right] \kappa ^8
  \notag\\
&& 
+8 \left[\left(18 \tilde{\beta }_4-5\right) \tilde{\beta }_3^4+5 \tilde{\beta }_4 \left(40 \tilde{\beta }_4-1\right) \tilde{\beta }_3^2+\tilde{\beta }_4^2 \left(440 \tilde{\beta }_4+3\right)\right] \kappa ^6
  \notag\\
&& 
+\left[8 \tilde{\beta }_3^4+4 \left(44 \tilde{\beta }_4-1\right) \tilde{\beta }_3^2+\tilde{\beta }_4 \left(576 \tilde{\beta }_4+1\right)\right] \kappa ^4+\left(8 \tilde{\beta }_3^2+52 \tilde{\beta }_4\right) \kappa ^2+2 \Big\} \notag\\
&& \times\left(4 \kappa ^2 \tilde{\beta }_4+1\right)^{-5} \left[\kappa ^4 \left(-\tilde{\beta }_3^2\right)+2 \kappa ^2 \left(\tilde{\beta }_3^2+2 \tilde{\beta }_4\right)+1\right]^{-2}.
\label{aLap3M3}
\ea
Notice that, \eqref{aLap3M3} will reduce to \eqref{aLap3M1} by taking ${\tilde \beta}_4=0$. 

Interested readers can check out the mathematical expressions appearing in here with \cite{supplemental}.
\section{Supplemental materials for the numerical analysis in Sec.\ref{application}}
\label{plotsanalysis}

\begin{figure}[h]
	\includegraphics[width=0.6\linewidth]{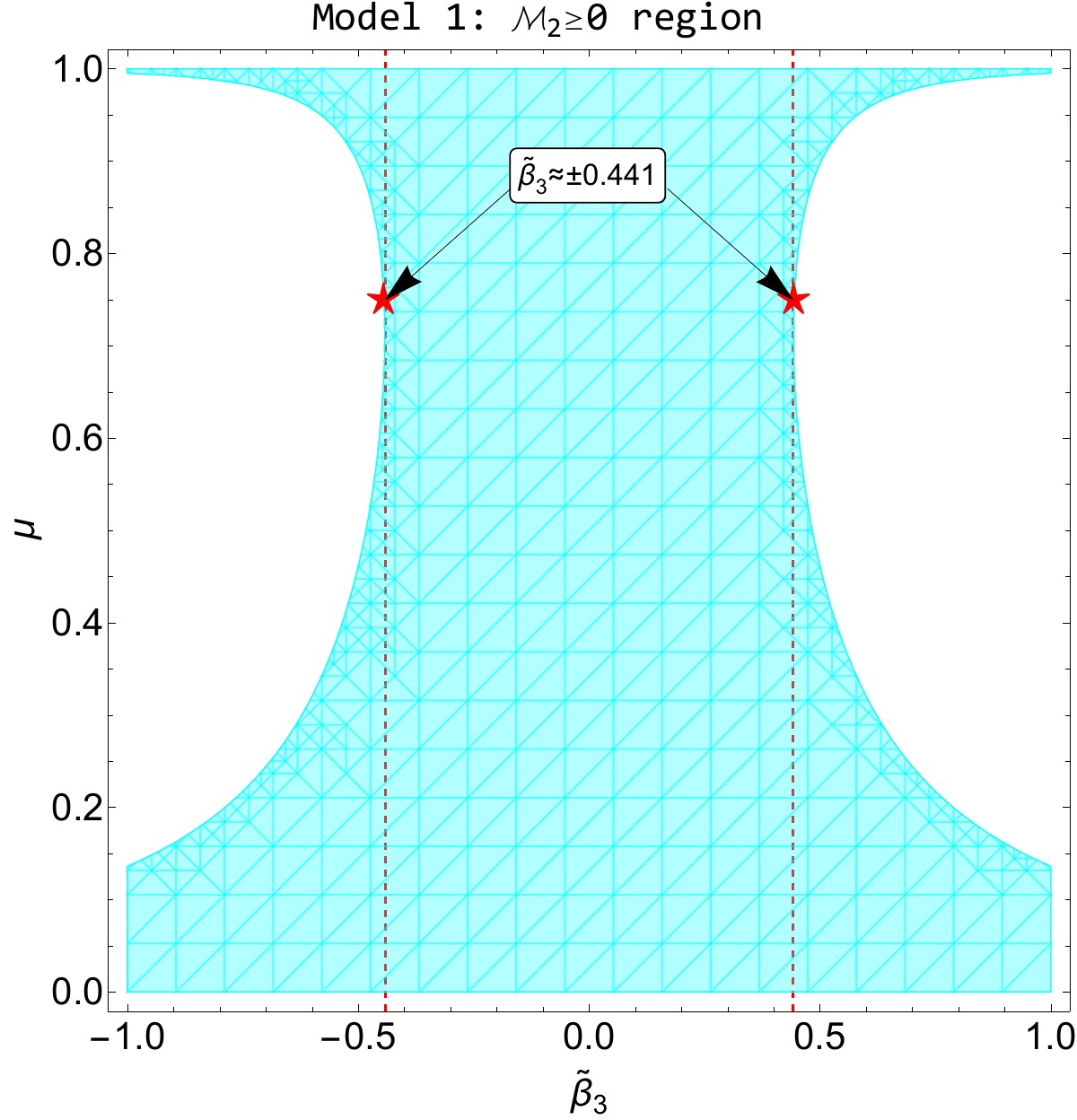} 
\caption{The behavior of $\mathcal{M}_2$ in phase space $\{\mu, {\tilde \beta}_3\}$ at the $r \to r_h$ limit for Model 1 [cf., \eqref{aLap3M1}]. Here the cyan shadowed region is where the second angular Laplacian stability condition gets satisfied, viz., $\mathcal{M}_2 \geq 0$. According to the contour of this shadowed region, we have marked the upper bound of allowed $|   {\tilde \beta}_3 |$ by red dashed vertical lines (which are tangent to the boundary of the shadowed region and are located at $|   {\tilde \beta}_3 | \approx 0.441$) and the red solid pentagrams. } 
	\label{plot1}
\end{figure}

\begin{figure}[h]
  	\begin{tabular}{cc}
  		\includegraphics[width=0.45\linewidth]{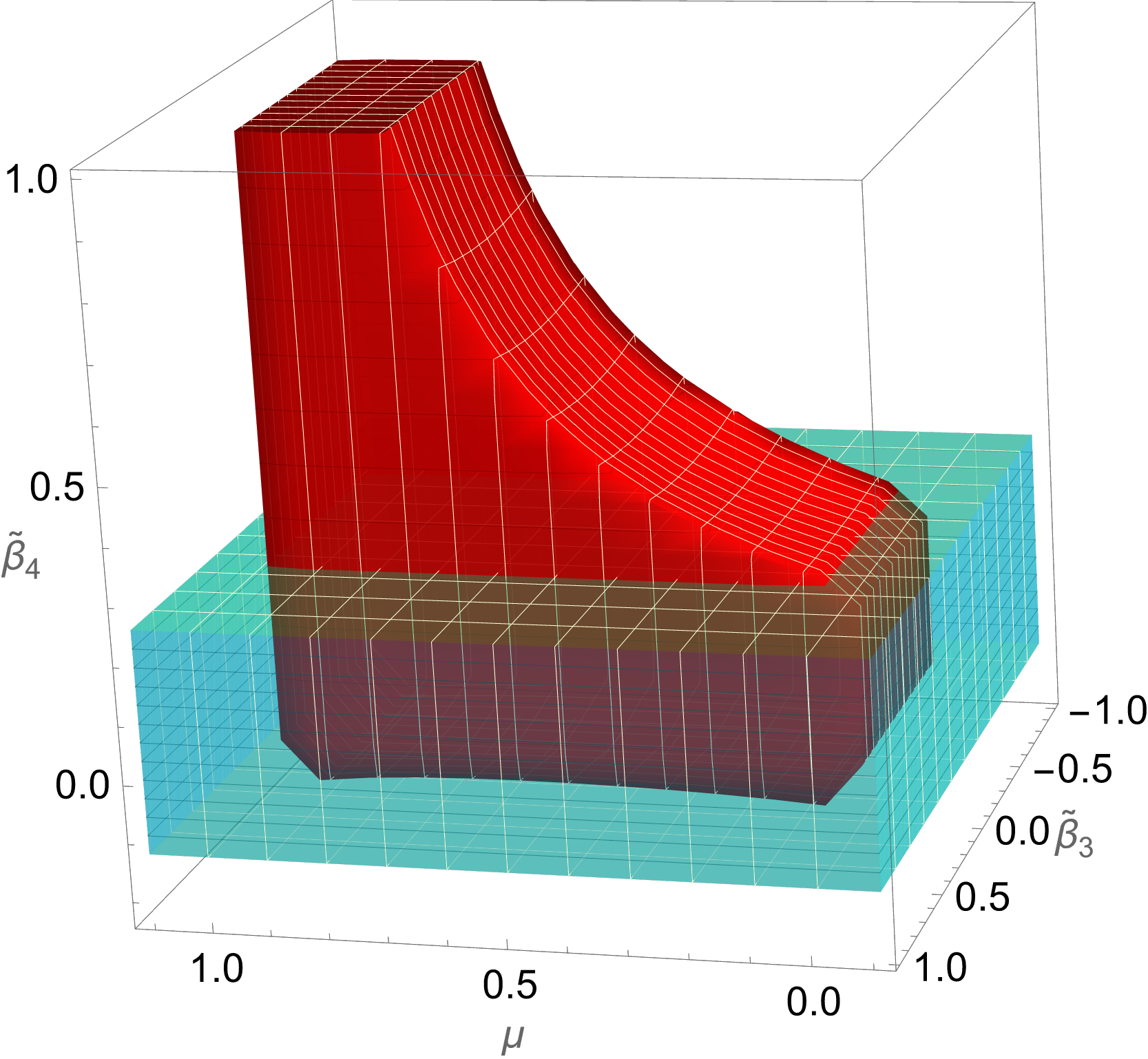} &   \includegraphics[width=0.45\linewidth]{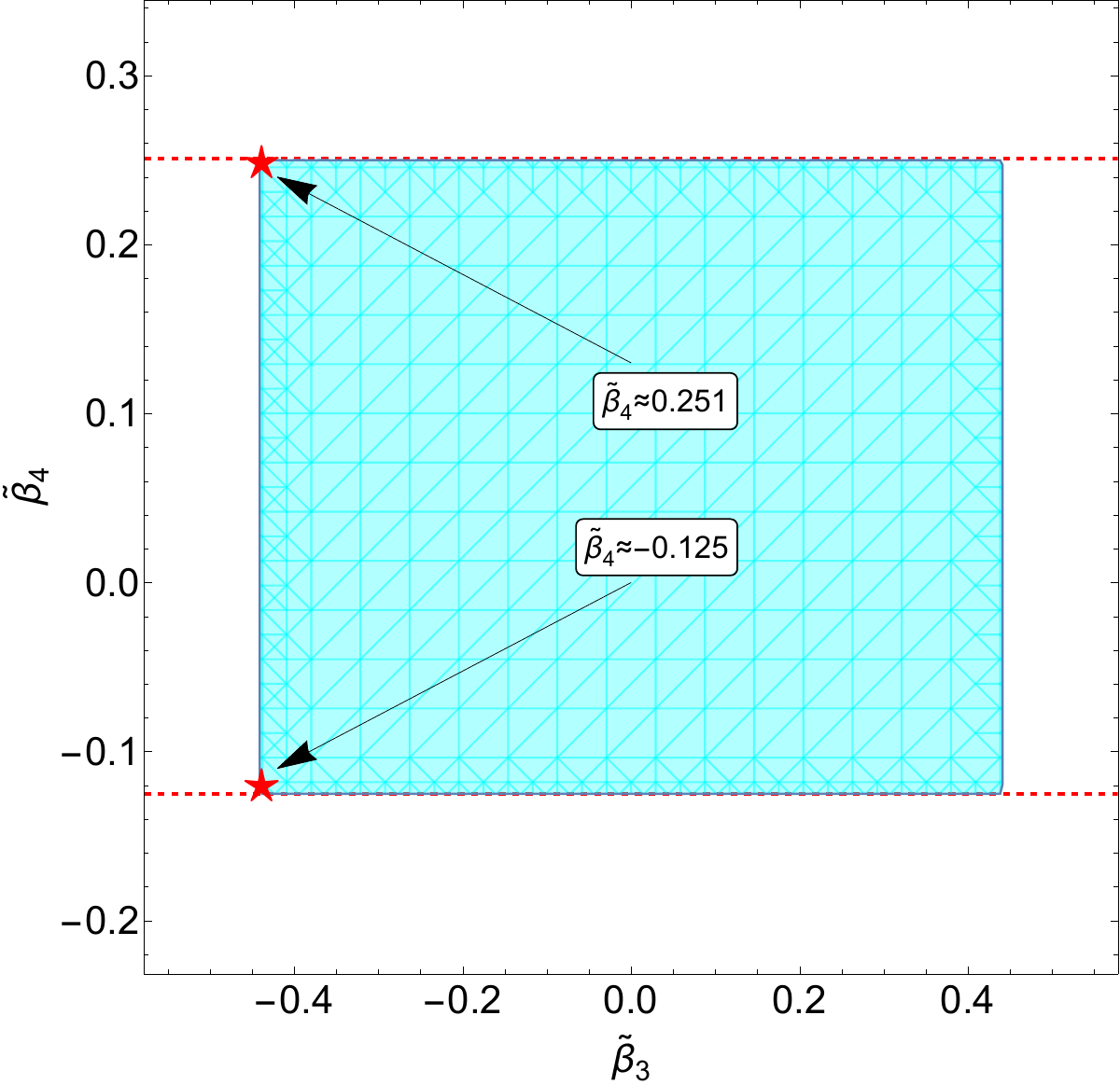} \\
  		(a)  & (b)  \\[6pt]
  	\end{tabular}
\caption{Panel (a): The  parameter phase space spanned by $\{{\tilde \beta}_3, {\tilde \beta}_4, \mu \}$. The red solid part represents the allowed region defined by Eqs.(4.12)-(4.14), (4.20) as well as (4.21) of \cite{Heisenberg:2018mgr} and by considering the constraints $\mu \in (0, 1)$ plus ${\tilde \beta}_3 \in (-0.441, 0.441)$. The cyan semi-transparent part indicates the bounds of ${\tilde \beta}_4$, within which the $\{{\tilde \beta}_3, {\tilde \beta}_4\}$ phase space is covered by the allowed region for arbitrary legal $\mu$, as required by the theory. Panel (b):  The  parameter phase space $\{{\tilde \beta}_3, {\tilde \beta}_4 \}$ by setting $\mu=0$. Here the cyan shadowed area represents the allowed region. According to the contour of the shadowed region, we have marked the upper and lower bounds of allowed ${\tilde \beta}_4$ by red dashed horizontal lines (which are tangent to the boundary of the shadowed region and are located at ${\tilde \beta}_4 \approx 0.251$ and ${\tilde \beta}_4 \approx -0.125$) and the red solid pentagrams.} 
	\label{plotbeta4}
\end{figure}

\begin{figure}[h]
  	\begin{tabular}{cc}
  		\includegraphics[width=0.5\linewidth]{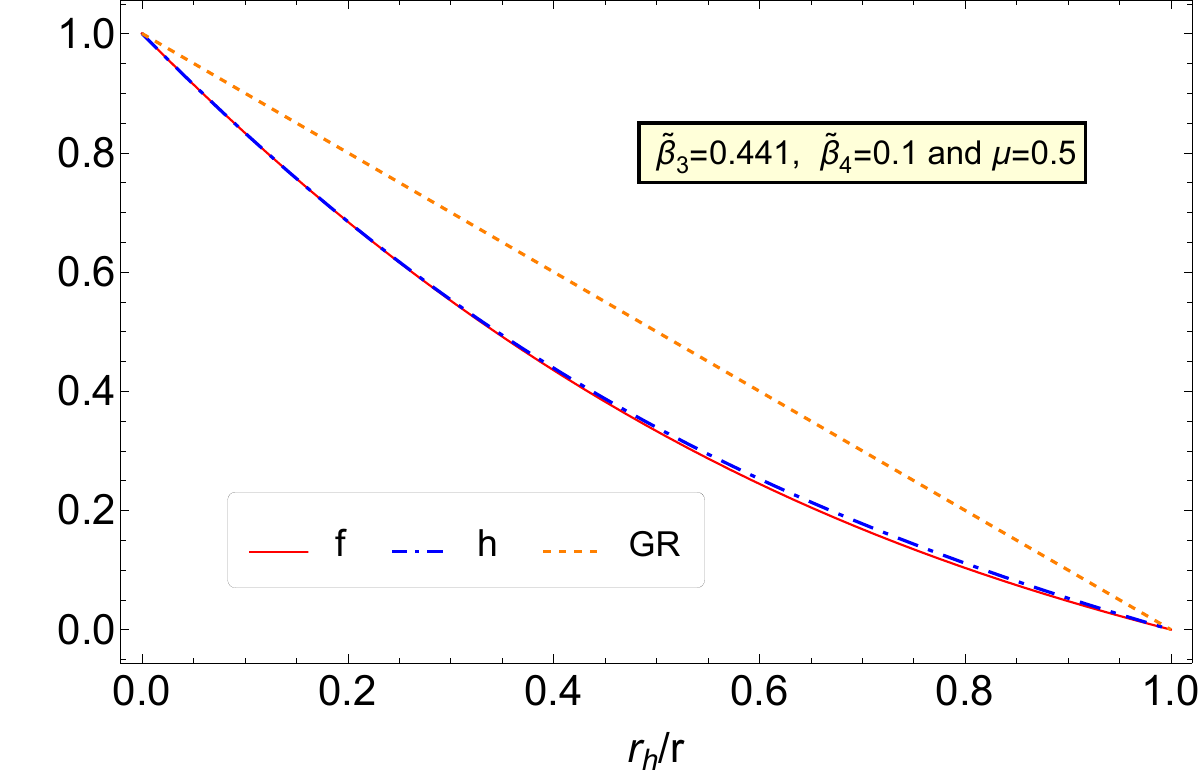} &   \includegraphics[width=0.5\linewidth]{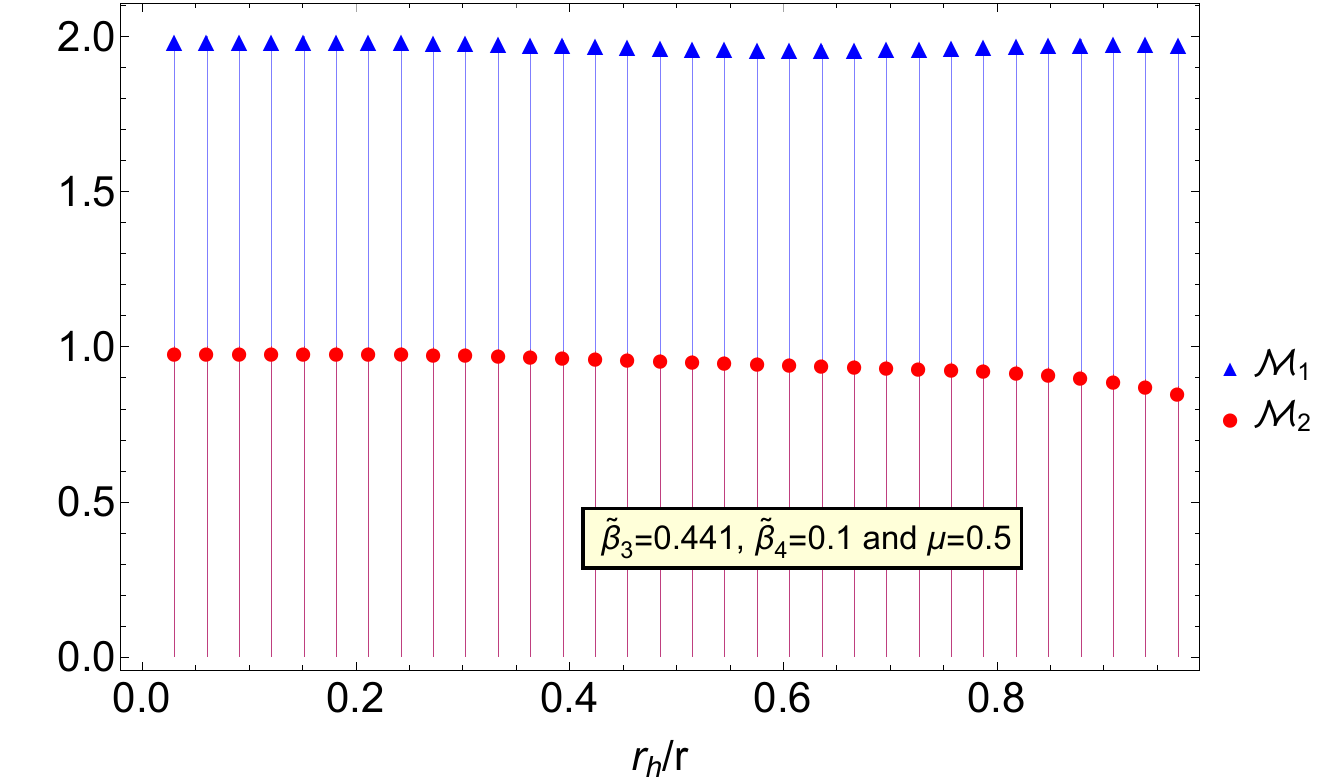} \\
  		(a)  & (b)  \\[6pt]
  	\end{tabular}
\caption{Panel (a): The numerical solutions for $f$ (red solid line) and $h$ (blue dot-dashed line) in Model 2, together with their GR limit (orange dotted line) as comparison (in GR we have $f=h$). Panel (b): The behavior of $\mathcal{M}_{1}$ and $\mathcal{M}_{2}$ in Model 2.  Notice that, the results for $\mathcal{M}_{1}$ are marked by blue triangles while those for $\mathcal{M}_{2}$ are marked by red dots.  These results are plotted as functions of $r_h/r$ (so that we have a larger scope) by choosing ${\mu}=0.5$, ${\tilde \beta}_3=0.441$ and ${\tilde \beta}_4=0.1$.} 
	\label{plot2}
\end{figure}

\begin{figure}[h]
  	\begin{tabular}{cc}
  		\includegraphics[width=0.5\linewidth]{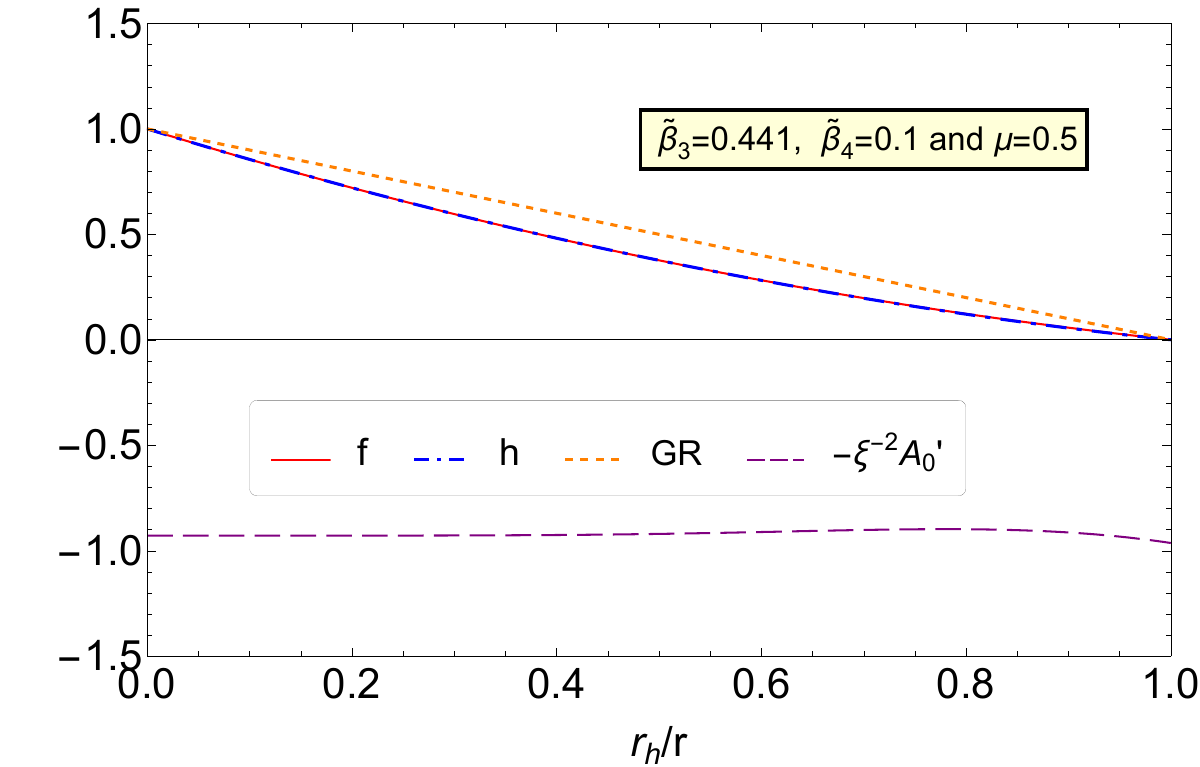} &   \includegraphics[width=0.5\linewidth]{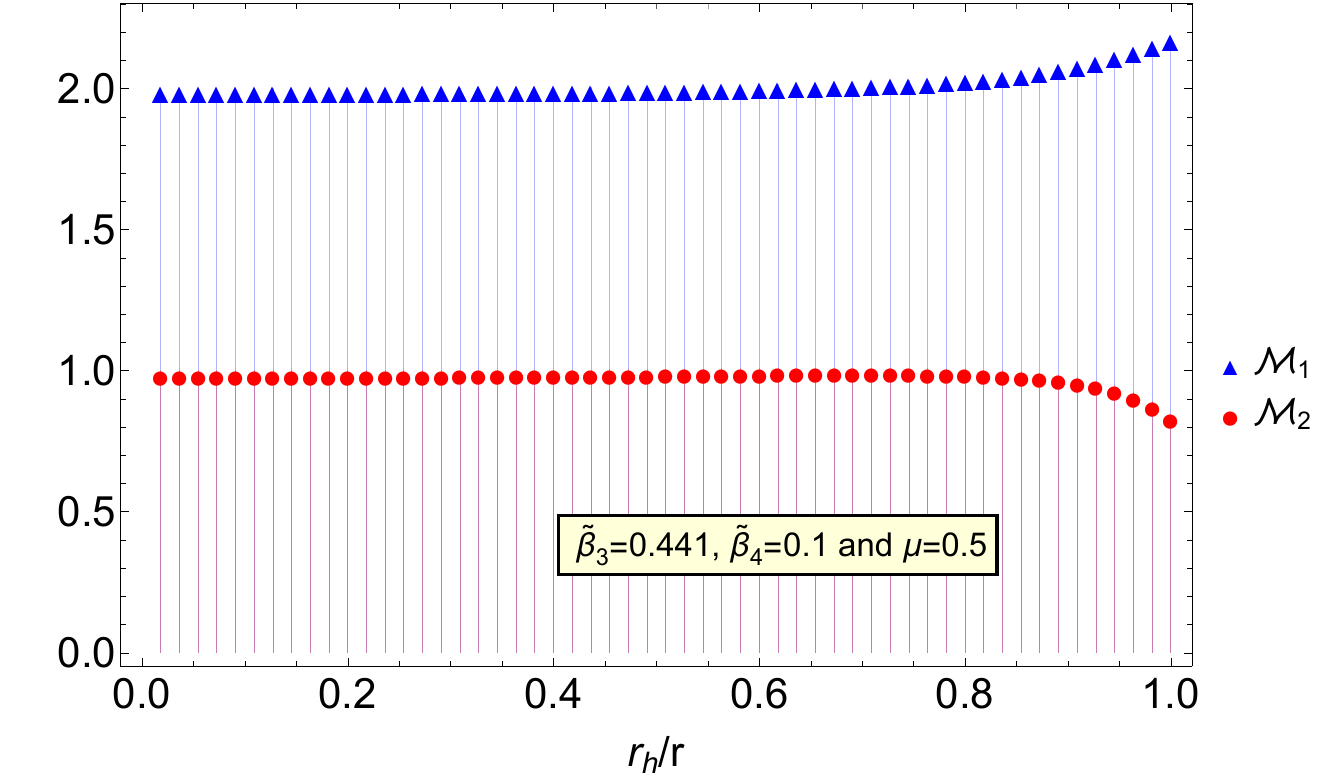} \\
  		(a)  & (b)  \\[6pt]
  	\end{tabular}
\caption{Panel (a): The numerical solutions for $f$ (red solid line), $h$ (blue dot-dashed line) and the dimensionless background quantity $r_h^{-1} dA_0/d\xi=-\xi^{-2} A_0'$ [purple dashed line, cf., \eqref{goZ}] in Model 3. The GR limit for $f$ and $h$ (in GR we have $f=h$) is also added as comparison  (orange dotted line). Panel (b): The behavior of $\mathcal{M}_{1}$ and $\mathcal{M}_{2}$ in Model 3.  Notice that, the results for $\mathcal{M}_{1}$ are marked by blue triangles while those for $\mathcal{M}_{2}$ are marked by red dots.  These results are plotted as functions of $r_h/r$ (so that we have a larger scope) by choosing ${\mu}=0.5$, ${\tilde \beta}_3=0.441$ and ${\tilde \beta}_4=0.1$.} 
	\label{plot4}
\end{figure}

\clearpage

\bibliographystyle{mybibstyle}
\bibliography{SVT_even}

\end{document}